\documentclass[12pt]{article}
\usepackage[utf8]{inputenc}
\usepackage{amsmath,amssymb}
\usepackage{slashed}
\usepackage{graphicx}
\usepackage{cite}
\usepackage{hyperref}
\usepackage{chngcntr}

\counterwithin*{equation}{section}

\begin{document}

\begin{titlepage}

\vspace*{-2cm}
\begin{flushright}
LU TP 20-47 \\
arXiv:2008.13487\\
revised October 2020
\vspace*{2mm}

\end{flushright}

\begin{center}
\vspace*{15mm}

\vspace{1cm}
{\LARGE \bf
Short-distance HLbL contributions to\\ the muon anomalous magnetic moment\\[2mm] beyond perturbation theory
} 
\vspace{1.4cm}

\renewcommand{\thefootnote}{\fnsymbol{footnote}}
{Johan Bijnens$^a$, Nils Hermansson-Truedsson$^b$, Laetitia Laub$^{b}$, Antonio Rodr\'iguez-S\'anchez$^{a}$}
\renewcommand{\thefootnote}{\arabic{footnote}}
\setcounter{footnote}{0}

\vspace*{.5cm}
\centerline{$^a${\it Department of Astronomy and Theoretical Physics,
Lund University,}} 
\centerline{\it S\"olvegatan 14A, SE 223-62 Lund, Sweden}
\centerline{${}^b$ \it Albert Einstein Center for Fundamental Physics, Institute for Theoretical Physics, }
\centerline{{\it Universit\"{a}t Bern, Sidlerstrasse 5, CH–3012 Bern, Switzerland}}

\vspace*{.2cm}

\end{center}

\vspace*{10mm}
\begin{abstract}\noindent\normalsize

The hadronic light-by-light contribution to the muon anomalous magnetic moment depends on an integration over three off-shell momenta squared ($Q_i^2$) of the correlator of four electromagnetic currents and the fourth leg at zero momentum. We derive the short-distance expansion of this correlator in the limit where all three $Q_i^2$ are large and in the Euclidean domain in QCD. This is done via a systematic operator product expansion (OPE) in a background field which we construct. The leading order term in the expansion is the massless quark loop. We also compute the non-perturbative part of the next-to-leading contribution, which is suppressed by quark masses, and the chiral limit part of the next-to-next-to leading contributions to the OPE. We build a renormalisation program for the OPE. The numerical role of the higher-order contributions is estimated and found to be small.
\end{abstract}

\end{titlepage}
\newpage 

\renewcommand{\theequation}{\arabic{section}.\arabic{equation}}

\section{Introduction}

The Standard Model (SM) is the theoretical framework developed to describe particle physics at its most fundamental level, and is able to predict the anomalous magnetic moments of the leptons with a high number of significant digits. At the current level of precision, all the building blocks of the SM leave sizable numerical imprints for the muon anomalous magnetic moment, or, $ a_{\mu} = (g-2)_{\mu}/2 $. Summing all the contributions, one finds \cite{Aoyama:2020ynm}
\begin{equation}
\label{eq:amutheo}
a_{\mu}^{\textrm{SM}}=
116 \, 591 \, 810(43) \times 10^{-11} \, ,
\end{equation}
showing a $3.7\sigma$ tension with the very precise experimental value \cite{Bennett:2006fi,Tanabashi:2018oca},
\begin{equation}
\label{eq:amuexp}
a^{\textrm{exp}}_{\mu}= 116 \, 592\,  091(63) \times 10^{-11} \, .
\end{equation}
The experimental value is expected to be significantly improved~\cite{Grange:2015fou,Abe:2019thb}. In case the discrepancy grows this could be a sign of new physics beyond the SM.

The current uncertainties in the theoretical prediction are dominated by contributions from the hadronic sector. Since the relevant energy scale, i.e.~the muon mass, is far below the region of applicability of perturbative QCD, the assessment of these contributions resorts to the use of non-perturbative tools. Further improvements are needed in order to find a SM value at the level of precision competing with that of the future experimental one. Decreasing the errors on the hadronic contributions would therefore shed some light on whether or not the current tension is a hint of new physics. An overview and assessment of the current theoretical situation is the white-paper~\cite{Aoyama:2020ynm}.

In this paper, we focus on the hadronic light-by-light (HLbL) scattering contribution, represented by the diagram in Fig.~\ref{fig:hlbl}. The calculation of the $(g-2)_{\mu}$ requires the integration of the HLbL tensor over $q_{1}$, $q_{2}$ and $q_{3}$, with the fourth photon in the static limit, i.e.~$q_{4}\rightarrow 0$.
Working with three Euclidean squared loop momenta $q_{i}^2=-Q_{i}^2$, this means one has to consider different kinematic regions of $Q_{i}^2$. We consider the short-distance regime with photon virtualities $Q_{1}^2 \sim Q_{2}^2 \sim Q_{3}^2 \gg \Lambda_{\mathrm{QCD}}^2$, and derive so-called short-distance constraints by means of an operator product expansion (OPE). The second important regime is with mixed virtualities, namely $Q_{3}^2, \Lambda_{\mathrm{QCD}}^2 \ll Q_{1}^2 \sim Q_{2}^{2}$, and has been considered in Ref.~\cite{Melnikov:2003xd}. There has been a lot of recent work in the latter regime Refs.~\cite{Colangelo:2019lpu,Colangelo:2019uex,Melnikov:2019xkq,Leutgeb:2019gbz,Cappiello:2019hwh,Knecht:2020xyr,Masjuan:2020jsf,Ludtke:2020moa}. 

The first full calculations of the HLbL were made using models in Refs.~\cite{Bijnens:1995cc,Bijnens:1995xf,Hayakawa:1997rq}. More recently a dispersion theory based approach as in Refs.~\cite{Colangelo:2015ama,Colangelo:2017fiz} has allowed for better control of the low-energy region. In the latter approach one considers individual intermediate states, for which short-distance constraints such as those derived herein can be used, examples are Refs.~\cite{Bijnens:1995xf,Knecht:2001qf,Colangelo:2019lpu,Colangelo:2019uex,Masjuan:2020jsf,Ludtke:2020moa}. One should of course be careful in comparing at the correct kinematics.

The naive OPE in the vacuum for the HLbL tensor, which is valid for $Q_{1}^2\sim Q_{2}^2\sim Q_{3}^3\sim Q_{4}^2\gg \Lambda_{\mathrm{QCD}}^2$, has the perturbative quark loop as its first term. The quark loop has always been used as an estimate for the whole contribution, using constituent quarks and in various models see e.g.~Ref.~\cite{Kinoshita:1984it,Goecke:2010if,Boughezal:2011vw,Greynat:2012ww,Masjuan:2012qn,Dorokhov:2015psa}
and for the contributions from heavy quarks~\cite{Kuhn:2003pu}.
However, the naive OPE breaks down for the $(g-2)_{\mu}$ kinematics with $q_{4}\rightarrow 0$ \cite{Bijnens:2019ghy}. The OPE of the HLbL tensor in this kinematics must be performed by taking into account that the static photon needs to be formulated as a soft degree of freedom. It was shown in Ref.~\cite{Bijnens:2019ghy} (see also Ref.~\cite{Melnikov:2003xd}) how this could be done by factoring out the soft photon as a background field. The background field can originate either from the hard degrees of freedom (e.g.~the massless quark loop) or the soft ones (e.g.~the so-called di-quark magnetic susceptibility contribution).
The resulting OPE, originally formulated for baryon magnetic moment sum rules in Refs.~\cite{Ioffe:1983ju} and \cite{Balitsky:1983xk}, and whose application to other hadronic $(g-2)_{\mu}$ contributions was introduced in Ref.~\cite{Czarnecki:2002nt}, has the massless quark loop as the leading term and the di-quark magnetic susceptibility of the vacuum as the leading quark-mass suppressed contribution. In this work we extend the results of Ref.~\cite{Bijnens:2019ghy} by computing the leading non-perturbative corrections not suppressed by masses. We also provide some more details about the calculations of Ref.~\cite{Bijnens:2019ghy}. Our result should be useful to help reducing the error coming from the intermediate and short-distance regime~\cite{Aoyama:2020ynm}.

In Sec.~\ref{sec:HLbLtensor} we briefly recapitulate the definitions of the four-point function of four electromagnetic currents, its decomposition in scalar functions and how it can be used for the muon $g-2$ HLbL contribution. This follows the conventions of Refs.~\cite{Colangelo:2015ama,Colangelo:2017fiz}. It also gives the relation between the needed derivative of the four-point function and the three-point function in a constant field background that is used in the remainder of this paper.

In Sec.~\ref{sec:OPEdesc} we give a complete description of the OPE in a constant background field and compare it to the usual vacuum OPE~\cite{Shifman:1978bx} and the one used in flavour-breaking transitions. We in addition comment on the physical meaning of the obtained matrix elements and build a renormalisation program. The renormalisation program is needed to systematically separate short-distance and long-distance effects while cancelling divergences. Both infrared and ultraviolet divergences are addressed. We also estimate the values of the matrix elements. The content of this section can in the future be used to obtain predictions for other Green functions in different kinematic regions.

Details on the calculation of the different non-perturbative pieces are provided in Sec.~\ref{sec:comp}, in particular we explain the different tools used, existing and newly developed, to obtain our analytic results and how the different infrared divergences systematically cancel. Finally, making use of the results of that section and the estimates of the matrix elements, in Sec.~\ref{sec:num} we calculate the numerical contribution of the different pieces for the $(g-2)_{\mu}$. Final remarks and conclusions are made in Sec.~\ref{sec:pros}. Several intermediate derivations are relegated to the appendices as well as the full formulae.

\section{The HLbL tensor}
\label{sec:HLbLtensor}

\begin{figure}[tb]\centering
\includegraphics[width=0.25\textwidth]{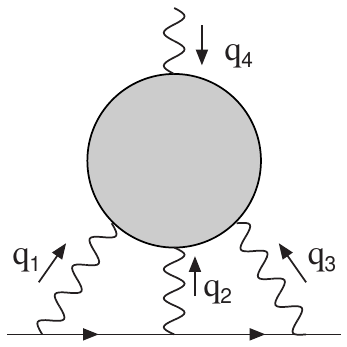}
\caption{\label{fig:hlbl}HLbL contribution to the $(g-2)_{\mu}$.}
\end{figure}

As can be seen in Fig.~\ref{fig:hlbl}, the HLbL process involves a four-point correlation function of electromagnetic quark currents, i.e.~
\begin{equation}\label{eq:hlbltensor}
\Pi^{\mu_{1}\mu_{2}\mu_{3}\mu_{4} } (q_{1},q_{2},q_{3})\equiv-i\int \frac{d^{4}q_{4}}{(2\pi)^{4}}\left(\prod_{i=1}^{4}\int d^{4}x_{i}\, e^{-i q_{i} x_{i}}\right)  \langle 0 | T\left(\prod_{j=1}^{4}J^{\mu_{j}}(x_{j})\right)|0\rangle \, .
\end{equation}
The currents are given by $J^{\mu}(x) = \bar{q}\, Q_{q}\gamma ^{\mu}q$
with the quark fields $q=(u,d,s)$ and
charge matrix $Q_{q}=\textrm{diag}(e_q) =\textrm{diag}(2/3,-1/3,-1/3)$. The tensor in~(\ref{eq:hlbltensor}) is the same as $\Pi ^{\mu \nu \lambda \sigma}$ in  Ref.~\cite{Bijnens:2019ghy}, but with the internal notation slightly changed in the definition in order to make manifest some of its symmetry properties. This will be systematically exploited in the following sections. Moreover, the integral over $q_{4}$ is introduced to remove the $\delta$-function of conservation of momentum\footnote{The integral could be performed instead in any other of the four momenta, leaving the HLbL tensor as a function of the other three.}, i.e.
\begin{equation}\label{eq:consmom}
\sum^{4}_{i=1}q_{i}=0 \, .
\end{equation}
This defines $q_{4}$ as the negative of that in  Ref.~\cite{Bijnens:2019ghy}, which again is a choice to maximise the number of explicit symmetries. 

The conservation of the electromagnetic current implies that the HLbL tensor satisfies the following Ward identities,
\begin{equation}\label{eq:wardiden}
q_{i,\, \mu_{i}} \, \Pi^{\mu_{1}\mu_{2}\mu_{3}\mu_{4}}(q_{1},q_{2},q_{3})=0  , \;\forall i \in [1,4] \, ,
\end{equation}
where $q_{4}$ must be rewritten in terms of the other three momenta through~(\ref{eq:consmom}). Note that the last Ward identity implies that all the information on the HLbL tensor is contained in its derivative \cite{Aldins:1970id},
\begin{equation}\label{eq:wardcons}
\Pi^{\mu_{1}\mu_{2}\mu_{3}\mu_{4}}(q_{1},q_{2},q_{3})=-q_{4,\, \nu_{4}}\frac{\partial \Pi^{\mu_{1}\mu_{2}\mu_{3}\nu_{4}}}{\partial q_{4,\, \mu_{4}}}(q_{1},q_{2},q_{3}) \, .
\end{equation}

In the $(g-2)_{\mu}$ kinematics, the loop integral over the loop momenta $q_{1}, q_{2}$ and $q_{3}$ can be rewritten as an integral over the Euclidean momenta $Q_{i}^2\equiv -q_{i}^2>0$. The fourth photon, i.e.~with momentum $q_{4}$, is taken in the static limit. This corresponds to taking the $q_{4}=-q_{1}-q_{2}-q_{3}\longrightarrow 0$ limit after doing the derivative in~(\ref{eq:wardcons}). Notice how in this limit
\begin{equation}
\label{eq:derPi}
\lim_{q_{4}\rightarrow 0}\frac{\partial \Pi^{\mu_{1}\mu_{2}\mu_{3}\nu_{4}}}{\partial q_{4,\, \mu_{4}}}(q_{1},q_{2},q_{3})=-\lim_{q_{4}\rightarrow 0}\frac{\partial \Pi^{\mu_{1}\mu_{2}\mu_{3}\mu_{4}}}{\partial q_{4,\, \nu_{4}}}(q_{1},q_{2},q_{3}) \, ,
\end{equation}
i.e.~it is anti-symmetric in the indices $\mu_{4}\nu _{4}$. This can be proven by multiplying both sides of~(\ref{eq:wardcons}) by $q_{4,\, \mu_{4}}$, then taking the derivative with respect to $q_{4,\alpha}$ and setting $\alpha = \nu _{4}$. The resulting linear and anti-symmetric structure of the HLbL tensor is directly related to the fact that $F_{\mu\nu}\equiv\partial_{\mu}A_{\nu}-\partial_{\nu}A_{\mu}$ is the lowest dimension gauge invariant photon operator.

Let us take advantage of the work of Refs.~\cite{Colangelo:2015ama,Colangelo:2017fiz} to find general relations between the tensor $\frac{\partial \Pi^{\mu_{1}\mu_{2}\mu_{3}\nu_{4}}}{\partial q_{4,\, \mu_{4}}}$ and its explicit contribution to the $(g-2)_{\mu}$. Making use of the Ward identities above, one can rewrite in full generality the HLbL tensor as a linear combination of $54$ scalar functions $\Pi_{i}(q_1,q_2,q_3)$ according to~\cite{Colangelo:2015ama,Colangelo:2017fiz}
\begin{equation}
  \Pi ^{\mu_{1} \mu_{2} \mu_{3} \mu_{4}}(q_{1},q_{2},q_{3}) = \sum _{i=1}^{54}\, T_{i}^{\mu_{1} \mu_{2} \mu_{3} \mu_{4}} \Pi _{i} (q_{1},q_{2},q_{3})\,.
\end{equation}
Expressions for the $T_{i}^{\mu_{1} \mu_{2} \mu_{3} \mu_{4}}$ in terms of the Lorentz basis built with $q_{1},q_{2},q_{3}$ and the metric $g_{\mu \nu}$ can be found in Refs.~\cite{Colangelo:2015ama,Colangelo:2017fiz}. In particular, the $T_{i}^{\mu_{1} \mu_{2} \mu_{3} \mu_{4}}$ satisfy the same Ward identities as the HLbL tensor. As a consequence, in the static limit $q_4\to0$
\begin{equation}\label{eq:derq4btt}
\lim_{q_4\to 0} \frac{\partial \Pi^{\mu_{1} \mu_{2} \mu_{3} \nu_{4}}}{\partial q_{4}^{\mu_{4}}}=\sum_{i=1}^{54}\frac{\partial T_i^{\mu_{1} \mu_{2} \mu_{3} \nu_{4}}}{\partial q_{4}^{\mu_{4}}}\,\Pi_{i}(q_1,q_2,q_3) \, .
\end{equation}
In the static limit $q_{4}\rightarrow 0$, the remaining tensor can thus be written as a function of $19$ linear combinations $\hat{\Pi}_{i}$ of the original $\Pi_{i} (q_{1},q_{2},q_{3})$. Only $6$ $\hat{\Pi}_{i}$ functions contribute to the $(g-2)_{\mu}$ and one finds
\begin{align}\label{eq:amuint}
    a_{\mu}^{\textrm{HLbL}} = \frac{2\alpha ^{3}}{3\pi ^{2}} 
    & \int _{0}^{\infty} dQ_{1}\int_{0}^{\infty} dQ_{2} \int _{-1}^{1}d\tau \, \sqrt{1-\tau ^{2}}\, Q_{1}^{3}Q_{2}^{3}
    \nonumber \\
    & \times \sum _{i=1}^{12} T_{i}(Q_{1},Q_{2},\tau)\, \overline{\Pi}_{i}(Q_{1},Q_{2},\tau)\, ,
\end{align}
where the integration variable $\tau$ is defined via $Q_3^2=Q_1^2+Q_2^2+2\tau Q_1Q_2$. The functions $T_{i}(Q_{1},Q_{2},\tau)$ can be found in Refs.~\cite{Colangelo:2015ama,Colangelo:2017fiz} and 
\begin{align}\label{eq:pibarfcns}
    & \overline{\Pi}_{1} = \hat{\Pi}_{1} \, , \; \overline{\Pi}_{2} = C_{23}\left[  \hat{\Pi}_{1}\right] \, , \; \overline{\Pi}_{3} = \hat{\Pi}_{4} \, , \; \overline{\Pi}_{4} = C_{23}\left[\hat{\Pi}_{4}\right]\, , 
    \nonumber \\
    & \overline{\Pi}_{5} = \hat{\Pi}_{7} \, , \; \overline{\Pi}_{6} = C_{12}\left[ C_{13}\left[  \hat{\Pi}_{7}\right] \right] \, , \; \overline{\Pi}_{7} = C_{23}\left[\hat{\Pi}_{7}\right] \, , \; 
    \nonumber  \\
    & \overline{\Pi}_{8} = C_{13}\left[\hat{\Pi}_{17}\right]\, , \; 
     \overline{\Pi}_{9} = \hat{\Pi}_{17} \, , \; \overline{\Pi}_{10} = \hat{\Pi}_{39} \, , \; 
     \nonumber  \\
    & \overline{\Pi}_{11} = -C_{23}\left[ \hat{\Pi}_{54}\right]  \, , \; \overline{\Pi}_{12} = \hat{\Pi}_{54}\, .
\end{align}
The exact definition of the $\hat{\Pi}_{i}$ functions is given in Refs.~\cite{Colangelo:2015ama,Colangelo:2017fiz}. The $C_{ij}$ in~(\ref{eq:pibarfcns}) represent interchanging $q_i$ and $q_j$.
In order to find general Lorentz projectors from the derivative of the tensor in the static limit,
i.e.~
\begin{equation}
\lim_{q_4\rightarrow 0}\frac{\partial \Pi^{\mu_{1} \mu_{2} \mu_{3} \nu_{4}}}{\partial q_{4}^{\mu_{4}}} \, ,
\end{equation}
to the $\hat{\Pi}_{i}$ functions, we start by taking $19$ independent projectors in the $\{q_1,q_2,q_3,g\}$ basis. Any other projector can be related to that set through the Ward identities given above. Applying them to~(\ref{eq:derq4btt}) returns a system of $19$ equations dependent on the $19$ $\hat{\Pi}_{i}$. A solution to that system of equations for the relevant $\hat{\Pi}_{i}$ is given in App.~\ref{app:proj}. In practice, this means that for any contribution to the HLbL tensor in any basis, one can compute the associated $(g-2)_{\mu}$ contribution by taking the derivative with respect to $q_{4}$, then taking the static limit $q_{4}\rightarrow 0$, applying the $6$ Lorentz projectors given in App.~\ref{app:proj} to find the associated $\hat{\Pi}_{i}$ and finally using them in the integral of~(\ref{eq:amuint}).

As explained above, the integral of~(\ref{eq:amuint}) requires the knowledge of the HLbL tensor with three Euclidean momenta $Q_{i}$ at different kinematic regions and the fourth in the static limit $q_{4}\rightarrow 0$. In this work, which extends the results of Ref.~\cite{Bijnens:2019ghy}, we focus on the kinematic region where the three loop momenta are large. As was shown in Ref.~\cite{Bijnens:2019ghy}, if one defines
\begin{align}
\label{eq:backdyson1}
&\Pi ^{\mu_{1} \mu_{2} \mu_{3} }(q_{1},q_{2})\equiv-\frac{1}{e}\int\frac{d^4 q_{3}}{(2\pi)^4}  \left(\prod_{i=1}^{3}\int d^{4}x_{i}\, e^{-i q_{i} x_{i}}\right)  \langle 0 | T\left(\prod_{j=1}^{3}J^{\mu_{j}}(x_{j})\right) | \gamma(q_4) \rangle \, ,
\end{align}
then in the static limit for the studied kinematic region, one can factor out the soft photon contributions according to
\begin{equation}
\Pi^{\mu_{1}\mu_{2}\mu_{3}}(q_{1},q_{2})= \Pi^{\mu_{1}\mu_{2}\mu_{3}\mu_{4}\nu_{4}}_{F}(q_{1},q_{2})\langle 0 |e_{q} F_{\nu_{4}\mu_{4}}|\gamma(q_{4})\rangle \, .
\end{equation}
As mentioned in the introduction, the soft background field $F_{\mu\nu}$ can originate from the hard degrees of freedom or the soft ones. One then finally has \cite{Bijnens:2019ghy}
\begin{equation}
  \lim_{q_{4}\rightarrow 0}  \frac{\partial \Pi^{\mu_{1} \mu_{2} \mu_{3} \mu_{4}}}{\partial q_{4}^{\nu_{4}}}=-i\,\Pi^{\mu_{1}\mu_{2}\mu_{3}[\mu_{4}\nu_{4}]}_{F}(q_{1},q_{2}) \, .
\end{equation}
For this OPE the massless quark loop is the leading term and the di-quark magnetic susceptibility of the vacuum is the leading, quark-mass suppressed, non-perturbative contribution. In this work we compute the leading non-perturbative (not mass-suppressed) corrections.

\section{The operator product expansion: a theoretical description}\label{sec:OPEdesc}
In this section we describe the OPE and the associated renormalisation program. From this we systematically separate the long-distance effects from the short-distance ones. The general framework and operators involved are presented in Sec.~\ref{sec:genopeframework} whereas the mixing of these operators is elaborated on in Sec.~\ref{sec:opeoperatormixing}. Finally, the OPE developed as well as operators and corresponding matrix elements involved are discussed.

\subsection{General framework}\label{sec:genopeframework}
Perturbative calculations are known to provide a huge predictive power in the framework of Quantum Field Theory. However, when the calculation of a given observable involves the interplay of two (or more) very different scales, large logarithms between them slow down, if not spoil, the convergence of the series. These large logarithms can be avoided in many cases through the OPE~\cite{Wilson:1969zs,Zimmermann:1972tv}, which integrates out the heavy degrees of freedom leaving the low-energy (long-distance) dependence encoded in effective operators, in such a way that the contributions from higher-dimension operators become suppressed by extra powers of the high-energy scale \cite{Buchalla:1995vs,Pich:1998xt,Manohar:2018aog}. There are cases in which this logarithmic re-summation is not enough, since one (or several) relevant couplings of the theory diverge when its running is performed. This is the case for QCD, where a low-energy description in terms of approximately free quarks and gluons does not hold and the matrix elements between initial and final states cannot be computed within perturbative QCD. The contributions from the operators whose quantum numbers are compatible with the transition must be fitted to data, computed with effective field theories or other non-perturbative methods, such as lattice QCD, dispersion relations or model estimates.

In the OPE of two-point correlation functions \cite{Shifman:1978bx}, all the local operators with the same quantum numbers as the QCD vacuum, such as the identity or $\bar{q} q$, can give a contribution to the Green functions\footnote{The words $n$-point function, correlation function and Green function are all used but have the same meaning.}. In the OPE used in flavour-breaking transitions (e.g.~see Ref.~\cite{Buchalla:1995vs}) all the local operators with quantum numbers compatible with the studied transition among hadrons can give a contribution. In the OPE we are working with \cite{Ioffe:1983ju,Balitsky:1983xk}, applied to~(\ref{eq:backdyson1}), any local operator with the same quantum numbers as $F_{\mu\nu}$ can absorb the remaining soft static photon and, as a consequence, give a contribution \cite{Czarnecki:2002nt,Bijnens:2019ghy}. Higher-dimensional operators are suppressed by extra powers of $\left(\frac{\Lambda_{\mathrm{QCD}}}{Q_{i}}\right)^{d}$, providing a hierarchy of contributions with a systematic counting. Up to dimension $6$ and order $\alpha_{s}$ a basis of those operators is\footnote{Notice how the order in $\alpha_{s}$ depends on the short-distance structure of the studied Green function. For example in baryon sum rules, the four-quark operators $S_{\{8\}}$ do not enter $\alpha_{s}$-suppressed. In our calculation, at order $g_{s}^{3}$ one may have a contribution from a three-gluon operator \cite{Ioffe:1983ju}, but it enters suppressed by $g_{s}$, loops and flavour (in the $SU(3)_{V}$ limit its contribution vanishes, since it transforms as a singlet and the photon field transforms as an octet).}

\begin{align}
S_{1,\,\mu\nu}&\equiv e\,   e_{q}  F_{\mu\nu} \;  ,\label{eq:firstop}
\\ 
S_{2,\,\mu\nu}&\equiv  \bar{q}\sigma_{\mu\nu}q   \; \label{eq:secondop},
\\ 
S_{3,\,\mu\nu}&\equiv  i \,  \,\bar{q} G_{\mu\nu}q  \; ,
\\ 
S_{4,\,\mu\nu}&\equiv  i \, \bar{q} \bar{G}_{\mu\nu}\gamma_{5} q \; ,
\\
S_{5,\,\mu\nu}&\equiv  \bar{q} q\; e\,   e_{q}F_{\mu\nu} \; ,
\\ 
S_{6,\,\mu\nu}&\equiv \frac{\alpha_{s}}{\pi}\, G_ {a}^{\alpha\beta}G^{a}_{\alpha\beta}\; e\,   e_{q}F_{\mu\nu}  \; ,
\\
S_{7,\,\mu\nu}&\equiv  \bar{q}(G_{\mu\lambda}D_{\nu}+D_{\nu}G_{\mu\lambda})\gamma^{\lambda}q-(\mu\leftrightarrow\nu) \; ,
\\
S_{\{8\},\,\mu\nu}&\equiv \alpha_{s}\, (\bar{q}\, \Gamma \,q \; \bar{q}\Gamma q)_{\mu\nu} \; .\label{eq:lastop}
\end{align}

We use here the notation of Ref.~\cite{Pascual:1984zb}. In particular, $G_{\mu\nu}= i g_S\lambda^a G^a_{\mu\nu}$, $\bar{G}^{\mu\nu}\equiv\frac{i}{2}\epsilon^{\mu\nu\lambda\rho} G_{\lambda\rho}$, covariant derivatives act on all objects to their right and $\mathrm{tr}\left(\gamma_5\gamma^\mu\gamma^\nu\gamma^\alpha\gamma^\beta\right)=-4i\varepsilon^{\mu\nu\alpha\beta}$. For $S_{1\ldots7,\mu\nu}$ the quark field $q$ refers to a given flavour and there are in principle different operators for different flavours $q$. Notice, however, that taking into account that the (massless) QCD vacuum preserves $SU(3)_{\mathrm{V}}$, the contributions of the operators to the studied transition depend, in the chiral limit ($m_{u}=m_{d}=m_{s}=0$), on a common constant multiplied by the corresponding quark electric charge. Note that with the conventions of Ref.~\cite{Pascual:1984zb}, $G^{\mu\nu}$ is order $g_{s}$.\footnote{In fact, the gluon tensor is named $F^{\mu\nu}$ in that reference. We rename it as $G^{\mu\nu}$ to avoid ambiguity with the electromagnetic field strength $F_{\mu\nu}$. Similar renamings should be obvious.} The four-quark operators are only indicated generically in~(\ref{eq:lastop}). A decomposition valid in the chiral limit into twelve operators is given in App.~\ref{app:fourquark}. In fact, only two combinations of four-quark operators contribute as discussed in Sec.~\ref{sec:comp}.
The adopted notation is analogous to the one in Ref.~\cite{Buchalla:1995vs}.

In order to perform the OPE of the tensor in~(\ref{eq:backdyson1}), one applies Wick's theorem with any number of needed (suppressed) extra (QCD or QED) vertices coming from the Dyson series. The uncontracted operators must then be Taylor expanded (e.g.~see Ref.~\cite{Pascual:1984zb}), so that the resulting expression is of the form\footnote{Further technical details on the computation of the different pieces are given in Sec. \ref{sec:comp}.}
\begin{equation}\label{eq:opeprior}
\Pi^{\mu_{1}\mu_{2}\mu_{3}}(q_{1},q_{2})=\frac{1}{e}\vec{C}^{T,\mu_{1}\mu_{2}\mu_{3}\mu_{4}\nu_{4}}(q_{1},q_{2})\langle \vec{S}_{\mu_{4}\nu_{4}}\rangle=\vec{C}^{T,\mu_{1}\mu_{2}\mu_{3}\mu_{4}\nu_{4}}(q_{1},q_{2})\,\vec{X}_{S}\,\langle e_{q}F_{\mu_{4}\nu_{4}}\rangle \, ,
\end{equation}
where $\vec{X}_{S}$ contains the magnetic susceptibilities of the operators,\footnote{We will define the susceptibilities more precisely later.} $\langle S_{i,\mu\nu}\rangle = e e_q X_S^i \langle F_{\mu\nu} \rangle$.

\begin{figure}[tb]\centering
\includegraphics[width=0.6\textwidth]{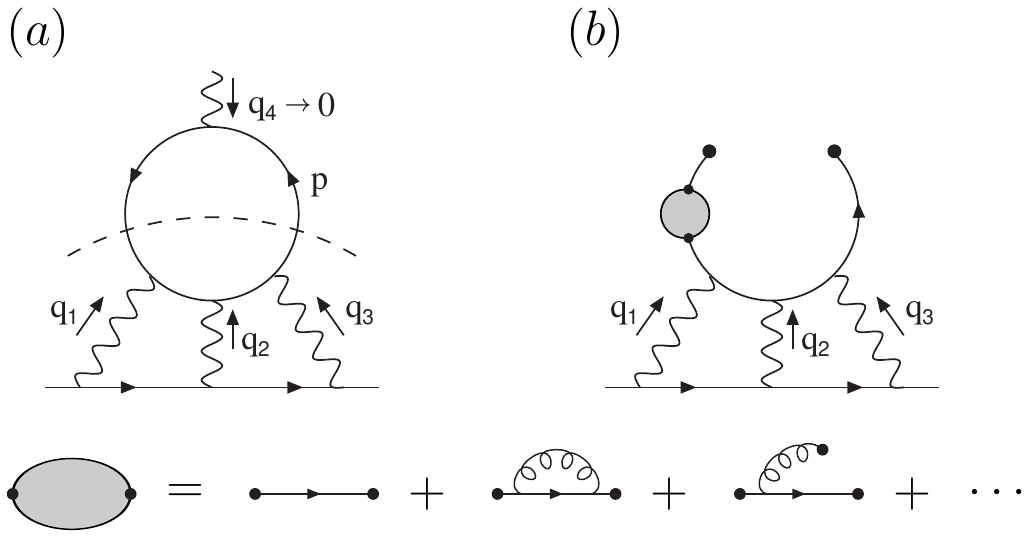}
\caption{\label{fig:twoexamples} Example of contributions of HLbL in the studied kinematic region. (a) The quark loop. (b) The di-quark magnetic susceptibility. A complete separation between short-distance and long-distance effects should subtract possible divergent low-momenta contributions from the quark loop and possible divergent perturbative series arising from soft lines.}
\end{figure}

Even when this was a first step to achieve the separation between short-distance and long-distance effects, such a separation is not yet complete. Let us illustrate this with the simplest contributions: the quark loop and the (di-quark) magnetic susceptibility, represented in Fig.~\ref{fig:twoexamples}. In Fig.~\ref{fig:twoexamples}b one has short-distance contributions that arise from expanding the Dyson series and introducing vertices in the soft lines, represented by a blob. Since there is no momentum flowing, the resulting series, $\sum c_n\alpha_{s}^n(0)$, is manifestly divergent. This kind of effects must be subtracted, since they belong to the non-perturbative domain. A slightly different problem arises with Fig.~\ref{fig:twoexamples}a. The quark loop runs over all possible momenta. If the low-momenta contributions do not vanish, they must somehow be subtracted and reabsorbed into the long distance matrix elements. This is the case for the $\mathcal{O}(m_{q}^{2})$ mass corrections, whose (not regulated) Wilson coefficient, $C_{m_q^{2}}$, leads to divergent series $\sum_n c_{n}\, \alpha_{s}^{n}(Q^{2})\log^{n} \left(Q^2/m_{q}^2\right)$, i.e. the coefficients are not well defined in the chiral limit. The $m_q^2$ dependence in the coefficients originates from the low-energy domain and must be included in the operator expectation values as well.

The way of achieving these subtractions is by dressing and renormalising the $\vec{S}^{\mu\nu}$ operators (normal-ordered operators in the notation of Ref.~\cite{Jamin:1992se}, tree-level operators in the notation of Ref.~\cite{Buchalla:1995vs}). Following a notation close to the one in Ref.~\cite{Buchalla:1995vs}, the dressed operators $\vec{Q}_{0}^{\mu\nu}$ in terms of the tree-level ones $\vec{S}^{\mu\nu}$ are obtained by reinserting them into the Dyson series, leading to a result of the form
\begin{equation}
\vec{Q}_{0}^{\mu\nu}=\hat{M}(\epsilon)\vec{S}^{\mu\nu} \, ,
\end{equation}
where the $\epsilon=-\frac{d-4}{2}$ dependence appears in dimensional regularization as a consequence of ultraviolet divergences in the loop diagrams. The infrared divergences that can appear are regularised using the quark mass. The resulting redefinition of the matrix elements\footnote{We need here the condensates induced by the external field. We refer to those as matrix elements to distinguish them from the usual (vacuum) condensates.} involves two steps, calculating the relevant contributions leading to an expression at one-loop order of the form
\begin{equation}\label{eq:dressingmat}
\hat{M}(\epsilon)=\mathbb{I}+\frac{m_{q}^{-2a\hat{\epsilon}}}{\hat{\epsilon}}\hat{M}_{\hat{\epsilon}}+\hat{M}_{\text{rem}} \, ,
\end{equation}
where $\hat{M}_{\hat{\epsilon}}$ and $\hat{M}_{\text{rem}}$ are perturbations either in $e$, in $g_{s}$ or in powers of $\frac{m_{q}}{\Lambda_{\mathrm{QCD}}}$ and $\frac{1}{\hat\epsilon}=\frac{1}{\epsilon}-\gamma_E+\log(4\pi)$. $a$ depends on the dimension of the operators when $d\ne4$. The ultraviolet divergences are unphysical, and are removed via renormalisation. A convenient and simple renormalisation scheme is by performing the operator renormalisation in the $\overline{MS}$ scheme, which basically removes the $\frac{1}{\hat{\epsilon}}$ factors and takes out from the bare operators the non-canonical part of their dimension, proportional to $2a\epsilon$:
\begin{align}\label{eq:renorm1}
\vec{Q}_{0}^{\mu\nu}&=\hat{Z}_{\overline{MS}}(\mu,\epsilon)\,\vec{Q}^{\mu\nu}_{\overline{MS}}(\mu) \, ,\\ \label{eq:renorm2}
\hat{Z}_{\overline{MS}}(\mu,\epsilon)&=\mathbb{I}+\frac{\hat{M}_{\hat{\epsilon}}\,\mu^{2 a \epsilon}}{\hat{\epsilon}} \, ,\\
\label{eq:renorm3}
\vec{Q}^{\mu\nu}_{\overline{MS}}(\mu)&=\hat{U}_{\overline{MS}}(\mu)\,\vec{S}^{\mu\nu}=\left(\mathbb{I}-a \log\left(\frac{m_{q}^{2}}{\mu^2}\right)\hat{M}_{\hat{\epsilon}}+\hat{M}_{\text{rem}} \right)\vec{S}^{\mu\nu}\,.
\end{align}
Putting this back into~(\ref{eq:opeprior}) one finds
\begin{align}\label{eq:opepost}
\Pi^{\mu_{1}\mu_{2}\mu_{3}}(q_{1},q_{2})&=\frac{1}{e}\vec{C}^{T,\mu_{1}\mu_{2}\mu_{3}\mu_{4}\nu_{4}}(q_{1},q_{2})\hat{U}_{\overline{MS}}^{-1}(\mu)\langle \vec{Q}_{\overline{MS}, \mu_{4}\nu_{4}}(\mu)\rangle\nonumber\\
&\equiv\frac{1}{e}\vec{C}^{T,\mu_{1}\mu_{2}\mu_{3}\mu_{4}\nu_{4}}_{\overline{MS}}(q_{1},q_{2})\langle \vec{Q}_{\overline{MS}, \mu_{4}\nu_{4}}(\mu)\rangle
 \, .
\end{align}
This equation defines the regularized and renormalised Wilson coefficients. The renormalised Wilson coefficients,
\begin{equation}\label{eq:wilsreg}
\vec{C}_{\overline{MS}}^{\mu_{1}\mu_{2}\mu_{3}\mu_{4}\nu_{4}}(q_{1},q_{2},\mu)=\left.\hat{U}_{\overline{MS}}^{-1}\!\right.^T\!\!(\mu)\vec{C}^{\mu_{1}\mu_{2}\mu_{3}\mu_{4}\nu_{4}}(q_{1},q_{2}) \, ,
\end{equation}
become free from long-distance effects and infrared divergences, completing the desired separation. For the matrix elements we define the magnetic susceptibilities $\vec X = \left(X_1,\ldots\right)$.\footnote{The four-quark operators have a slightly different definition of the susceptibilities to get the charge behaviour correctly, see Sec.~\ref{sec:Q8i}. $X_2$ is often referred to as the (di-quark) magnetic susceptibility.} 
\begin{equation}
\label{eq:defsuscept}
\langle\vec{Q}_{\overline{MS}, \mu\nu}(\mu)\rangle =
e\vec{X}\langle e_{q}F_{\mu\nu}\rangle
\end{equation}

The contributions we calculate explicitly in Sec.~\ref{sec:comp} are:
The leading contribution stems from $Q_1^{\mu\nu}$ at $d=2$ and corresponds to the massless quark loop. This operator receives mass corrections suppressed by powers of $\frac{m_{q}^{2}}{\Lambda^{2}}$. The first correction from a different operator comes from the $d=3$ di-quark operator, $Q_{2}^{\mu\nu}$, which happens to enter suppressed by one power of the quark mass becoming effectively $d=4$. This contribution using the mixing as defined in~(\ref{eq:dressingmat}--\ref{eq:renorm3}) removes the $m_q^2\log(m_q^2)$ part of the quark loop and together with that forms the $d=4$ contribution. The $d=5$ contributions from $Q_{3-5}^{\mu\nu}$ are also suppressed by one power of the quark mass. $Q_{6-8}^{\mu\nu}$ give the first contributions that are not suppressed by the quark masses. The mixing here again removes contributions proportional to $\log(m_q^2)$ and other infrared divergences. We therefore calculate with respect to the massless quark loop the corrections of orders $g_{s}\,\frac{\Lambda_{\mathrm{QCD}}^4}{Q^4}$, $\frac{m_q^2}{Q^2}$, $g_s^2 \,\frac{\Lambda_{\mathrm{QCD}}^4}{Q^4}$, $m_q\frac{\Lambda_{\mathrm{QCD}}}{Q^2}$,$m_q\frac{\Lambda_{\mathrm{QCD}}^3}{Q^4}$, $m_q^3\frac{\Lambda_{\mathrm{QCD}}}{Q^4}$. The computation of the last three is needed, since they give contributions to the $g_{s}^2 \, \frac{\Lambda_{\mathrm{QCD}}^4}{Q^4}$ through operator mixing as defined above and calculated in Sec.~\ref{sec:opeoperatormixing}. This is analogous to the mixing between bi-linear quark condensate and gluon condensate in the usual vacuum OPE \cite{Novikov:1983gd,Generalis:1983hb,Broadhurst:1984rr}.

\subsection{The operator mixing}\label{sec:opeoperatormixing}

\begin{figure}[tb]\centering
\includegraphics[width=0.7\textwidth]{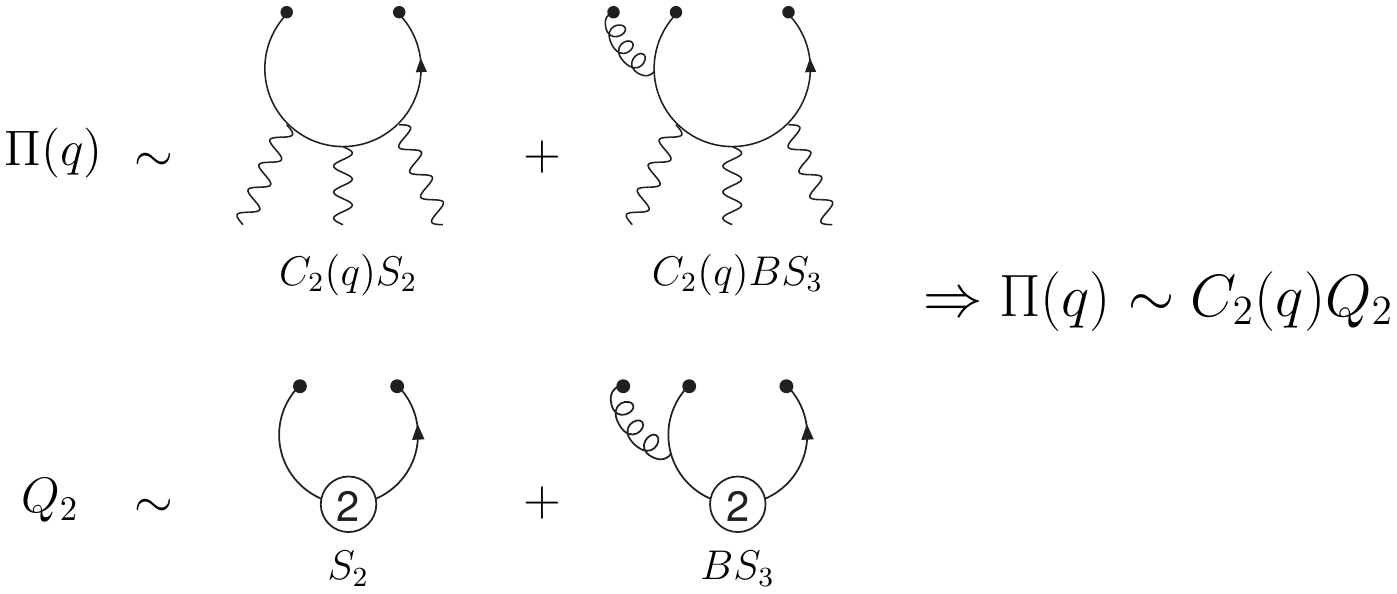}
\caption{\label{fig:softcanc}An example of a cancellation for contributions arising from attaching extra vertices to soft, at zero momentum, lines. The circle with ``2'' inside refers to the $S_2$ operator defined in~(\ref{eq:secondop}), $B$ indicates the extra factor compared to the left diagrams. All Lorentz indices have been suppressed for clarity.}
\end{figure}

In this section we calculate the mixing matrix $\hat U_{\overline{MS}}$. However, there are some parts that can be ignored. These we discuss first.
The original Wilson coefficients contain in principle infrared divergent parts that arise from attaching extra vertices to the soft zero-momentum lines. These long-distance contributions, explicitly independent of momenta, systematically cancel with analogous terms in the dressing procedure, which in addition are independent of the studied physical process, so we simply ignore them in both sides of the calculation. An example of this is sketched in Fig.~\ref{fig:softcanc}.
The top graphs show the contribution from $S_{2,\mu\nu}$ to the correlator we calculate and the bottom line contributions to the mixing matrix. The contribution from the top right is via the diagram in the bottom right absorbed into the definition of the operator $Q_{2,\mu\nu}$.

As a consequence, only mixing terms with loops need to be considered for the calculation of $\hat U_{\overline{MS}}$. This limits the type of terms that can show up. Then, at this stage, we have two possibilities. 1) If no more (QCD or QED) vertices are added, the original operator can only mix with lower dimensional ones, since loops imply connecting the fields of both operators with propagators. The only remaining scale to compensate dimensions is $m_q$ and as a consequence $\hat{M}(\epsilon)$ becomes a $m_{q}^n$ perturbation, with $n$ positive. The calculation for $\hat U^{21}_{\overline{MS}}$ is of this type. 2) Alternatively, one can have mixing terms with operators with equal or higher dimensions that show up by adding fields through introducing extra vertices. However, these new vertices come with an extra cost ($g_{s}$ or $e$), and then they can also be regarded as perturbations, an example of this type is $\hat U^{76}_{\overline{MS}}$. Finally, when studying the mixing of lower dimensional operators with higher dimensional ones, one finds terms that go as $m_{q}^{-n}$. A well-known case of this in the usual vacuum OPE is the gluon condensate mixing with the quark condensate
\begin{align}
\label{eq:qbarqgluon}
    \langle \bar q q\rangle = -\frac{1}{12m_q}\left\langle\frac{\alpha_S}{\pi}G^a_{\mu\nu}G^{a\mu\nu}\right\rangle + \cdots \, .
\end{align}
However this kind of mixing is simply absorbing in the renormalised operator unphysical (infrared) divergences contained in the perturbative Wilson coefficients. The quark mass is used as an infrared regulator as well as for calculating genuine quark mass corrections.\footnote{The procedure is equivalent to what is used in the usual vacuum OPE~\cite{Novikov:1983gd,Generalis:1983hb,Braaten:1991qm,Broadhurst:1984rr,Jamin:1992se}.}

\begin{figure}[tb]
\centering
\includegraphics[width=0.95\textwidth]{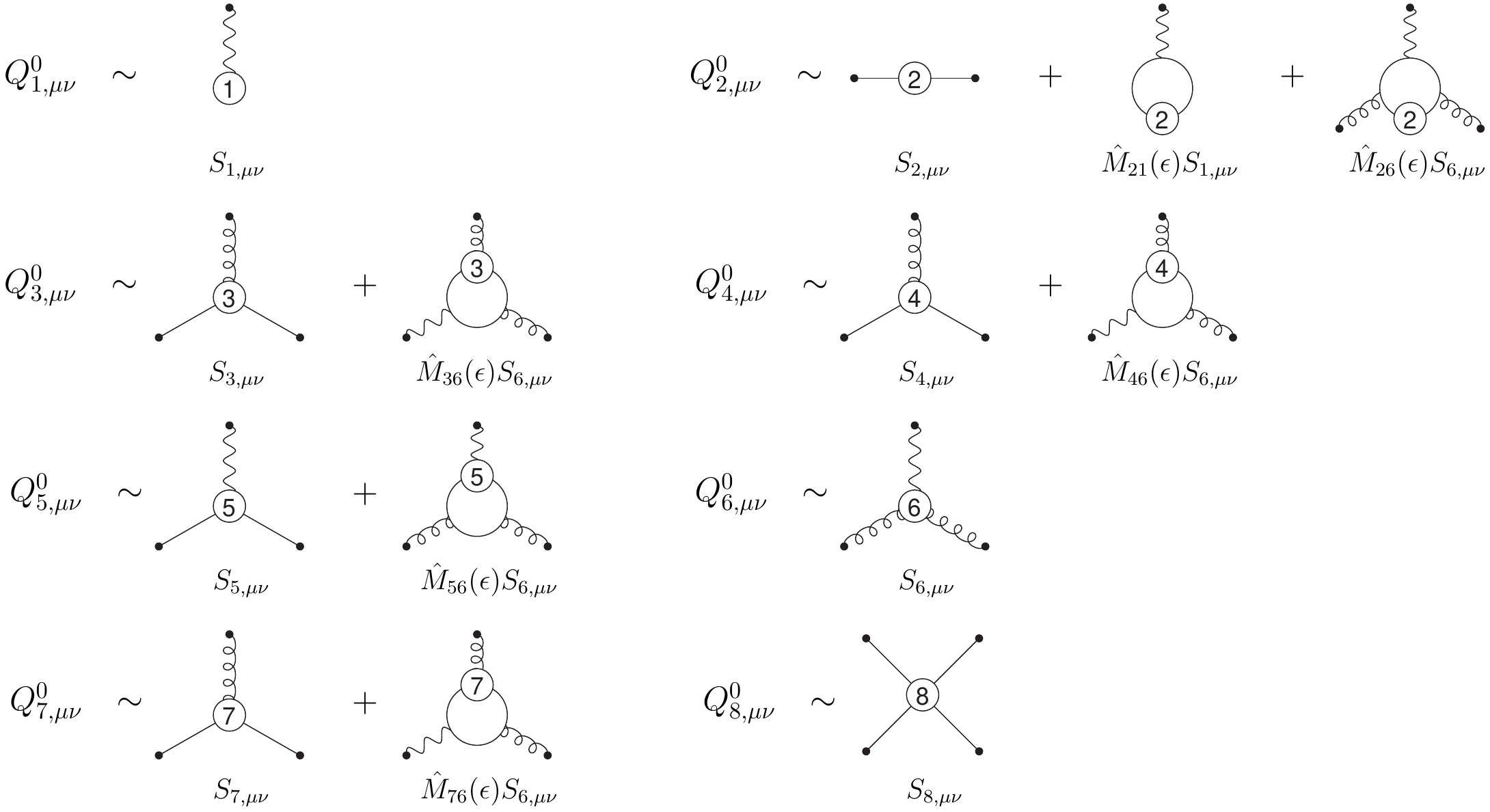}
\caption{\label{fig:Mixing} Topologies involved in the computation of the matrix $\hat{M}(\epsilon)$. All the possible permutations of the bosonic lines attached to the fermion loops must be considered. The numbered circles refer to the operators defined in~(\ref{eq:firstop})-(\ref{eq:lastop}). The ``outside'' black dots form the operators that mix in.}
\end{figure}

The first step is thus the dressing of the tree level operators and the computation of the associated matrix $\hat{M}(\epsilon)$. All the diagrams involved are shown in Fig.~\ref{fig:Mixing}. The first diagram for each operator always corresponds to identity term in~(\ref{eq:dressingmat}). The procedure should be done, as is also the case for the usual vacuum OPE, only including operators up to the dimension considered.

The only way $Q_{1}^{\mu\nu}$ mixes with other operators is by introducing extra electromagnetic vertices. Since $Q_{1}^{\mu\nu}$ is already $\mathcal{O}(e)$, the resulting corrections are $\mathcal{O}(e^2)$, and thus do not need to be considered. The first non-trivial case is that of the operator $Q_2^{\mu \nu}$. The non-zero element of $\hat M(\varepsilon)$ and thus $\hat U_{\overline{MS}}$ are calculated in the next subsubsections. As in Sec.~\ref{sec:comp} the calculations in this section benefit very much from using the radial gauge.

\subsubsection{$Q_{2}^{\mu\nu}$}

The mixing with $Q_{1}^{\mu\nu}= e e_qF^{\mu\nu}$ requires an extra electromagnetic vertex and closing the quark lines. This corresponds to the second $Q_2$-diagram in Fig.~\ref{fig:Mixing}. Let us sketch it in some detail to illustrate the procedure:
\begin{align}\begin{split}
Q^{0}_{2,\mu\nu}=\bar{q}\sigma_{\mu\nu}q_0&=\bar{q}\sigma_{\mu\nu}q\, e^{iS}=
\bar{q}\sigma_{\mu\nu}q \left(1+ie\int d^{d}x \, e_{q}\, \bar{q}(x)\gamma^{\nu_{1}}A_{\nu_{1}}(x) q(x)+\ldots\right)
\\ \
&=\bar{q}\sigma_{\mu\nu}q \left(1+ie\, e_{q}\frac{i F_{\mu_1\nu_1}}{2}\lim_{p_{1}\rightarrow 0}\partial_{p_1}^{\mu_1}\int d^{d}x \, e^{-i p_{1}x} \bar{q}(x)\gamma^{\nu_{1}} q(x)+\cdots\right)
\\ 
&=S_{2,\, \mu\nu}+S_{1,\, \mu\nu}\frac{N_{c}\, m_{q}}{4\pi^{2}}\left(-\frac{1}{\hat{\epsilon}} + \log(m_{q}^2)\right)+\cdots \, ,
\end{split}\end{align}
where we have used that we are working in the radial gauge, $A_{\mu}(x)=\frac{1}{2}x^{\nu}F_{\nu\mu}$ \cite{Pascual:1984zb,Novikov:1983gd}.\footnote{The results are of course gauge-independent.} Taking into account~(\ref{eq:dressingmat}) and~(\ref{eq:renorm3}), one finds
\begin{equation}
\label{eq:uhat21}
\hat{U}^{21}_{\overline{MS}}(\mu)=\frac{N_{c}m_{q}}{4\pi^2}\log\left(\frac{m_{q}^{2}}{\mu^2} \right) \, .
\end{equation}
Through the $\overline{MS}$ renormalisation we have introduced a subtraction point, $\mu$, which is the scale of separation between long-distance and short-distance effects. In particular this mixing term, when plugged into~(\ref{eq:wilsreg}), removes the (infrared divergent) long-distance parts from the loops associated to the perturbative $\mathcal{O}(m_{q}^{2})$ corrections by effectively replacing $\log Q_i^2/m_{q}^2\rightarrow \log Q_i^2/\mu^2$.\footnote{In Ref.~\cite{Bijnens:2019ghy} this replacement was not done for the $m_q^2$ quark loop numerics.} Notice how, in contrast with the usual vacuum OPE, where these divergences can only start at $\mathcal{O}(m_{q}^{4})$ (and be regulated by the quark condensate), the divergences in this OPE with respect to the quark loop start at $\mathcal{O}(m_q^{2})$. They become regulated by $Q_{2}^{\mu\nu}$.

At the order we are working with, $Q_{2}^{\mu\nu}$ also mixes with $Q_{6}^{\mu\nu}$ through the third $Q_2$-diagram in Fig.~\ref{fig:Mixing}, leading to
\begin{equation}
\hat{U}^{26}_{\overline{MS}}(\mu)=-\frac{1}{72 m_{q}^{3}} \, .
\end{equation}
The Wilson coefficient proportional to $\frac{S_{2}^{\mu\nu}}{m_{q}^{2}}$ combined with this matrix element cancels the power divergence of the (unregulated) Wilson coefficient associated to $S_{6}^{\mu\nu}$, while the mass correction to the Wilson coefficient associated to $S_{1}^{\mu\nu}$ gives a finite contribution that needs to be included. Once again, this interplay is analogous to the corresponding one in the vacuum OPE \cite{Novikov:1983gd,Broadhurst:1984rr,Generalis:1983hb,Jamin:1992se}.

\subsubsection{$Q_{3-6}^{\mu\nu}$}

Since $Q_{3}^{\mu\nu}$ and $Q_{4}^{\mu\nu}$ are already $\mathcal{O}(g_{s})$ and two lines need to be closed to form a loop to give a non-zero contribution, they only mix with the gluon matrix element at the order we are working with (see Fig.~\ref{fig:Mixing}), giving a finite contribution to its associated regulated Wilson coefficient. One finds,
\begin{align}
U^{36}_{\overline{MS}}(\mu)&=\frac{1}{36 m_{q}}\, ,\\
U^{46}_{\overline{MS}}(\mu)&=\frac{1}{24 m_{q}}\, .
\end{align}

Since $Q_{5}^{\mu\nu}$ is already order $e$, it can only mix with $Q_{1}^{\mu\nu}$ and $Q_{6}^{\mu\nu}$ at the order we are working with. Dimensional analysis shows that the mixing with $Q_{1}^{\mu\nu}$ only modifies the very tiny and safely neglected $\mathcal{O}(m_{q}^{4})$ contribution. The mixing with $Q_{6}^{\mu\nu}$ is the same as the mixing of the quark condensate with the gluon condensate, as in~(\ref{eq:qbarqgluon}), since at the order in $e$ we are working with the photon does not see strong interactions\cite{Shifman:1978bx,Novikov:1983gd,Broadhurst:1984rr,Generalis:1983hb,Jamin:1992se},
\begin{equation}
U^{56}_{\overline{MS}}(\mu)=-\frac{1}{12 m_{q}} \, .
\end{equation}

\subsubsection{$Q_{7}^{\mu\nu}$}

The last operator to treat is $Q_7^{\mu\nu}$ which through the topologies shown in Fig.~\ref{fig:Mixing} only mixes with $Q_6^{\mu\nu}$. One has
\begin{equation}
U^{76}_{\overline{MS}}(\mu)=-\frac{1}{12}\left(1-2\log\frac{m_q^2}{\mu^2}\right) \, .
\end{equation}

\subsubsection{Full matrix $\hat U_{\overline{MS}}(\mu)$}

Putting all the elements together one finds for $N_{c}=3$
\begin{equation}
  \hat{U}_{\overline{MS}}(\mu)=  \left(
\begin{array}{cccccccc}
 1 & 0 & 0 & 0 & 0 & 0 & 0 & 0  \\
 \frac{3 \, m_{q}}{4 \pi ^2} \log\frac{m_{q}^{2}}{\mu^{2} } & 1 & 0 & 0 & 0 & -\frac{1}{72 m_{q}^{3}} & 0 &
   0  \\
 0 & 0 & 1 & 0 & 0 & \frac{1}{36 m_{q}} & 0 & 0  \\
 0 & 0 & 0 & 1 & 0 & \frac{1}{24 m_{q}} & 0 & 0  \\
0 & 0 & 0 & 0 & 1 & -\frac{1}{12 m_{q}} & 0
   & 0  \\
 0 & 0 & 0 & 0 & 0 & 1 & 0 & 0  \\
 0 & 0 & 0 & 0 & 0 & \frac{1}{6} \left(\log\frac{m_{q}^{2}}{\mu^{2} } -\frac{1}{2}\right) & 1 & 0
    \\
 0 & 0 & 0 & 0 & 0 & 0 & 0 & 1  \\
\end{array}
\right) \, .
\end{equation}

\subsection{Values of the matrix elements}
\label{sec:meests}

Apart from providing model-independent information on the kinematic dependence of the non-perturbative corrections for the short-distance HLbL, in principle this OPE can be used to study different Green functions with all its (Euclidean) momenta large except for one soft photon. This might also allow to obtain more information on the expectation values (or the susceptibilities). However in absence of the latter we need to determine the values in a different way. We can find values for all of them with a number of assumptions that  should at least give the correct order of magnitude.

\subsubsection{$Q_5^{\mu\nu}$ and $Q_6^{\mu\nu}$}

The matrix elements associated to $Q_{5}^{\mu\nu}$ and $Q_{6}^{\mu\nu}$ are directly related to the quark and the gluon condensates, respectively. The former one is well-known, since it is the order parameter of the spontaneous chiral symmetry breaking of QCD, both from chiral perturbation theory and the lattice. Updated lattice values can be found in Ref.~\cite{Aoki:2019cca}. The gluon condensate is not so well-known, since separating its effect from those of the perturbative series is non-trivial. However its order of magnitude is known, namely $X_{6}\sim 0.02\, \mathrm{GeV}^{4}$ \cite{Shifman:1978bx}.

\subsubsection{$Q_2^{\mu\nu}$, $Q_3^{\mu\nu}$ and $Q_4^{\mu\nu}$}

The matrix elements $\langle 0| Q_i^{\mu\nu}|\gamma(q_{4})\rangle=X_{i}\langle 0|e\,e_{q}F^{\mu\nu}|\gamma(q_{4})\rangle$ are directly related to the values of QCD two-point functions at zero momentum, $\Pi_{VQ_{i}}(q^{2})$. In order to see that, let us take a generic dressed operator $Q_{i,0}^{\mu\nu}$. Its contribution to a matrix element with a final static photon can only arise through an extra electromagnetic vertex. Then, in the static limit,
\begin{equation}\label{eq:condlow}
\langle 0 | Q^{\mu\nu}_{i,0}|\gamma(q_{4})\rangle=\langle 0 | Q^{\mu\nu}_{i,0(\mathrm{QCD})} ie  \int d^{4}x\,e_q \, A_{\beta}(x)J^{\beta}(x) |\gamma(q_{4}) \rangle=-\langle 0 | e\,e_{q}\,F^{\mu\nu}|\gamma(q_{4})\rangle\,\Pi_{JQ_i}^{\mathrm{QCD}}(0) \, ,
\end{equation}
where
\begin{equation}\label{eq:twopointfcns}
\Pi^{\mathrm{QCD},\alpha\mu\nu}_{JQ_{i}}(q)=\int d^{4}x \, e^{-iqx}\,T(J^{\alpha}(x)Q_{i,0(\mathrm{QCD})}^{\mu\nu}(0))=(q^{\mu}g^{\alpha\nu}-q^{\nu}g^{\alpha\mu})\Pi_{JQ_i}^{\mathrm{QCD}}(q^2)  \, . \end{equation}
These two-point functions can be computed at large Euclidean momenta through the OPE in the vacuum \cite{Shifman:1978bx}. 

One could compute these matrix elements in chiral perturbation theory. In fact, promoting the global $SU(3)_{\mathrm{V}}$ symmetries to local ones lead to trivial transformations for the operators $Q_{2}^{\mu\nu}, Q_{3}^{\mu\nu}, Q_{4}^{\mu\nu}$ and $Q_{8}^{\mu\nu}$. The resulting effective low-energy Lagrangians are given in Ref.~\cite{Cata:2007ns}.\footnote{Since the symmetry transformation of $Q_{3}^{\mu\nu}, Q_{4}^{\mu\nu}$ and $Q_{8}^{\mu\nu}$ are identical, their Lagrangians are functionally equivalent and only the low-energy couplings are different.} At the lowest order the $X^{\overline{MS}}_{i}(\mu)$ are directly proportional to the low-energy constants $\Lambda_{1}^{i,\overline{MS}}(\mu)$. However, this gives no extra information by itself in the $SU(3)_{V}$ limit, since the $\Lambda_{1}^{i,\overline{MS}}(\mu)$ are not known.

For $X_{2}$, $X_{3}$ and $X_{4}$ we obtain an educated guess by making use of~(\ref{eq:condlow}) and extrapolating the argument for $X_{2}$ from Ref.~\cite{Craigie:1981jx,Balitsky:1985aq,Knecht:2001xc,Mateu:2007tr}. First of all, in the large-$N_{c}$ limit the QCD spectrum is made of an infinite number of free, stable meson states \cite{tHooft:1973alw,Witten:1979kh,Pich:2002xy}. The two-point functions in that limit are then saturated by the exchange of resonances. Owing to the quantum numbers of the studied two-point functions, the corresponding resonances must be vector mesons \cite{Cata:2008zc}. The low-energy part of the $N_c=3$ QCD spectrum is actually close to the sum of narrow width resonances predicted by the large-$N_{c}$ limit, while at higher energies a transition towards the flat perturbative QCD spectrum is observed. Taking all this into account, and that in the chiral limit the $VT$ two-point function vanishes in the perturbative regime, it is reasonable to assume that the physical spectrum is saturated by the contribution of the lowest vector meson, i.e.~the $\rho$ meson. Using the formalism developed in Ref.~\cite{Ecker:1988te} and adding a tensor source~\cite{Cata:2007ns,Mateu:2007tr}, one can write the two-point functions of~(\ref{eq:twopointfcns}) on the form
\begin{equation}\label{eq:VTansatz}
\Pi_{JQ_{i}}(q^2)=\frac{C_{Ti}}{q^{2}-M_{\rho}^2} \, .
\end{equation}
Here, the $C_{T_{i}}$ are constants. It then follows that
\begin{equation}
X_{i}=\frac{C_{T_{i}}}{M_{\rho}^2} \, .
\end{equation}
In order to estimate the $C_{T_{i}}$, we can match the ansatz of~(\ref{eq:VTansatz}) with the short-distance OPE of~(\ref{eq:twopointfcns}). This is sometimes referred to as a vector-meson-dominance (VMD) estimate.
We find
\begin{align}\label{eq:shortVT}
\Pi_{JQ_{2}}(q^{2})&=\frac{2\langle \bar{q}q\rangle}{q^2} \, ,\\
\Pi_{JQ_{3}}(q^{2})&=-\frac{\langle \bar{q}G^{\mu\nu}_{a}\frac{\lambda^{a}}{2}  \sigma_{\mu\nu} q\rangle}{6q^2}\,\equiv -m_{0}^{2}\frac{\langle \bar{q}{q} \rangle}{6q^2} ,\\
\Pi_{JQ_{4}}(q^{2})&=-\frac{\langle \bar{q}G^{\mu\nu}_{a}\frac{\lambda^{a}}{2}  \sigma_{\mu\nu} q\rangle}{6q^2}\,\equiv -m_{0}^{2}\frac{\langle \bar{q}{q} \rangle}{6q^2} .
\end{align}
This leads to
\begin{align}
X_{2}&=\frac{2}{M_{\rho}^2}\langle\bar{q}q\rangle , \\
X_{3}&=-\frac{m_{0}^{2}}{6M_{\rho}^2}\langle \bar{q} q\rangle \, ,\\
X_{4}&=-\frac{m_{0}^{2}}{6M_{\rho}^2}\langle \bar{q} q\rangle \, .
\end{align}
The obtained value for $X_{2}$ is, in fact, in very good agreement with a precise lattice determination~\cite{Bali:2011qj,Bali:2020bcn}.\footnote{The sign differs from the one quoted there, presumably because of a number of non-trivial convention differences. They give the VMD estimate also with the opposite sign.} No precise modern evaluation of $m_{0}^2$ can be found in the literature. However, once again, some numerical estimates are available~\cite{Belyaev:1982sa}. The estimates of $X_3$ and $X_4$ are new to the best of our knowledge.

Notice how, once again, the mass correction to~(\ref{eq:shortVT}) leads to an infrared divergent term which is regularised by the mixing of the tensor current with $F_{\mu\nu}$ of~(\ref{eq:uhat21}), 
\begin{align}\nonumber
&\langle 0 |Q_{2,0}^{\mu\nu}  |\gamma(q)\rangle=ie e_q\epsilon_{\alpha}\Pi^{\alpha\mu\nu,0}_{\mathrm{QCD}}(q)=i e e_q
\left(
q^{\mu}\epsilon ^{\nu}
-q^{\nu}\epsilon^{\mu}
\right)
\frac{m_{q}N_c}{2\pi^2}
\left[
1+\frac{1}{2\hat{\epsilon}}-\frac{1}{2}\log\left(\frac{-q^2}{m_q^2}\right)-\frac{1}{2}\log m_{q}^{2}
\right]
\\
&=\hat{Z}_{\overline{MS}}(\mu)\, \langle 0 |Q_{2,R}^{\mu\nu}(\mu) |\gamma(q)\rangle=i\, e\, e_q\epsilon_{\alpha}\Pi_{R,\overline{MS}}^{\alpha\mu\nu}(q,\mu)-\frac{N_c\mu^{2\epsilon} m_{q}}{4\pi^2\hat{\epsilon}}e e_q (-i(q^{\mu}\epsilon ^\nu-q^{\nu}\epsilon^\mu)) \, ,
\end{align}
with
\begin{equation}
\Pi_{R,\overline{MS}}(q^2,\mu)=\frac{m_{q}N_{c}}{2\pi^2}\left(1-\frac{1}{2}\log\frac{-q^{2}}{\mu^2}\right) \, .
\end{equation}

\subsubsection{$Q_{7}^{\mu\nu}$}

Our largest uncertainty comes from $Q_{7}$, where we simply perform a dimensional guess inspired in its mixing with the gluon matrix element. One has
\begin{equation}
|X_{7}|\sim\frac{1}{6}\Big\langle \frac{\alpha_{s}}{\pi}GG\Big\rangle \, .
\end{equation}
Notice how the derivative term of this operator makes non-trivial its low-energy effective realization when trying to promote invariance under the global symmetry to a local one.

\subsubsection{$Q_{8,1}^{\mu\nu}$ and $Q_{8,2}^{\mu\nu}$}
\label{sec:Q8i}

For the four-quark operators there are only two combinations that contribute. These are
\begin{align}
    {S}_{8,1}^{\mu\nu} &= -\frac{g_s^2}{2}\epsilon^{\mu\nu\lambda\sigma}\sum_{A,B}\bar{q}_{A}\gamma_{\lambda}\frac{\lambda_{a}}{2}q_{A}e_{q_{B}}^{3}\bar{q}_{B}\gamma_{\sigma}\gamma_{5}\frac{\lambda_{a}}{2}q_{B}\,,
\\
S_{8,2}^{\mu\nu}&= -\frac{g_s^{2}}{2}\epsilon^{\mu\nu\lambda\sigma}
\sum_{A,B}e_{q_A}^2 \bar{q}_{A}\frac{\lambda_{a}}{2}\gamma_{\lambda}q_{A}
e_{q_B}\bar{q}_{B}\frac{\lambda_{a}}{2}\gamma_{\sigma}\gamma^{5}q_{B} \,.
\end{align}
These do not mix with other operators at the order we work so the $Q_{8,i}^{\mu\nu}$ are the same. Note that the two operators only differ in the way the quark charges appear.

For both operators we define a generalised magnetic susceptibility
\begin{equation}
\label{eq:X82bardef}
    \left\langle Q_{8,i}^{\mu\nu}\right\rangle = e \overline{X}_{8,i} F^{\mu\nu}\,.
\end{equation}
In the massless limit we can also use
\begin{equation}
    \left\langle Q_{8,1}^{\mu\nu}\right\rangle = e {X}_{8,1} F^{\mu\nu}\sum_B e_q^4\,,
\end{equation}
which is a definition more similar to~(\ref{eq:defsuscept}). This is not possible for $Q_{8,2}^{\mu\nu}$.

The operators ${Q}_{8,1}^{\mu\nu}$ and ${Q}_{8,2}^{\mu\nu}$ can be decomposed in a basis of $12$ four-quark operators containing different flavour and Dirac matrices (see App.~\ref{app:fourquark} for details on the reduction). However, in the large-$N_{c}$ limit not all of these survive. In particular, one finds that only two are non-vanishing due to the factorisation of two colour singlet currents in this limit. These are 
\begin{equation}
\label{eq:largeNccond}
 \langle0|\bar{q}\sigma_{\mu\nu}\lambda_{8}q\,\bar{q}q \pm \bar{q}\sigma_{\mu\nu}q\,\bar{q}\lambda_{8}q|\gamma(q)\rangle_{N_{c}\rightarrow \infty}=3\, 
 \langle0|\bar{q}\sigma_{\mu\nu}\lambda_{8}q|\gamma(q)\rangle\langle \bar{q}q\rangle=3\sum _{i} \lambda_{8,ii}X_{2}\langle \bar{q}q\rangle e_{q,i}F_{\mu\nu} \, .
\end{equation}
In the first equality we have performed the sum over flavour indices for the quark condensate. In the second step we have projected out the flavour matrix $\lambda _{8}$ which is traced with the quark charge matrix. Here one sees why only these two matrix elements can survive. First of all, the quark charge matrix is a linear combination of $\lambda _{3}$ and $\lambda _{8}$, so the relation $\textrm{Tr} (\lambda _{a}\lambda _{b}) = 2\, \delta _{ab}$ implies that only matrix elements  with $\lambda _{3}$ or $\lambda _{8}$ are non-zero. In addition, the only non-vanishing two-quark condensates are $\langle \bar{q}q\rangle $ and the di-quark matrix element $\bar{q}\sigma _{\mu\nu}q$. To leading order in $N_{c}$, one therefore finds 
\begin{align}
\label{eq:condX8}
&
\overline{X}_{8,1}^{N_c\rightarrow \infty}
= \overline{X}_{8,2}^{N_c\rightarrow \infty}
=-2\frac{\pi\alpha_{s}}{9}X_{2}\langle \bar{q}{q}\rangle  
\, .
\end{align}
Alternatively one can directly evaluate the large-$N_c$ limit of the two matrix elements needed by using
$\lambda_{a\alpha\beta}\lambda_{a\gamma\delta}=2\delta_{\alpha\delta}\delta_{\gamma\beta}$. Then use Fierzing and
the charge matrix equivalent of~(\ref{eq:largeNccond}). The result agrees with~(\ref{eq:condX8}). This way one sees also directly that $q_A=q_B$ in the non-zero part of the matrix elements at large $N_c$. The entire contribution to HLbL is then proportional to $\sum_A e_{q_A}^4$.

\section{Calculation of the HLbL contributions}\label{sec:comp}

In this section we consider the analytic calculations of the various contributions in the OPE discussed above. They are the fully connected quark loop, diagram topologies with one quark line non-contracted (related to two-quark operator matrix elements), diagram topologies with two quark lines non-contracted (giving rise to four-quark operator matrix elements) as well as the gluon matrix contribution. One check we have performed on all contributions is that
\begin{align}
q_{1\mu_1} \Pi^{\mu_1\mu_2\mu_3} = q_{2\mu_2} \Pi^{\mu_1\mu_2\mu_3} = q_{3\mu_3} \Pi^{\mu_1\mu_2\mu_3} = 0\,,
\end{align}
where $\Pi^{\mu_1\mu_2\mu_3}$ is defined in~(\ref{eq:backdyson1}).

This work relied heavily on \textsc{FORM}~\cite{Vermaseren:2000nd,Kuipers:2012rf}. The Feynman integral reduction for the quark loop and the gluon matrix element contributions was done with Reduze 2~\cite{vonManteuffel:2012np} and Kira~\cite{Maierhoefer:2017hyi}. In the supplementary material we provide the analytic results as \textsc{FORM} output in the file results.txt.

\subsection{The quark loop}
The quark loop contribution arises from allowing for a soft emission from one hard vertex, which is equivalent to modifying a quark propagator by an external background field \cite{Ioffe:1983ju,Novikov:1983gd}, as shown explicitly in Ref.~\cite{Bijnens:2019ghy}. Starting from~(\ref{eq:backdyson1}), the needed background field, $F_{\nu_{4}\mu_{4}}$, is simply obtained by Taylor expanding the photon field appearing in the Dyson series. In the static limit in the radial gauge one has
\begin{equation}\label{eq:backgauge}
A^{\mu_{4}}(x_{4})=\frac{1}{2}x^{\nu_{4}}F_{\nu_{4}\mu_{4}}=\lim_{q_{4}\rightarrow 0}\frac{i}{2}\, \partial^{\nu_{4}}\;e^{-iq_{4}x} F_{\nu_{4}\mu_{4}} \, .
\end{equation}
Define the quark propagator for a quark of mass $m_{q}$ as
\begin{align}
    S(p) = \frac{\slashed{p}+m_{q}}{p^{2}-m^{2}_{q}+i\varepsilon} \, .
\end{align}
One may then write $\Pi^{\mu_{1} \mu_{2} \mu_{3}\mu_{4} \nu_{4}}_{F}$ in a very compact way. After having contracted all the quark fields in the definition of the tensor in question one finds
\begin{align}\nonumber
&\Pi^{\mu_{1} \mu_{2} \mu_{3}\mu_{4} \nu_{4}}_{F}(q_{1},q_{2})= 
\int \frac{d^{4}p}{(2\pi)^d}
\\&-\frac{N_{c}e_{q}^4}{2}\lim_{q_{4}\rightarrow 0}\frac{\partial}{\partial q_{4}^{\nu_{4}}} \left[\sum_{\sigma(1,2,4)} \mathrm{Tr}\Big( \gamma^{\mu_{3}} S(p+q_{1}+q_{2}+q_{4}) \gamma^{\mu_{4}} S(p+q_{1}+q_{2}) \gamma^{\mu_{1}} S(p+q_{2})\gamma^{\mu_{2}}S(p)
\Big)  \right] \, ,\label{eq:dertrick}
\end{align}
where $\sigma (i,j,k)$ denotes a member of the permutation group acting on the set $\{i,j,k\} = \{ (q_{\ell},\mu _{\ell})\}_{\ell \in \{ i,j,k\} }$. In other words, $ \sigma (i,j,k)$ simply states that we sum over all permutations of momentum and Lorentz index pairs. Using iteratively the relation
\begin{equation}
\frac{\partial}{\partial q_{4}^{\nu_{4}}}S(p+q_{4})=-S(p+q_{4})\gamma^{\nu_{4}}S(p+q_{4}) \, ,
\end{equation}
allows for a systematic computation of the quark loop. Applying the projectors given in App.~\ref{app:proj} and reducing the (ultraviolet finite) integrals, the result is left in terms of scalar tadpole, self-energy and triangle integrals. Expanding these in the masses,\footnote{Taking into account that the mass dependence goes as $\sim A+B\,  m_q^{2}\log(m_q^{2})+C\,  m_q^{2}+ \mathcal{O}(m_q^{4})$, one needs to be careful in order to obtain the correct coefficients as the naive Taylor expansion does not hold.} one finds
\begin{equation}
\hat{\Pi}_{i,S}(Q_{1}^{2},Q_{2}^{2},Q_{3}^{2})=\hat{\Pi}^{0}_{i,S}(Q_{1}^{2},Q_{2}^{2},Q_{3}^{2})+m_{q}^{2}\,\hat{\Pi}^{m_{q}^{2}}_{i,S}(Q_{1}^{2},Q_{2}^{2},Q_{3}^{2},m_{q}^{2})+\mathcal{O}(m_{q}^{4}) \, ,
\end{equation}
where 
\begin{align}
\hat{\Pi}_{m,S}^{0}=\frac{N_{c}\,e_{q}^{4}}{\pi^{2}}\sum_{i,j,k,n}\left[c_{i,j,k}^{(m,n)}+f_{i,j,k}^{(m,n)}F+g^{(m,n)}_{i,j,k}\log\left(\frac{Q_{2}^2}{Q_{3}^{2}}\right)+h^{(m,n)}_{i,j,k}\log\left(\frac{Q_{1}^2}{Q_{2}^{2}}\right)\right]\lambda^{-n}\, Q_{1}^{2i}\,Q_{2}^{2j}\,Q_{3}^{2k} \, , 
\end{align}
\begin{align}
\begin{split}
\hat{\Pi}^{m_{q}^{2}}_{m,S}&=\frac{N_{c}\,e_{q}^{4}}{\pi^{2}}\sum_{i,j,k,n}\lambda^{-n}\, Q_{1}^{2i}\,Q_{2}^{2j}\,Q_{3}^{2k}
\\&
\times \left[d_{i,j,k}^{(m,n)}+p_{i,j,k}^{(m,n)}F+q^{(m,n)}_{i,j,k}\log\left(\frac{Q_{2}^2}{Q_{3}^{2}}\right)+r^{(m,n)}_{i,j,k}\log\left(\frac{Q_{1}^2}{Q_{2}^{2}}\right)+s^{(m,n)}_{i,j,k}\log\left(\frac{Q_{3}^2}{m_{q}^{2}}\right)\right] \, . 
\end{split}
\end{align}
Here, $\lambda=\lambda(q_1^2,q_2^2,q_3^2)$ is the Käll\'{e}n function defined through
\begin{equation}
 \lambda(q_1^2,q_2^2,q_3^2)=q_1^4+ q_2^4+ q_3^4 -2 q_1^2 q_2^2-2q_1^2q_3^2-2q_2^2q_3^2\, ,
\end{equation}
and $F=F(Q_{1}^2,Q_{2}^2,Q_{3}^2)$ the massless triangle integral
\begin{equation}
F(Q_{1}^2,Q_{2}^2,Q_{3}^2)\equiv (4\pi)^2 i \int \frac{d^4k}{(2\pi)^4}\frac{1}{k^2(k-q_1)^2(k-q_1-q_2)^2} \, .
\end{equation}
The different coefficients ($c, d, f, g, h, p, q, r, s$) can be found in App.~\ref{app:explicit}.
Explicit analytic formulas for $F(Q_{1}^2,Q_{2}^2,Q_{3}^2)$ in terms of Clausen, Glaisher and $L$ functions can be found in Ref.~\cite{Lu:1992ny}.
Spurious singularities in the $\lambda \rightarrow 0$ limit cancel against the zeros of the triangle function $F$ for the different $\hat{\Pi}_{i}$. 

As explained above, one finds logarithmic infrared divergences for $\hat{\Pi}^{m_{q}^{2}}_{i,S}(Q_{1}^{2},Q_{2}^{2},Q_{3}^{2})$. Rearranging the logarithms, they can be expressed on the form $\log \frac{Q_{3}^2}{m_{q}^2}$. Once the operator renormalisation is performed through~(\ref{eq:wilsreg}), all the infrared divergences exactly cancel. This yields the finite result
\begin{equation}
\hat{\Pi}^{\overline{MS}}_i(Q_{1}^{2},Q_{2}^{2},Q_{3}^{2},\mu^2)=\hat{\Pi}^{0}_{i,S}(Q_{1}^{2},Q_{2}^{2},Q_{3}^{2})+m_{q}^{2}\,\hat{\Pi}^{m_{q}^{2}}_{i,\overline{MS}}(Q_{1}^{2},Q_{2}^{2},Q_{3}^{2},\mu^2)+\mathcal{O}(m_{q}^{4}) \, . \label{eq:quarkloopdefs}
\end{equation}
As a consequence, while the massless quark loop corresponds to the leading term in the short-distance regime, the naive mass correction does not. Infrared divergent logarithms must be substracted first.

\subsection{Contributions from diagrams with one cut quark line}
\begin{figure}[tb]\centering
\includegraphics[width=0.8\textwidth]{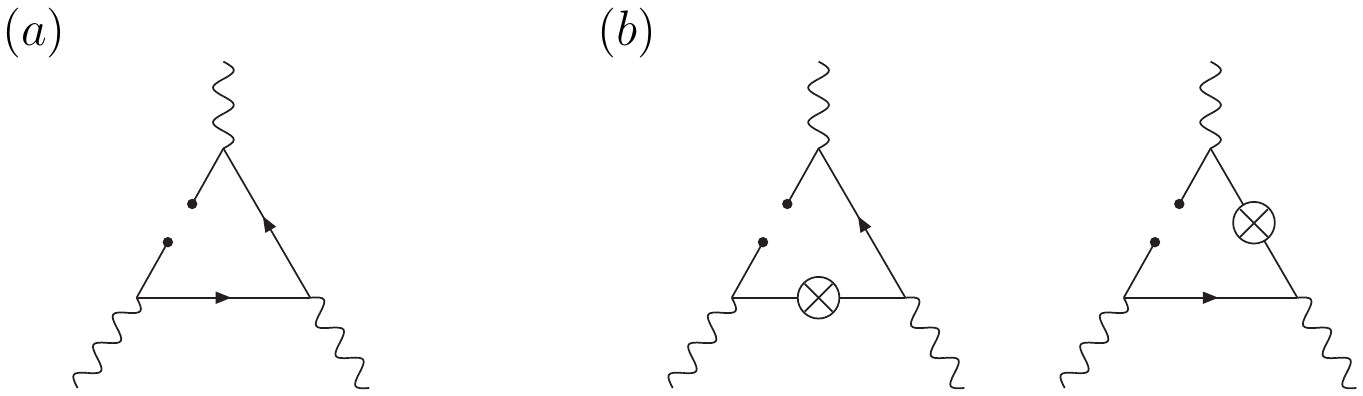}
\caption{\label{fig:onecutdiag} Diagrams with one cut quark line without gauge boson (a), and with gauge boson (b). The crossed vertices represent a gauge boson insertion on a propagator.}
\end{figure}
Several kinds of contributions need to be taken into account up to the computed order from topologies in which, starting from~(\ref{eq:backdyson1}), one quark line is left uncontracted: see Fig.~\ref{fig:onecutdiag}. There are several expansions involved. First, the uncontracted quark fields must be Taylor expanded. Working in the radial gauge both for the gluon and for the photon, the Taylor expansion \cite{Novikov:1983gd,Pascual:1984zb} of the quark bilinears can be written as:
\begin{equation}\label{eq:quarexp}
\bar{q}_{a}(x_{i})q_{b}(x_{j})=\sum_{m,n}\frac{(-1)^{n}}{n!m!}x_{i,\mu_{1}}\cdots  x_{i,\mu_{n}}\, x_{j,\nu_{1}}\cdots  x_{j,\nu_{m}}
\, \bar{q}_{a}(0) D^{\mu_{1}}\cdots D^{\mu_{n}}D^{\nu_{1}}\cdots D^{\nu_{m}} q_{b}(0) \, .
\end{equation}
Since our computation goes up to dimension $D=6$, we need to expand up to three derivatives. A lower number of derivatives in that expansion can give contributions (apart from the di-quark magnetic susceptibility, which, as shown in Ref.~\cite{Bijnens:2019ghy}, gives a contribution already at $D=4$ when combined with masses) when combined with mass terms from the hard propagators or from soft gluons or photons coming from hard propagators. A first simplification consists in realizing that one can put the gluons and photons together with covariant derivative terms. Extending the results of Refs.~\cite{Novikov:1983gd,Pascual:1984zb}, one finds
\begin{equation}\label{eq:phplusgl}
\bar{q}(x_{i})\Big( B^{\epsilon}(u)+ie_{q}A^{\epsilon}(u)\Big) q(x_{j})=\sum_{p=1}^{\infty}\frac{1}{(p-1)!(p+1)}u^{\omega_{1}} \cdots   u^{\omega_{p}}
\, 
\bar{q}(x_{i}) \Bigg[ D^{\omega_{1}},\Big[ D ^{\omega _{2}}, \ldots,[D^{\omega_{p}},D^{\epsilon}] \Big]  \Bigg] q(x_{j})
\, .
\end{equation}
In fact, it can be shown that for a given flavour the sum of all the contributions that enter into our computation can be reduced to a compact form. Define $\Gamma^{A}$ to be an element in the set of Clifford matrices according to
\begin{equation}\label{eq:spinor}
\Gamma^{A}\in \Bigg\{\mathbb{I},\gamma^{5},\gamma^{\mu},\gamma^{\mu}\gamma_{5},\sigma^{\mu\nu}\equiv\frac{i}{2}[\gamma^{\mu},\gamma^{\nu}] \Bigg\} \, .
\end{equation}
The compact expression for $\Pi^{\mu_{1}\mu_{2}\mu_{3}}$ is then
\begin{align}\nonumber
\Pi^{\mu_{1}\mu_{2}\mu_{3}}(q_{1},q_{2})&=-e_{q}^3\lim_{q_{3}\rightarrow -q_{1}-q_{2}}\sum_{A,p,n,\sigma(1,2,3)}(-1)^{n}\langle 0|\bar{q} D_{\nu_{1}}\ldots D_{\nu_{n}}c_{A}\Gamma^{A}q| \gamma(q_{4})\rangle\,\\
&
\times 
\mathrm{Tr}\left\{\gamma^{\mu_{3}}\Gamma^{A}\gamma^{\mu_{1}}iS(-q_{1})\gamma^{\nu_{1}}iS(-q_{1})\ldots\gamma^{\nu_{p}}iS(-q_{1})\gamma^{\mu_{2}}iS(q_{3})\gamma^{\nu_{p+1}}iS(q_{3})\ldots\gamma^{\nu_{n}}iS(q_{3}) \right\} \, ,\label{eq:onecut}
\end{align}
where $\sigma (1,2,3)$ again denotes a pairwise permutation over $q_{i}$ and $\mu _{i}$. The coefficients $c_{A}$ are defined as
\begin{align}\label{eq:spinornorm}
    c_{A} = \Bigg[ \mathrm{Tr} \Big( \Gamma^{A}\Gamma^{A}\Big) \Bigg] ^{-1} \, ,
\end{align}
such that one in a standard fashion can decompose in a spinor basis according to
\begin{equation}\label{eq:Ddec}
\bar{q}_{i} q_{j}=
\sum_{A}c_{A} \Gamma^{A}_{ji} \, \bar{q}\, \Gamma^{A}   q  \, . 
\end{equation}
Here, all dependence on the other quantum numbers such as colour or flavour has been suppressed. 

The proof of~(\ref{eq:onecut}) up to the order that we need, i.e.~up to $p=3$, can be found in App.~\ref{app:dertrick}. Note that already for $p\leq 3$ the proof involves a very large cancellation of contributions, and the compact form allows for a much simplified calculation of the diagram topologies with one cut quark line.

The reduction of the matrix elements $\langle 0|\bar{q} D_{\nu_{1}}\ldots D_{\nu_{n}}\Gamma^A q| \gamma(q_{4})\rangle$ into the matrix elements of Sec.~\ref{sec:OPEdesc} is rather involved. One needs to recursively exploit spinor algebra relations, symmetry transformations under Lorentz, parity and charge conjugation as well as the equations of motion of the quarks and the gluons. The resulting non-zero matrix elements are of eight types. With zero derivatives we have 
\begin{equation}
\frac{1}{e e_{q}}\langle \bar{q}\sigma^{\alpha_ {1}\alpha_{2}}q\rangle =X^{2}_{S}\langle F^{\alpha_{1}\alpha_{2}}\rangle\, .
\end{equation}
With one derivative one has
\begin{equation}
\frac{1}{e e_{q}}\langle \bar{q}D^{\nu_{1}}\gamma^{\alpha_{1}}\gamma_{5}q\rangle=-\frac{i m_q}{4}X^{2}_{S}\epsilon^{\nu_{1}\alpha_{1}\alpha\beta}\langle F_{\alpha\beta}\rangle \, ,
\end{equation}
and with two they are
\begin{equation}
\frac{1}{e e_{q}}\langle \bar{q}D^{\nu_{1}}D^{\nu_{2}}q\rangle =-\frac{i }{2}\langle F^{\nu_1\nu_2} \rangle \left( X_{S}^{5}-X_{S}^{3}\right)
\, ,
\end{equation}

\begin{equation}
\frac{1}{e e_{q}}\langle \bar{q}D^{\nu_{1}}D^{\nu_{2}}\gamma^{5}q\rangle =-\frac{1 }{4}X^{4}_{S}\epsilon^{\nu_{1}\nu_{2}\alpha\beta}\langle F_{\alpha\beta} \rangle \, ,
\end{equation}

\begin{equation}
\frac{1}{e e_{q}}\langle \bar{q}D^ {\nu_{1}}D^{\nu_{2}}\sigma ^{\alpha_{1}\alpha_{2}} q\rangle =A_{1}g^{\nu_{1}\nu_{2}}\langle F^{\alpha_{1}\alpha_{2}}\rangle 
+A_{2} \Big(
g^{\nu_{1}\alpha_{1}}\langle F^{\nu_{2}\alpha_{2}}\rangle
+g^{\nu_{2}\alpha_{1}}\langle F^{\nu_{1}\alpha_{2}}\rangle
-g^{\nu_{1}\alpha_{2}}\langle F^{\nu_{2}\alpha_{1}}\rangle
-g^{\nu_{2}\alpha_{2}}\langle F^{\nu_{1}\alpha_{1}}\rangle 
\Big)
\, .
\end{equation}
Here, we have defined the linear combinations
\begin{align}
&A_{1}=\frac{-\left(
m_q^2 X_{S}^{2}+X_{S}^{4}\right)+\frac{X^{5}_{S}-X^{3}_{S}}{2}}{3} \, ,
 \\
&A_{2}=\frac{\left(
m_q^2 X_{S}^{2}+X_{S}^{4}\right)+X^{5}_{S}-X^{3}_{S}}{12} \, .
\end{align}
For three derivatives there are two contributions. These are
\begin{align}\label{eq:threederiv}
\frac{1}{e e_{q}}  \,
\langle \bar{q}D^{\nu_{1}}D^{\nu_{2}}D^{\nu_{3}}\gamma^{\nu_{4}}q \rangle 
 =
 &\, 
A_3 \Big(
    g ^{\nu _{1} \nu _{2}} \langle F^{\nu _{3} \nu _{4}} \rangle
    -g ^{\nu _{2} \nu _{3}} \langle  F^{\nu _{1} \nu _{4}} \rangle
\Big) \nonumber
\\ \nonumber
&
+A_4
\Big(
    g^{\nu _{1} \nu _{4}} \langle F^{\nu _{2} \nu _{3}} \rangle
    +g^{\nu _{3} \nu _{4}}  \langle F^{\nu _{1} \nu _{2}} \rangle
\Big)
\\ 
&
+A_5 \, g^{\nu _{2} \nu _{4}} \langle F^{\nu _{1} \nu _{3}} \rangle
\, ,
\\ \nonumber
\frac{1}{e e_{q}}  \, \langle \bar{q}D^{\nu_{1}}D^{\nu_{2}}D^{\nu_{3}}\gamma^{\nu_{4}}\gamma _{5} q \rangle
= 
&  \, 
 A_6\, g^{\nu _{1} \nu _{3}}\langle \bar{F}^{\nu _{2} \nu _{4}} \rangle
\\ \nonumber
&
+A_7
\Big(
    g^{\nu _{1} \nu _{2}}  \langle \bar{F}^{\nu _{3} \nu _{4}} \rangle 
    + g^{\nu _{2} \nu _{3}} \langle \bar{F}^{\nu _{1} \nu _{4}} \rangle
\Big)
\\
&
+A_8
\Big(
    g^{\nu _{1} \nu _{4}} \langle \bar{F}^{\nu _{2} \nu _{3}} \rangle
    -g^{\nu _{3} \nu _{4}}  \langle  \bar{F}^{\nu _{1} \nu _{2}} \rangle
\Big) \, .
\end{align}
Here, $\bar{F}^{\mu\nu}\equiv \frac{i}{2}\epsilon^{\mu\nu\alpha\beta}F^{\alpha\beta}$ and the $A_{i}$ are given by
\begin{align}
& 
A_3 =  
\frac{1}{24}\Big(
- 5 X_S^{8,1} + 2 X_S^{7} - 5 m_q X_S^{4} + 2 m_q X_S^{3}
\Big)
\, ,
\\
&
A_4 = 
\frac{1}{24}\Big(
-  X_S^{8,1} +  X_S^{7} - 3 m_q X^5_S -  m_q X_S^{4} + 4 m_q X_S^{3}
\Big)
\, ,
\\
&
A_5 = 
\frac{1}{24}\Big(
- 2 X_S^{8,1} -  X_S^{7} - 3 m_q X_S^{5} - 2 m_q X_S^{4} + 2 m_q X_S^{3}
\Big)
\, ,
\\
& 
A_6 = 
\frac{1}{24}\Big( 
- 6 X_S^{8,1} +  X_S^{7} -  m_q X_S^{5} - 2 m_q X_S^{4} + 2 m_q X_S^{3} +  2 m_q^3 X_S^{2}
\Big)
\, ,
\\
& 
A_7 =  
\frac{1}{24}\Big(
-  X_S^{8,1} +  X_S^{7} -  m_q X_S^{5} +  m_q X_S^{4} + 2 m_q X_S^{3} + 2 m_q^3 X_S^{2}
\Big)
\, ,
\\
&
A_8 = 
\frac{1}{24}\Big(
 X_S^{8,1} + 3 m_q X_S^{4}
\Big) 
\, .
\end{align}
The operator
\begin{equation}
S^{\mu\nu}_{q8,1}\equiv-\frac{g_s^2}{2}\epsilon^{\mu\nu\lambda\sigma}\sum_A\bar{q}_A\gamma_{\lambda}\frac{\lambda_a}{2}q_A\bar{q}\gamma_{\sigma}\gamma_{5}\frac{\lambda_a}{2}q
\end{equation}
enters from using the gluon equation of motion. Its susceptibility is defined as
\begin{equation}
\langle S^{\mu\nu}_{q8,1}\rangle =e_{q}X^{8,1}_{S}\langle F^{\mu\nu}\rangle\, .
\end{equation}

Using the above decompositions in~(\ref{eq:onecut}) and rewriting the $S_i^{\mu\nu}$ into the $Q_i^{\mu\nu}$ and replacing the $X_{S}^{i}$ with the corresponding $X_i$, one finds 
\begin{align}
\hat{\Pi}_{m}=e^{4}_{q}\, \sum_{i,j,k,n,p}c^{m,n,p}_{i,j,k}\, m_{q}^{n}\, X_{p} \ Q_{1}^{-2i}Q_{2}^{-2j}Q_{3}^{-2k} \, .
\end{align}
The numerical coefficients $c^{m,n,p}_{i,j,k}$ can be found in App.~\ref{app:explicit}.

\subsection{Contributions from four-quark operators}\label{sec:fourquark} 
\begin{figure}[tb]\centering
\includegraphics[width=0.2\textwidth]{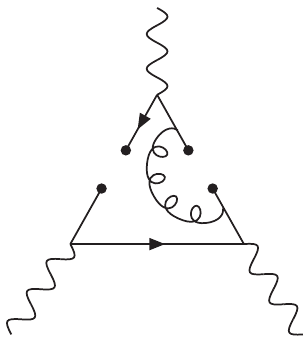}
\caption{\label{fig:fourquark}Contributions from four-quark operators obtained cutting two quark lines. All possible ways to connect the gluon to the quark lines must be considered.}
\end{figure}

The four-quark operator contributions arise from cutting two quark lines.\footnote{Afterwards one needs to add the one coming from the one cut line and the gluon equation of motion.} The resulting diagrams become split in two parts and, as a consequence, introduce flavour mixing. An extra gluon propagator needs to be included from the Dyson series expansion to connect the quark lines. The resulting contribution, up to permutations of $\sigma(1,2,3)$, is shown in Fig.~\ref{fig:fourquark} where the gluon can be connected in three different positions in the quark line above and in two different positions in the one below. These diagram contributions can be compactly written as
\begin{align}\begin{split}
\Pi^{\mu_1\mu_2\mu_3}(q_{1},q_{2})&=-\frac{1}{16}\sum _{A,B}
\Big\langle e_{q_A}^{2}\, \bar{q}_{A}\frac{\lambda_{a}}{2}\Gamma^{\omega_{1}P}q_{A}\, e_{q_B} \bar{q}_{B}\frac{\lambda_{a}}{2}\Gamma^{\omega_{2}Q}q_{B}\Big\rangle
\\
&
\times \lim_{q_{3}\rightarrow-q_{1}-q_{2}}\sum_{\sigma(1,2,3)}\frac{1}{q_{3}^{2}}
\mathrm{Tr}\Bigg[\Gamma_{Q\omega_{2}}
\Big(
\gamma^{\mu_{3}}S(-q_{3})\gamma^{\epsilon}+\gamma^{\epsilon}S(q_{3})\gamma^{\mu_{3}}
\Big)
\Bigg]
\\
&
\times \mathrm{Tr}
\Bigg[
-\Gamma_{P,\omega_{1}}
\Big(
\gamma^{\mu_{1}}S(-q_{1})\gamma^{\mu_{2}}S(q_{3})\gamma^{\epsilon}+\gamma^{\epsilon}S(-q_{3})\gamma^{\mu_{1}}S(q_{2})\gamma^{\mu_{2}}+\gamma^{\mu_{1}}S(-q_{1})\gamma^{\epsilon}S(q_{2})\gamma^{\mu_{2}}
\Big)  
\Bigg]
\, ,
\end{split}\end{align}
where $\Gamma^{\omega P}\in \{\gamma^{\omega},\gamma^{\omega}\gamma_{5}\}$ and $\Gamma_{P}^{\omega}\in \{\gamma^{\omega},-\gamma^{\omega}\gamma_{5}\}$. In fact charge conjugation requires that $A$ and $B$ must be different to get a non-zero matrix element and one of the remaining two possible contributions vanishes when taking the traces. Recalling the definition of $\overline{X}_{8,2}$ in~(\ref{eq:X82bardef}) we find
\begin{align}
& \hat{\Pi}_{1}=\hat{\Pi}_{4}=8\overline{X}_{8,2}\frac{Q_{1}^{2}+Q_{2}^{2}}{Q_{1}^{4}Q_{2}^{4}Q_{3}^{2}}
\, ,\\
& \hat{\Pi}_{54}=8\overline{X}_{8,2}\frac{Q_{2}^{4}-Q_{1}^{4}}{Q_{1}^{6}Q_{2}^{6}Q_{3}^{2}} 
\, ,\\
& \hat{\Pi}_{7}=\hat{\Pi}_{17}=\hat{\Pi}_{39}=0\, .
\end{align}
A reduction of all possible four-quark matrix elements into a basis of $12$ independent ones is given in App.~\ref{app:fourquark} for the chiral limit, i.e.~$m_u=m_d=m_s=0$.

\subsection{The gluon matrix element}
\begin{figure}[t!]\centering
\includegraphics[width=0.2\textwidth]{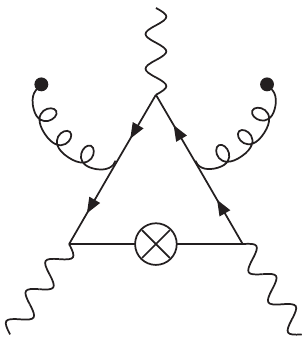}
\caption{\label{fig:gluonconddiag} An example of a topology of the gluon matrix element contributions.}
\end{figure}

The gluon matrix element contribution arises from all the possible combinations in which, starting from~(\ref{eq:backdyson1}), one extra QED and two extra QCD vertices are added (see Fig.~\ref{fig:gluonconddiag}). The gauge boson fields $F_{\mu\nu}$, $G^{a}_{\mu\nu}$ and $G^{b}_{\mu\nu}$ are then Taylor expanded according to~(\ref{eq:backgauge}). Since all the quark fields are connected, the colour chain always leads to the same colour trace, namely
\begin{equation}
\mathrm{Tr}\left(\frac{\lambda^{a}}{2}\frac{\lambda^{b}}{2} \right)=\frac{\delta^{ab}}{2} \, .
\end{equation}
Once the colour and the space-time terms from the Taylor expansions have been factored out, the remaining six-point function is fully symmetric under the exchange of indices. Taking advantage of this symmetry, one can rewrite the whole contribution as a sum of permutations according to
\begin{align}\begin{split}
\Pi^{\mu_{1}\mu_{2}\mu_{3}}_{GG}(q_1,q_2)
&
=e_q^4F_{\nu_{4}\mu_{4}}\frac{4\pi^{2}\langle \alpha_{s}G^{\mu\nu}_{a}G_{\mu\nu}^a\rangle}{32 d(d-1)}
\left(
g_{\nu_{5}\nu_{6}}g_{\mu_{5}\mu_{6}}-g_{\mu_{5}\nu_{6}}g_{\nu_{5}\mu_{6}}
\right)
\left(\prod_{i=4}^{6}\lim_{q_{i}\rightarrow 0} \frac{\partial}{\partial q_{i}^{\nu_{i}}}\right)
\int \frac{d^{4}p}{(2\pi)^d} 
\\
&
\times \sum_{\sigma(1,2,4,5,6)} \mathrm{Tr}\Bigg(\gamma^{\mu_{3}}S(p+q_1+q_2+q_4+q_5+q_6)\gamma^{\mu_1}S(p+q_2+q_4+q_5+q_6)
\\
&
\qquad \,\times \gamma^{\mu_2}S(p+q_4+q_5+q_6)\gamma^{\mu_4}S(p+q_5+q_6)\gamma^{\mu_5}S(p+q_6)\gamma^{\mu_6}S(p)\Bigg)\, .
\end{split}
\end{align}
Here, $\sigma (1,2,4,5,6)$ is the set of pairwise permutations of $\mu_{i}$ and $q_{i}$ for $i=1,\, 2,\, 3, \, 5, \, 6$. Also, the equation has been written in terms of $d=4-2\epsilon$ for renormalisation purposes. Using~(\ref{eq:dertrick}) iteratively in the above equation, then taking the momentum limits, calculating the Dirac trace and projecting the results into the $\hat{\Pi}_{i}$, one finds the results in terms of ultraviolet finite integrals. We do this by reducing the loop integrals to combinations of triangle, self-energy and tadpole integrals with the help of the package \texttt{KIRA}, all the time carefully performing both the expansions in $\epsilon$ and in the quark masses. Without including operator mixing the result takes the form
\begin{equation}
\hat{\Pi}_{GG m,S}=X_{6,S}\,e_{q}^{4}\sum_{i,j,k}\left[c_{i,j,k}^{(m)}+f_{i,j,k}^{(m)}\, m_{q}^{-2}+g^{(m)}_{i,j,k}\log\left(\frac{Q_{1}^2}{Q_{2}^{2}}\right)+h^{(m)}_{i,j,k}\log\left(\frac{Q_{3}^2}{m_{q}^{2}}\right)\right] Q_{1}^{-2i}\,Q_{2}^{-2j}\,Q_{3}^{-2k} \, , 
\end{equation}
where $c,f,g$ and $h$ are numerical coefficients given in App.~\ref{app:explicit}. 

When operator mixing is taken into account, the found divergences, which scale as $ \frac{1}{m_{q}^{2}}$ and $\log\frac{Q^2}{m_{q}^2}$, exactly cancel respectively with the $X_{1}$ and $X_{7}$ contributions. The dependence on the triangle integral also cancels, leading to a fully analytic gluon matrix element contribution. One should note that the final expressions for this contribution are very simple, even compared to the quark loop, and there are substantial cancellations along the way that lead to this simple form.

\section{Numerical results}\label{sec:num}

In this section we present numerical results for the full $a_{\mu}^{\textrm{HLbL}}$ integral in~(\ref{eq:amuint}). Using the relations in~(\ref{eq:pibarfcns}) between the set of functions $\bar{\Pi}_{i}$ and the $\hat{\Pi}_{i}$ given in the appendices as well as Sec.~\ref{sec:fourquark}, we evaluate $a_{\mu}^{\textrm{HLbL}}$ for the matrix element as well as loop contributions. We use the matrix elements as estimated in Sec.~\ref{sec:meests}. For this purpose, we use the {\sc Cuba} library~\cite{HAHN200578}, in particular the {\tt Vegas} integrator building on Monte Carlo sampling the three-dimensional integral. The results have been checked as well with an adaptive deterministic integrator implemented by us. Care has to be taken in the numerics since $\lambda$ can vanish or get very small and appears with rather high negative powers in some expressions. Those areas in the integration have to be treated by expanding the loop functions around the $\lambda=0$ points analytically, some of these limits are given explicitly in App.~\ref{app:quarklooplimits}.

We investigate the various contributions first at two benchmark values of the lower momentum cut-off, i.e.~$Q_{1,2,3}>Q_{\textrm{min}}\in\{ 1,2\} $ GeV. Then we proceed to investigate how the different pieces scale with $Q_{\textrm{min}}$ and compare the respective sizes. For notational convenience, we refer to the contributions with respect to the corresponding $X_{i}$.

\begin{figure}[tbh]\centering
\includegraphics[width=0.85\textwidth]{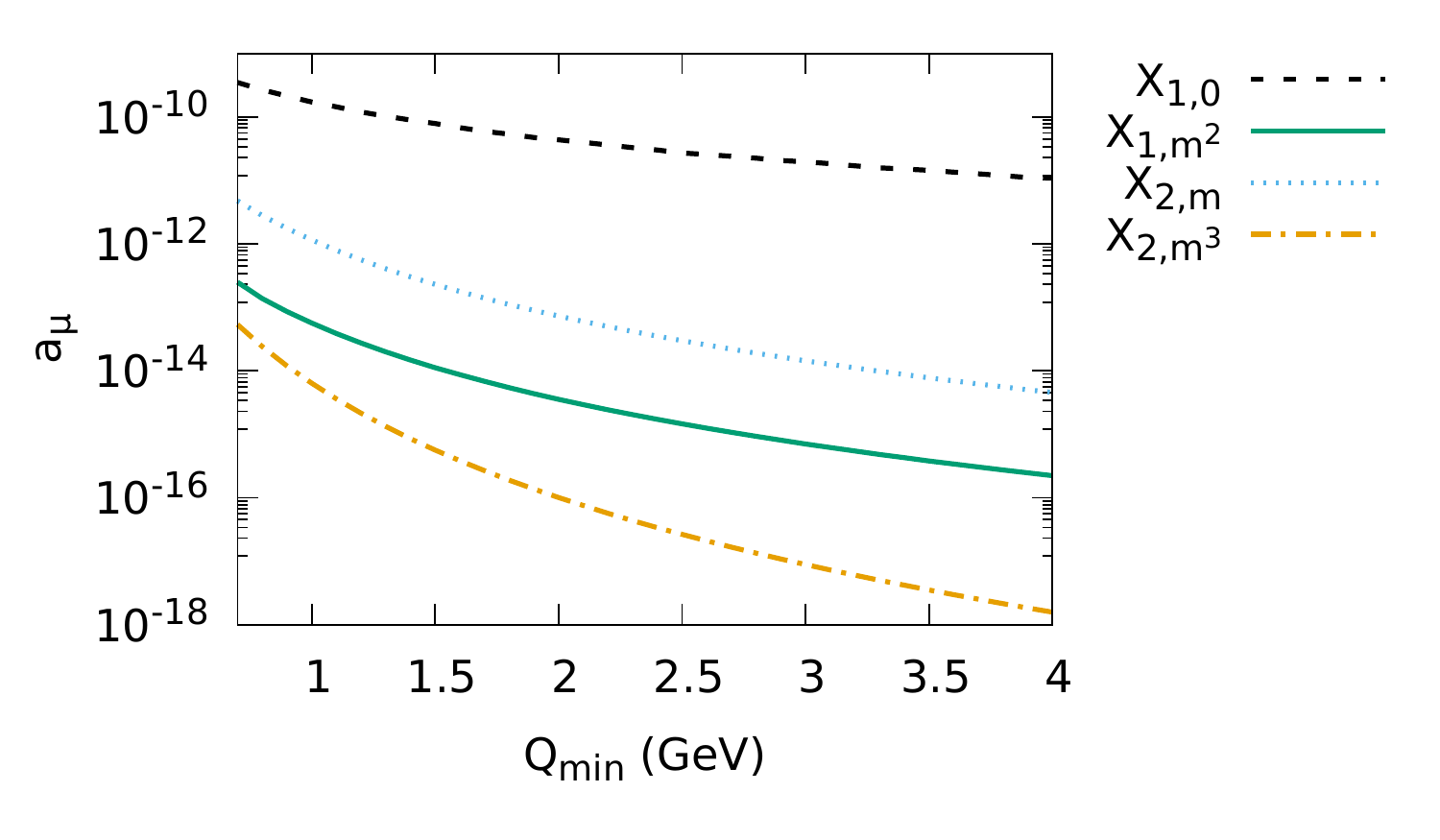}
\caption{\label{fig:num1}Numerical contributions from the $X_{1}$ and $X_{2}$ pieces in absolute value using the inputs from Table~\ref{tab:num}. As expected, the quark loop fully dominates.}
\end{figure}

\begin{figure}[tbh]\centering
\includegraphics[width=0.85\textwidth]{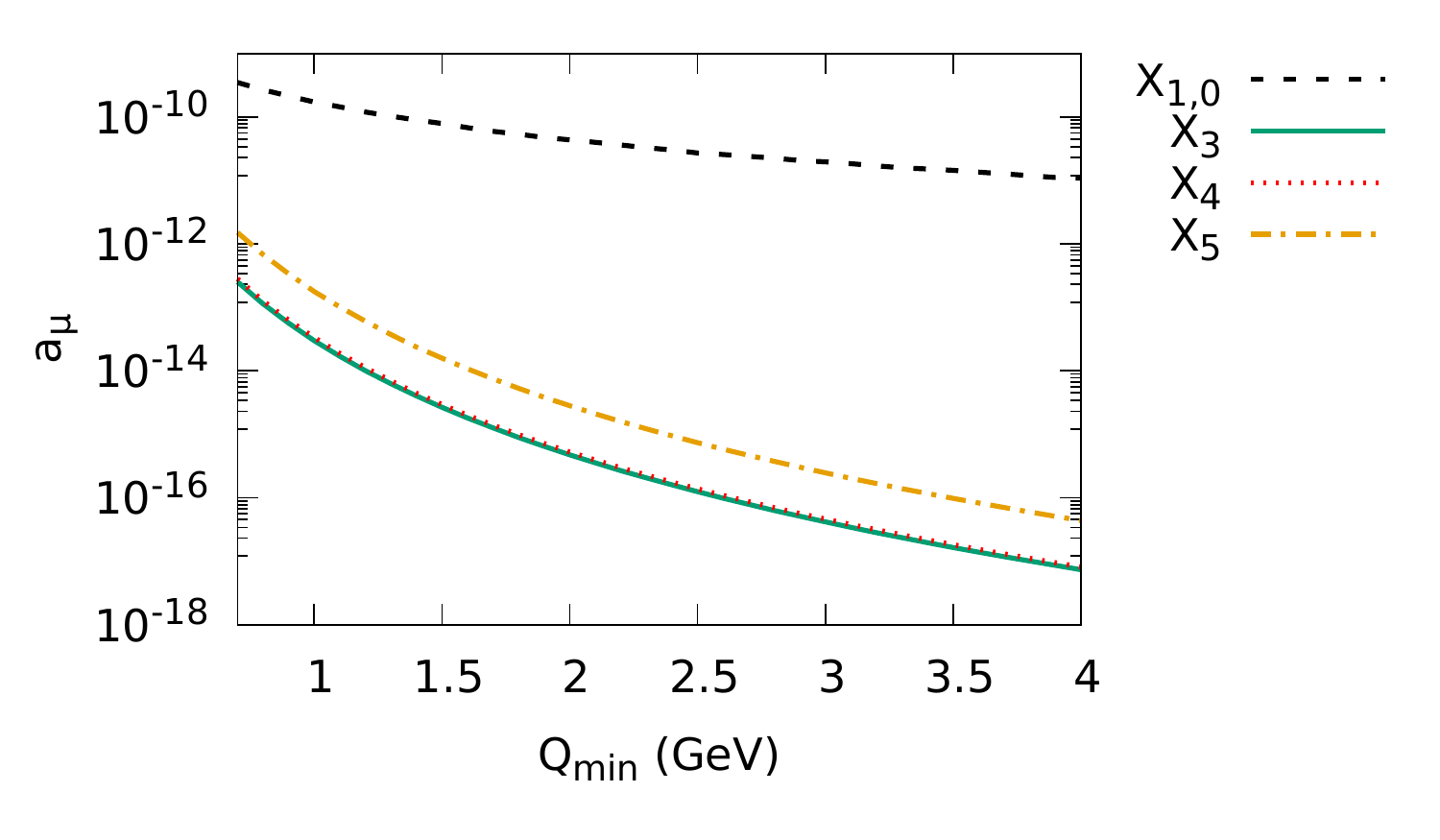}
\caption{\label{fig:num2}Numerical contributions from the $X_{3-5}$ pieces in absolute value using the inputs from Table~\ref{tab:num}. The massless quark loop is shown for comparison.}
\end{figure}

\begin{figure}[tbh]\centering
\includegraphics[width=0.85\textwidth]{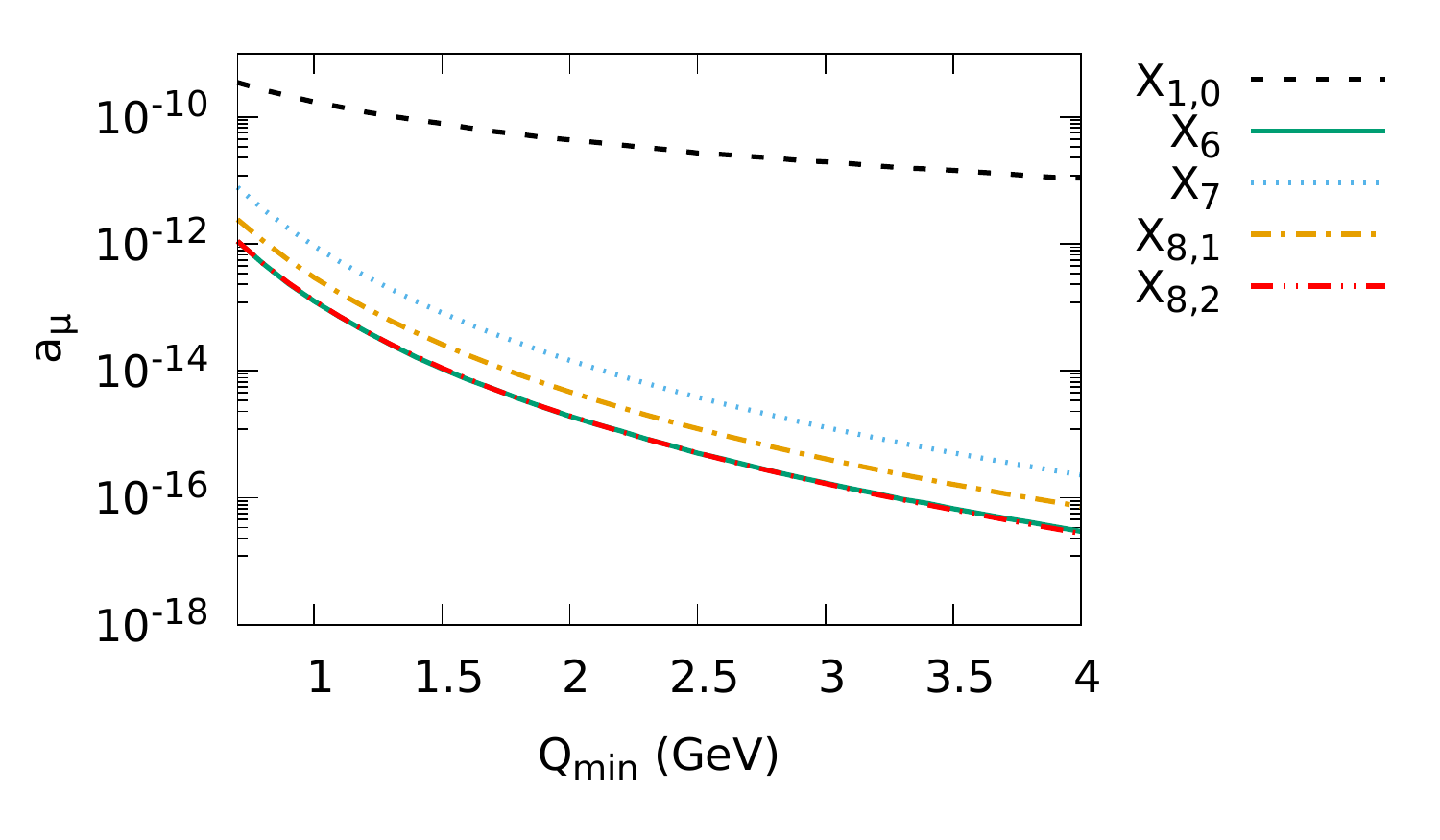}
\caption{\label{fig:num3}Numerical contributions from the $X_{6-8}$ pieces in absolute value using the inputs from Table~\ref{tab:num}. Even when they are not suppressed by the quark mass size, they are found to be small compared with the massless quark loop contribution.}
\end{figure}

At the sought precision level it is sufficient to assess the order of magnitude of these corrections. Therefore, in want of precise input we resort to simplified input as discussed in the previous section. For this purpose, we use
\begin{align}
    & m_{u}=m_{d}= 5 \, \mathrm{MeV} \, , 
    & m_{s}=\mathrm{100}\,\mathrm{MeV} \,  , 
    & \qquad \quad \mu = Q_{\mathrm{min}} \, , 
    & \alpha_{s}=0.33 \, . &
\end{align}
Given the smallness of the matrix element and quark mass correction contributions we did not take into account the running with $\mu$ of the various inputs but kept them fixed.
The benchmark points for $Q_{\textrm{min}}\in\{ 1,2\} $ GeV are presented in Table~\ref{tab:num}. Since the contributions from the matrix element $X_2$ come in both suppressed at order $m_{q}$ and at order $m_{q}^3$, we here present the respective contributions, labelled $X_{2,m}$ and $X_{2,m^3}$, of these. The table shows that power correction are suppressed by at least two orders of magnitude with respect to the quark loop. This is also visible in Figs.~\ref{fig:num1}--\ref{fig:num3} where we consider the scaling with $Q_{\textrm{min}}$. 
The $T_i\left(Q_1,Q_2,\tau\right)$ in~(\ref{eq:amuint}), when expanded for large $Q_i$, are of order $m_\mu^2$, except for $T_1$ which is $m_\mu^4$. The variation with $Q_{\textrm{min}}$ from dimensions is thus $1/Q_{\textrm{min}}^2$ for the massless quark loop, $1/Q_{\textrm{min}}^4$ for the $d=4$ contributions and $1/Q_{\textrm{min}}^6$ for the $d=6$ contributions. The scaling is found to agree with naive dimensional counting.

The power corrections not suppressed by quark masses, i.e.~$X_{6}$, $X_{7}$, $X_{8,1}$ and $X_{8,2}$, are found to be numerically suppressed. This is partially explained by their extra suppression in powers of $\Lambda_{\mathrm{QCD}}$. Their numerical impact is similar to the one of the di-quark magnetic susceptibility, $X_{2}$, and clearly more important than the perturbative mass contributions. We find that even at $1$ GeV, all these power corrections are suppressed by at least two orders of magnitude with respect to the massless quark loop. Even though this result motivates studying whether the smallness of the corrections pinpoint a trend also for the purely perturbative ones, no strong conclusions should be derived from it.

\begin{table}[tbh]\begin{center}
\begin{tabular}{|l|l|r|r|}
\hline
   Contribution 
   & Inputs (GeV units)                                                   
   & $Q_{\textrm{min}}=1 \,\mathrm{GeV}$ 
   & $Q_{\textrm{min}}=2 \,\mathrm{GeV}$ 
   \\ \hline
   $X_{1,0}$   
   &                                                                
   & $1.73\cdot 10^{-10}$               
   & $4.35\cdot 10^{-11}$                
   \\ \hline
   $X_{1,m^2}$   
   &                                                                
   & $-5.7 \cdot 10^{-14}$               
   & $-3.6 \cdot 10^{-15}$                
   \\ \hline
   $X_{2,m}$    
   & $X_2=-4\cdot 10^{-2}$                                
   & $-1.2\cdot 10^{-12}$               
   & $-7.3\cdot 10^{-14}$                
   \\ \hline
   $X_{2,m^3}$    
   & $X_2=-4\cdot 10^{-2}$                                                
   & $6.4\cdot 10^{-15}$
   & $1.0\cdot 10^{-16}$                
   \\ \hline
   $X_{3}$      
   & $X_3=3.51\cdot 10^{-3}$ 
   & $-3.0\cdot 10^{-14}$
   & $-4.7\cdot 10^{-16}$             
   \\ \hline
   $X_{4}$      
   &  $X_4=3.51\cdot 10^{-3}$                         
   & $3.3\cdot 10^{-14}$ 
   & $5.3\cdot 10^{-16}$
   \\ \hline
   $X_{5}$      
   & $X_5=- 1.56 \cdot 10^{-2} $                                                   
   & $-1.8\cdot 10^{-13}$ 
   & $-2.8\cdot 10^{-15}$ 
   \\ \hline
   $X_{6}$      
   & $X_6=2\cdot 10 ^{-2}$                                          
   & $1.3 \cdot 10^{-13}$    
   & $2.0\cdot 10^{-15}$         
   \\ \hline
   $X_{7}$      
   &  $X_7=3.33\cdot 10 ^{-3} $                                
   & $9.2\cdot 10^{-13}$     
   & $1.5\cdot 10^{-14}$   
   \\ \hline
   $X_{8,1}$    
   & $\overline{X}_{8,1}=-1.44\cdot 10^{-4}$  & $3.0\cdot 10^{-13}$             &    $4.7\cdot 10^{-15}$         \\ \hline
$X_{8,2}$    & $\overline{X}_{8,2}=-1.44\cdot 10^{-4} $ &   $-1.3\cdot 10^{-13}$              &  $-2.0\cdot 10^{-15}$ \\  
\hline
\end{tabular}\end{center}
 \caption{\label{tab:num}Numerical results for $a_{\mu}^{\textrm{HLbL}}$ for the indicated inputs. }
\end{table}

\section{Conclusions and prospects}\label{sec:pros}

The leading asymptotic behaviour of the HLbL for the $(g-2)_{\mu}$ kinematics has been confirmed to be given by the massless quark loop contribution. Although this had been commonly assumed previously, this is by no means obvious due to the static limit associated to the $(g-2)_{\mu}$ definition with the external photon leg at zero momentum. The main result of this work and of Ref.~\cite{Bijnens:2019ghy} is to show how a proper short-distance expansion can be done in the limit of $Q_1^2\sim Q_2^2\sim Q_3^2 \gg \Lambda_{QCD}^2$.

In order to show that the quark loop is the first order of a well-defined expansion, the soft photon has been formulated as a long-distance, or, background, degree of freedom, following previous works of Refs.~\cite{Ioffe:1983ju,Balitsky:1983xk}. We stress that using the vacuum OPE valid for HLbL when all the four Euclidean momenta are large, one would for the $(g-2)_{\mu}$ kinematics have obtained a divergent expansion~\cite{Bijnens:2019ghy}.

A comprehensive description of the resulting OPE has been provided, including a detailed explanation on how to achieve a complete and systematic separation of short and long distance effects while cancelling internal divergences. The physical meaning of some of the resulting matrix elements is also given, and we have used those to present estimates of all of them. The obtained results could in the future be used to analyse other Green functions and their phenomenological applications.

The resulting OPE is applied to the HLbL in the $(g-2)_{\mu}$ kinematics for $Q_1,Q_2,Q_3\gg \Lambda_{\mathrm{QCD}}$. As a consequence of setting one of the momenta to zero, the long distance effects become functionally more important. The quark loop is still found to be the dominant contribution, but the first non-perturbative correction becomes suppressed by just one power of $\Lambda_{\mathrm{QCD}}$ (plus one power of $m_{q}$), in contrast with the $\Lambda_{\mathrm{QCD}}^3$ ($\bar{q}q$) suppression in the OPE applicable when all the Euclidean momenta are large. 

However, no operators allowed by the symmetries are found to enter without quark mass suppression below $\Lambda_{\mathrm{QCD}}^4$ with respect to the quark loop contribution. These $\Lambda_{\mathrm{QCD}}^4$ contributions are computed and their role for the $(g-2)_{\mu}$ is estimated. They are found to be very small. Whether or not this may be indicating that the quark loop gives a precise description of the HLbL at relatively low momenta (i.e.~$Q_{i}\sim 1 \, \mathrm{GeV}$) will only be known once the two-loop perturbative corrections have been computed. This calculation is already under way and will be presented in a future publication.

\section*{Acknowledgments}
N.~H.--T. and L.L.~are funded by the Albert Einstein Center for Fundamental Physics at Universit\"{a}t Bern and the Swiss National Science Foundation respectively. J.~B.~and A.~R.--S.~are supported in part by the Swedish Research Council grants contract numbers 2016-05996 and 2019-03779, and by the European Research Council (ERC) under the European Union’s Horizon 2020 research and innovation programme under grant agreement No 668679.

\newpage
\appendix
\renewcommand{\theequation}{\thesection.\arabic{equation}}
\section{A set of Lorentz projectors for the $\hat{\Pi}_{i}$}\label{app:proj}
In this appendix we present a set of projectors useful for projecting to the set of $\hat{\Pi}_{i}$. Note that due to gauge invariance this set is not unique. We have derived and used a second set of projectors. Obtaining the same results with both projectors is one of the checks we did. Below we only present one of the sets of projectors.
\begin{align}
P_{\hat{\Pi}_1}^{\mu \nu \lambda \rho \sigma}(q_1,q_2,q_3)=&-8\lambda ^{-2} q_1^{\nu} q_1^{\sigma} q_2^{\lambda} q_2^{\rho} q_3^{\mu}  + 2 \lambda^{-1} g^{\mu \sigma} g^{\nu \rho} q_2^{\lambda} \nonumber \\
& - 8 \lambda ^{-2 } q_2^2 g^{\mu \sigma} q_1^{\nu} q_2^{\lambda} q_3^{\rho}   - 4 \lambda ^{-2 }\left( q_3^2 + q_2^2 -  q_1^2 \right) g^{\mu \sigma} q_1^{\lambda} q_2^{\rho} q_3^{\nu}     \nonumber\\
& -8 \lambda ^{-2 }q_1^2 g^{\nu \sigma} q_1^{\lambda} q_2^{\mu} q_3^{\rho}-4\lambda ^{-2} \left( q_3^2 -  q_2^2 + q_1^2\right)g^{\nu \sigma} q_1^{\rho} q_2^{\lambda} q_3^{\mu}  \, ,
\end{align}

\begin{align}
    &P_{\hat{\Pi}_4}^{\mu \nu \lambda \rho \sigma}(q_1,q_2,q_3)=8\lambda ^{-4 } \left(  6 q_3^8 + 11 q_2^2 q_3^6 - 29 q_2^4 q_3^4 +  q_2^6 q_3^2 + 11 q_2^8 + 11 q_1^2 q_3^6 + 14 q_1^2 q_2^2 q_3^4 \right. \nonumber \\
    &\left. \qquad \qquad -  q_1^2 q_2^4 q_3^2 - 44 q_1^2  q_2^6 - 29 q_1^4 q_3^4 - q_1^4 q_2^2 q_3^2 + 66 q_1^4 q_2^4
    + q_1^6 q_3^2 - 44 q_1^6 q_2^2 + 11 q_1^8 \right) q_1^{\nu} q_1^{\sigma} q_2^{\lambda} q_2^{\rho} q_3^{\mu} \nonumber \\ 
    &+ \lambda ^{-2 } \left(q_3^4 - 6 q_2^2 q_3^2 -  q_2^4 + 2 q_1^2 q_2^2 - q_1^4\right) g^{\mu \nu} g^{\lambda \sigma} q_1^{\rho} +\lambda ^{-2}\left(q_3^4 - q_2^4 - 6 q_1^2 q_3^2 + 2 q_1^2 q_2^2 - q_1^4\right)g^{\mu \nu} g^{\lambda \sigma} q_2^{\rho} \nonumber \\  
    &-    4\lambda ^{-3 } \left( q_3^6 + 3 q_2^2 q_3^4 - 4 q_2^4 q_3^2 + 3 q_1^2 q_3^4 + 8 q_1^2 q_2^2 q_3^2 - 4  q_1^4 q_3^2 \right) g^{\mu \nu} q_1^{\lambda} q_1^{\sigma} q_2^{\rho} \nonumber \\    
    &+  2 \lambda ^{-2 }\left( q_2^2 q_3^2 -  q_1^2 q_3^2\right) g^{\mu \lambda} g^{\nu \sigma} q_1^{\rho} + \lambda ^{-2}\left( q_3^4 - q_2^4 - 2 q_1^2 q_3^2 + 2 q_1^2 q_2^2 - q_1^4\right)g^{\mu \lambda} g^{\nu \sigma} q_3^{\rho} \nonumber \\  
    &+  2\lambda ^{-3 }\left( 3 q_3^6 +  q_2^2 q_3^4 -  q_2^4 q_3^2 - 3 q_2^6 - 3 q_1^2 q_3^4 + 4 q_1^2 q_2^2 q_3^2 + 9 q_1^2 q_2^4 - 3 q_1^4 q_3^2 - 9 q_1^4 q_2^2 + 3 q_1^6 \right) g^{\mu \lambda} q_1^{\rho} q_3^{\nu} q_3^{\sigma}    \nonumber \\ 
    &+  2 \lambda ^{-2 }\left( q_3^4 - 4 q_2^2 q_3^2 - q_2^4 + 2 q_1^2 q_2^2 - q_1^4 \right)g^{\mu \rho} g^{\lambda \sigma} q_1^{\nu} \nonumber \\
    &-   2 \lambda ^{-2 }\left(q_2^2 q_3^2 -  q_1^2 q_3^2 \right)g^{\mu \sigma} g^{\nu \lambda} q_2^{\rho}  +\lambda ^{-2 } \left(q_3^4 - 2 q_2^2 q_3^2 - q_2^4 + 2 q_1^2 q_2^2 - q_1^4 \right)  g^{\mu \sigma} g^{\nu \lambda} q_3^{\rho}\nonumber \\    
    &-  10 \lambda ^{-3 }\left(q_3^6 - q_2^2 q_3^4 - 3 q_2^4 q_3^2 - q_2^6 - q_1^2 q_3^4 + 4 q_1^2 q_2^2 q_3^2 + 3 q_1^2 q_2^4 - q_1^4 q_3^2 - 3 q_1^4 q_2^2 + q_1^6 \right) g^{\mu \sigma} q_1^{\nu} q_2^{\lambda} q_3^{\rho} \nonumber \\ 
    &- 4\lambda ^{-3 }\left(2 q_3^6 - 9 q_2^2 q_3^4 - 3 q_2^4 q_3^2 + q_1^2 q_3^4 + 6 q_1^2 q_2^2 q_3^2 - 3 q_1^4 q_3^2 \right) g^{\mu \sigma} q_1^{\lambda} q_2^{\rho} q_3^{\nu}  \nonumber \\  
    &+  2\lambda ^{-3 }\left(3 q_3^6 - 3 q_2^2 q_3^4 - 3 q_2^4 q_3^2 + 3 q_2^6 + q_1^2 q_3^4 + 4 q_1^2 q_2^2 q_3^2- 9 q_1^2 q_2^4 - q_1^4 q_3^2 +9 q_1^4 q_2^2 - 3 q_1^6 \right)  g^{\nu \lambda} q_2^{\mu} q_2^{\sigma} q_3^{\rho}  \nonumber \\  
    &+  2 \lambda ^{-2 } \left( q_3^4 - q_2^4 - 4 q_1^2 q_3^2 +2 q_1^2 q_2^2 - q_1^4 \right)g^{\nu \sigma} g^{\lambda \rho} q_3^{\mu} \nonumber \\  
    &-   10 \lambda ^{-3 }\left( q_3^6 - q_2^2 q_3^4 - q_2^4 q_3^2 + q_2^6 - q_1^2 q_3^4 + 4 q_1^2 q_2^2  q_3^2 - 3 q_1^2 q_2^4 - 3 q_1^4 q_3^2 + 3 q_1^4 q_2^2 - q_1^6 \right) g^{\nu \sigma} q_1^{\lambda} q_2^{\mu} q_3^{\rho}  \nonumber \\ 
    &-   4\lambda ^{-3 }\left(2 q_3^6 +  q_2^2 q_3^4 - 3 q_2^4 q_3^2 - 9 q_1^2 q_3^4 + 6 q_1^2 q_2^2 q_3^2 - 3 q_1^4 q_3^2 \right) g^{\nu \sigma} q_1^{\rho} q_2^{\lambda} q_3^{\mu}   \nonumber \\ 
    &+ \lambda ^{-3 } \left(6 q_3^6 - 6 q_2^2 q_3^4 - 6 q_2^4 q_3^2 + 6 q_2^6 - 50 q_1^2 q_3^4 + 72 q_1^2 q_2^2 
    q_3^2 + 10 q_1^2 q_2^4 + 22 q_1^4 q_3^2 \right. \nonumber \\
    &\qquad \left. - 38 q_1^4 q_2^2 + 22 q_1^6 \right) g^{\lambda \sigma} q_1^{\nu} q_2^{\rho} q_3^{\mu} \qquad + \lambda ^{-3}\left( 6 q_3^6 - 50 q_2^2 q_3^4 + 22 q_2^4 q_3^2 + 22 q_2^6 - 6 q_1^2 q_3^4  \right. \nonumber \\
    &\qquad \left.+ 72 q_1^2 q_2^2 q_3^2- 38 q_1^2 q_2^4 - 6 q_1^4 q_3^2 + 10 q_1^4 q_2^2 + 6 q_1^6 \right) g^{\lambda \sigma} q_1^{\rho} q_2^{\mu} q_3^{\nu} \, , 
\end{align}

\begin{align}
    &P_{\hat{\Pi}_7}^{\mu \nu \lambda \rho \sigma}(q_1,q_2,q_3)= 80\lambda ^{-4 } \left(  - 2 q_3^6- q_2^2 q_3^4 +  q_2^4 q_3^2 + 2 q_2^6 + 2 q_1^2 q_3^4 - 3 q_1^2 q_2^2 q_3^2 - 6 q_1^2 q_2^4 + 2 q_1^4 q_3^2\right. \nonumber  \\
    &  \left.   + 6 q_1^4  q_2^2 - 2 q_1^6 \right) q_1^{\nu} q_1^{\sigma} q_2^{\lambda} q_2^{\rho} q_3^{\mu}-   2\lambda ^{-2 }\left( q_3^2 +  q_2^2 -  q_1^2 \right) g^{\mu \nu} g^{\lambda \sigma} q_1^{\rho} +  2 \lambda ^{-2 }\left(2 q_3^2 -  q_2^2 +  q_1^2\right)  g^{\mu \nu} g^{\lambda \sigma} q_2^{\rho} \nonumber \\ 
    &+  20\lambda ^{-3}\left( q_3^4 +  q_2^2 q_3^2 -  q_1^2 q_3^2 \right)  g^{\mu \nu} q_1^{\lambda} q_1^{\sigma} q_2^{\rho} +  2 \lambda ^{-2 } q_3^2 g^{\mu \lambda} g^{\nu \sigma} q_1^{\rho} \nonumber \\ 
    &-   2\lambda ^{-2 }\left( q_2^2 -  q_1^2 \right) g^{\mu \lambda} g^{\nu \sigma} q_3^{\rho} -  4\lambda ^{-3 } \left( 3 q_3^4 + 4 q_2^2 q_3^2 + 3 q_2^4 - 6 q_1^2 q_3^2 - 6 q_1^2 q_2^2 + 3 q_1^4 \right)g^{\mu \lambda} q_1^{\rho} q_3^{\nu} q_3^{\sigma} \nonumber \\ 
    &- 4 \lambda ^{-2 }\left( q_3^2 + q_2^2 - q_1^2\right)  g^{\mu \rho} g^{\lambda \sigma} q_1^{\nu} -  2 \lambda ^{-2 }q_3^2 g^{\mu \sigma} g^{\nu \lambda} q_2^{\rho} -   2 \lambda ^{-2 } \left(q_3^2 +  q_2^2 -  q_1^2 \right) g^{\mu \sigma} g^{\nu \lambda} q_3^{\rho} \nonumber \\ 
    &+  20 \lambda ^{-3 }\left( q_3^4 + 2 q_2^2 q_3^2 +  q_2^4 - 2 q_1^2 q_3^2 - 2 q_1^2 q_2^2 +  q_1^4\right)  g^{\mu \sigma} q_1^{\nu} q_2^{\lambda} q_3^{\rho} + 20 \lambda ^{-3 }\left(q_3^4 + q_2^2 q_3^2 - q_1^2 q_3^2 \right) g^{\mu \sigma} q_1^{\lambda} q_2^{\rho} q_3^{\nu} \nonumber \\ 
    &-  4\lambda ^{-3 }\left(2 q_3^4 +  q_2^2 q_3^2 - 3 q_2^4 +  q_1^2 q_3^2 + 6 q_1^2 q_2^2 - 3 q_1^4 \right) g^{\nu \lambda} q_2^{\mu} q_2^{\sigma} q_3^{\rho} +  4 \lambda ^{-2 } \left(q_3^2 - q_2^2 + q_1^2 \right) g^{\nu \sigma} g^{\lambda \rho} q_3^{\mu} \nonumber \\ 
    &+  20 \lambda ^{-3 }\left( q_2^2 q_3^2 - q_2^4 - q_1^2 q_3^2 + 2 q_1^2 q_2^2 - q_1^4 \right) g^{\nu \sigma} q_1^{\lambda} q_2^{\mu} q_3^{\rho} -  20 \lambda ^{-3 } \left( q_3^4 -  q_2^2 q_3^2 +  q_1^2 q_3^2 \right) g^{\nu \sigma} q_1^{\rho} q_2^{\lambda} q_3^{\mu} \nonumber \\ 
    &+  4\lambda ^{-3 }\left(6 q_3^4 - 7 q_2^2 q_3^2 +  q_2^4 + 3 q_1^2 q_3^2 + 8 q_1^2 q_2^2 - 9 q_1^4 \right) g^{\lambda \sigma} q_1^{\nu} q_2^{\rho} q_3^{\mu} \nonumber \\ 
    &-  4\lambda ^{-3 }\left(3 q_3^4 + 4 q_2^2 q_3^2 - 7 q_2^4 - 6 q_1^2 q_3^2 + 4 q_1^2 q_2^2 + 3 q_1^4 \right) g^{\lambda \sigma} q_1^{\rho} q_2^{\mu} q_3^{\nu} \, ,
\end{align}

\begin{align}
    &P_{\hat{\Pi}_{17}}^{\mu \nu \lambda \rho \sigma}(q_1,q_2,q_3)=80 \lambda ^{-3 } \left( q_3^2 - q_2^2 - q_1^2 \right) q_1^{\nu} q_1^{\sigma} q_2^{\lambda} q_2^{\rho} q_3^{\mu} - 2 \lambda ^{-2 }\left( q_3^2 - 3 q_2^2 - q_1^2 \right) g^{\mu \nu} g^{\lambda \sigma} q_1^{\rho} \nonumber \\ 
    &-  2\lambda ^{-2 }\left( q_3^2 -  q_2^2 - 3 q_1^2 \right) g^{\mu \nu} g^{\lambda \sigma} q_2^{\rho} -    8 \lambda ^{-2 }g^{\mu \nu} q_1^{\lambda} q_1^{\sigma} q_2^{\rho} \nonumber \\ 
    &-   2 \lambda ^{-2 }\left( q_3^2 +  q_2^2 -  q_1^2 \right) g^{\mu \lambda} g^{\nu \sigma} q_1^{\rho} -   2 \lambda ^{-2 } \left( q_3^2 - q_2^2 - q_1^2 \right)g^{\mu \lambda} g^{\nu \sigma} q_3^{\rho} \nonumber \\ 
    &-  4 \lambda ^{-2 }\left(q_3^2 - 2 q_2^2 - q_1^2 \right) g^{\mu \rho} g^{\lambda \sigma} q_1^{\nu} -  2 \lambda ^{-2 }\left(q_3^2 - q_2^2 + q_1^2 \right)g^{\mu \sigma} g^{\nu \lambda} q_2^{\rho} \nonumber \\ 
    &- 2 \lambda ^{-2 } \left( q_3^2 - q_2^2 - q_1^2 \right) g^{\mu \sigma} g^{\nu \lambda} q_3^{\rho} - 4  \lambda ^{-2 }q_3^2 g^{\mu \sigma} g^{\nu \rho} q_2^{\lambda} \nonumber \\ 
    &+  8 \lambda ^{-3 }\left(2 q_3^4 + q_2^2 q_3^2 - 3 q_2^4 - 4 q_1^2 q_3^2 + q_1^2 q_2^2 + 2 q_1^4 \right) g^{\mu \sigma} q_1^{\nu} q_2^{\lambda} q_3^{\rho} \nonumber \\ 
    &+  8 \lambda ^{-3 }\left(4 q_3^4 - 3 q_2^2 q_3^2 - q_2^4 - 3 q_1^2 q_3^2 + 2 q_1^2 q_2^2 - q_1^4 \right) g^{\mu \sigma} q_1^{\lambda} q_2^{\rho} q_3^{\nu} -  4 \lambda ^{-2 }\left(q_3^2 - q_2^2 - 2 q_1^2 \right)  g^{\nu \sigma} g^{\lambda \rho} q_3^{\mu} \nonumber \\ 
    &+  8\lambda ^{-3 }\left( 2 q_3^4 - 4 q_2^2 q_3^2 + 2 q_2^4 + q_1^2 q_3^2 + q_1^2 q_2^2 -3 q_1^4 \right) g^{\nu \sigma} q_1^{\lambda} q_2^{\mu} q_3^{\rho} \nonumber \\ 
    &+  8\lambda ^{-3 } \left( 4 q_3^4 - 3 q_2^2 q_3^2 - q_2^4 - 3 q_1^2 q_3^2 + 2 q_1^2 q_2^2 - q_1^4 \right) g^{\nu \sigma} q_1^{\rho} q_2^{\lambda} q_3^{\mu} \nonumber \\ 
    &-  8 \lambda ^{-3 } \left(2 q_3^4 - 4 q_2^2 q_3^2 + 2 q_2^4 - 9 q_1^2 q_3^2 + 11 q_1^2 q_2^2 + 7 q_1^4 \right) g^{\lambda \sigma} q_1^{\nu} q_2^{\rho} q_3^{\mu} \nonumber \\ 
     &-   8\lambda ^{-3 } \left( 2 q_3^4 - 9 q_2^2 q_3^2 + 7 q_2^4 - 4 q_1^2 q_3^2 + 11 q_1^2 q_2^2 + 2 q_1^4 \right)  g^{\lambda \sigma} q_1^{\rho} q_2^{\mu} q_3^{\nu} \, ,
\end{align}

\begin{align}
    &P_{\hat{\Pi}_{39}}^{\mu \nu \lambda \rho \sigma}(q_1,q_2,q_3)=160\lambda ^{-4 }\left(q_3^6 - q_2^2 q_3^4 - q_2^4 q_3^2 + q_2^6 - q_1^2 q_3^4 +  q_1^2 q_2^2 q_3^2 - q_1^2 q_2^4 - q_1^4 q_3^2 \right. \nonumber \\ 
    & \left. - q_1^4 q_2^2 + q_1^6 \right) q_1^{\nu} q_1^{\sigma} q_2^{\lambda} q_2^{\rho} q_3^{\mu}- 2 \lambda ^{-2 }q_2^2  g^{\mu \nu} g^{\lambda \sigma} q_1^{\rho} - 2 \lambda ^{-2 }q_1^2 g^{\mu \nu} g^{\lambda \sigma} q_2^{\rho} \nonumber \\ 
     &- 4\lambda ^{-3 } \left( 2 q_3^4 + q_2^2 q_3^2 - 3 q_2^4 + q_1^2 q_3^2 + 6 q_1^2 q_2^2 - 3 q_1^4 \right) g^{\mu \nu} q_1^{\lambda} q_1^{\sigma} q_2^{\rho} -  2\lambda ^{-2 } q_3^2  g^{\mu \lambda} g^{\nu \sigma} q_1^{\rho} \nonumber \\ 
    &-  2 \lambda ^{-2 }q_1^2 g^{\mu \lambda} g^{\nu \sigma} q_3^{\rho} +  4\lambda ^{-3 } \left( 3 q_3^4 - q_2^2 q_3^2 - 2 q_2^4 - 6 q_1^2 q_3^2 - q_1^2 q_2^2 + 3 q_1^4 \right) g^{\mu \lambda} q_1^{\rho} q_3^{\nu} q_3^{\sigma} \nonumber \\ 
    &-  4 \lambda ^{-2 }q_2^2  g^{\mu \rho} g^{\lambda \sigma} q_1^{\nu} - 2 \lambda ^{-2 } q_3^2g^{\mu \sigma} g^{\nu \lambda} q_2^{\rho} -  2 \lambda ^{-2 }q_2^2  g^{\mu \sigma} g^{\nu \lambda} q_3^{\rho} \nonumber \\ 
    &- 4 \lambda ^{-2 }q_3^2 g^{\mu \sigma} g^{\nu \rho} q_2^{\lambda} -  4\lambda ^{-3 } \left( q_3^4 - 7 q_2^2 q_3^2 - 4 q_2^4 - 2 q_1^2 q_3^2 + 3 q_1^2 q_2^2 +  q_1^4 \right) g^{\mu \sigma} q_1^{\nu} q_2^{\lambda} q_3^{\rho} \nonumber \\ 
    &+ 4\lambda ^{-3 }\left( 4 q_3^4 + 7 q_2^2 q_3^2 - q_2^4 - 3 q_1^2 q_3^2 + 2 q_1^2 q_2^2 - q_1^4 \right) g^{\mu \sigma} q_1^{\lambda} q_2^{\rho} q_3^{\nu}  \nonumber \\ 
    &+ 4\lambda ^{-3 }\left( 3 q_3^4 - 6 q_2^2 q_3^2 + 3 q_2^4 - q_1^2 q_3^2 - q_1^2 q_2^2 - 2 q_1^4 \right) g^{\nu \lambda} q_2^{\mu} q_2^{\sigma} q_3^{\rho} - 4 \lambda ^{-2 }q_1^2 g^{\nu \sigma} g^{\lambda \rho} q_3^{\mu} \nonumber \\ 
    &-  4 \lambda ^{-3 } \left( q_3^4 - 2 q_2^2 q_3^2 + q_2^4 - 7 q_1^2 q_3^2 + 3 q_1^2 q_2^2 - 4 q_1^4 \right) g^{\nu \sigma} q_1^{\lambda} q_2^{\mu} q_3^{\rho} \nonumber \\ 
    &+ 4\lambda ^{-3 }\left( 4 q_3^4 - 3 q_2^2 q_3^2 - q_2^4 + 7 q_1^2 q_3^2 + 2 q_1^2 q_2^2 - q_1^4 \right) g^{\nu \sigma} q_1^{\rho} q_2^{\lambda} q_3^{\mu} \nonumber \\ 
    &-   4\lambda ^{-3 }  \left( q_3^4 - 2 q_2^2 q_3^2 + q_2^4 + 3 q_1^2 q_3^2 - 7 q_1^2 q_2^2 - 4 q_1^4 \right) g^{\lambda \sigma} q_1^{\nu} q_2^{\rho} q_3^{\mu} \nonumber \\ 
    &-4 \lambda ^{-3 }\left( q_3^4 + 3 q_2^2  q_3^2 - 4 q_2^4 - 2 q_1^2 q_3^2 - 7 q_1^2 q_2^2 + q_1^4 \right) g^{\lambda \sigma} q_1^{\rho} q_2^{\mu} q_3^{\nu} \, ,
\end{align}

\begin{align}
    &P_{\hat{\Pi}_{54}}^{\mu \nu \lambda \rho \sigma}(q_1,q_2,q_3)=-  40\lambda ^{-3 } \left(q_2^2 -  q_1^2 \right) q_1^{\nu} q_1^{\sigma} q_2^{\lambda} q_2^{\rho} q_3^{\mu} +\lambda ^{-2 }  \left( q_3^2 + q_2^2 - q_1^2 \right) g^{\mu \nu} g^{\lambda \sigma} q_1^{\rho}\nonumber \\ 
    &- \lambda ^{-2 } \left( q_3^2 - q_2^2 + q_1^2 \right) g^{\mu \nu} g^{\lambda \sigma} q_2^{\rho} +   2 \lambda ^{-2 }\left( q_2^2 -  q_1^2 \right) g^{\mu \lambda} g^{\nu \sigma} q_1^{\rho} + \lambda ^{-2 }\left( q_3^2 - q_2^2 - 3 q_1^2 \right) g^{\mu \lambda} g^{\nu \sigma} q_3^{\rho} \nonumber \\ 
    &+ 2 \lambda ^{-2 } g^{\mu \lambda} q_1^{\rho} q_3^{\nu} q_3^{\sigma} +  4 \lambda ^{-2 }q_2^2  g^{\mu \rho} g^{\lambda \sigma} q_1^{\nu} +  2\lambda ^{-2 } \left( q_2^2 - q_1^2 \right)  g^{\mu \sigma} g^{\nu \lambda} q_2^{\rho} \nonumber \\
    &- \lambda ^{-2 }\left( q_3^2 - 3 q_2^2 - q_1^2\right) g^{\mu \sigma} g^{\nu \lambda} q_3^{\rho} + 4 \lambda ^{-2 }\left(q_2^2 -  q_1^2 \right) g^{\mu \sigma} g^{\nu \rho} q_2^{\lambda} \nonumber \\ 
    &+ 2 \lambda ^{-3 }\left(3 q_3^4 - 6 q_2^2 q_3^2 - 17 q_2^4 - 6 q_1^2 q_3^2 + 14 q_1^2 q_2^2 + 3 q_1^4 \right)  g^{\mu \sigma} q_1^{\nu} q_2^{\lambda} q_3^{\rho} \nonumber \\ 
    &+ 2\lambda ^{-3 }\left(2 q_3^4 - 14 q_2^2 q_3^2 - 8 q_2^4 + 6 q_1^2 q_3^2 + 16 q_1^2 q_2^2 - 8 q_1^4 \right) g^{\mu \sigma} q_1^{\lambda} q_2^{\rho} q_3^{\nu}- 2 \lambda ^{-2 } g^{\nu \lambda} q_2^{\mu} q_2^{\sigma} q_3^{\rho}  \nonumber \\ 
    &-  4 \lambda ^{-2 }q_1^2  g^{\nu \sigma} g^{\lambda \rho} q_3^{\mu} - 2\lambda ^{-3 }\left(3 q_3^4 - 6 q_2^2 q_3^2 + 3 q_2^4 - 6 q_1^2 q_3^2 + 14 q_1^2 q_2^2 - 17 q_1^4 \right) g^{\nu \sigma} q_1^{\lambda} q_2^{\mu} q_3^{\rho} \nonumber \\ 
    &-  2\lambda ^{-3 }\left( 2 q_3^4 + 6 q_2^2 q_3^2 - 8 q_2^4 - 14 q_1^2 q_3^2 + 16 q_1^2 q_2^2 - 8 q_1^4 \right) g^{\nu \sigma} q_1^{\rho} q_2^{\lambda} q_3^{\mu} \nonumber \\ 
    &- 2\lambda ^{-3 }\left(3 q_3^4 - 6 q_2^2 q_3^2 + 3 q_2^4 + 4 q_1^2 q_3^2 + 4 q_1^2 q_2^2 - 7 q_1^4 \right)  g^{\lambda \sigma} q_1^{\nu} q_2^{\rho} q_3^{\mu} \nonumber \\ 
       &+ 2\lambda ^{-3 }\left(3 q_3^4 + 4 q_2^2 q_3^2- 7 q_2^4 - 6 q_1^2 q_3^2 + 4 q_1^2 q_2^2 + 3 q_1^4 \right) g^{\lambda \sigma} q_1^{\rho} q_2^{\mu} q_3^{\nu} \, ,
\end{align}

where $\lambda=\lambda(q_1^2,q_2^2,q_3^2)$ is the Käll\'{e}n function defined through
\begin{equation}
 \lambda(q_1^2,q_2^2,q_3^2)=q_1^4+ q_2^4+ q_3^4 -2 q_1^2 q_2^2-2q_1^2q_3^2-2q_2^2q_3^2\, .
\end{equation}

\section{Four-quark reduction}\label{app:fourquark}
In this section we reduce the number of four-quark matrix elements from the basis $\bar{q}_{iA\bar{\alpha}}q_{jB\bar{\beta}}\bar{q}_{kC\bar{\gamma}}q_{lD\bar{\delta}}$, where the barred Greek indices denote colour, the capital ones flavour and the latin ones spinor, into one with only twelve non-zero elements. 

First of all, due to confinement only colour singlet operators can give contributions. From this one has
\begin{align}\nonumber
\bar{q}_{iA\bar{\alpha}}q_{jB\bar{\beta}}\bar{q}_{kC\bar{\gamma}}q_{lD\bar{\delta}}=\frac{1}{N_{c}^{2}-1}\bar{q}_{i'A'}q_{j'B'}\bar{q}_{k'C'}q_{l'D'}&\bigg[ \delta_{\bar{\alpha}\bar{\beta}}\delta_{\bar{\gamma}\bar{\delta}}\left(\delta_{iA}^{i'A'}\delta_{jB}^{j'B'}\delta_{kC}^{k'C'}\delta_{lD}^{l'D'}+\frac{1}{N_{c}}\delta_{iA}^{i'A'}\delta_{kC}^{k'C'}\delta_{jB}^{l'D'}\delta_{lD}^{j'B'}\right)\\&- 
\delta_{\bar{\alpha}\bar{\delta}}\delta_{\bar{\beta}\bar{\gamma}}\left(\frac{1}{N_{c}}\delta_{iA}^{i'A'}\delta_{jB}^{j'B'}\delta_{kC}^{k'C'}\delta_{lD}^{l'D'}+\delta_{iA}^{i'A'}\delta_{kC}^{k'C'}\delta_{jB}^{l'D'}\delta_{lD}^{j'B'}\right) \bigg] \, . \label{eq:colorred}
\end{align}
Next, taking into account that the QCD vacuum preserves $SU(3)_{\mathrm{V}}$ in the flavour sector up to small quark mass corrections, the contributing four-quark operators must break $SU(3)_{\mathrm{V}}$ in the same direction as the octet charge operator. There are therefore four independent flavour structures which contribute and can be taken to be
\begin{align}
    O_1&=Q_{AB}\delta_{CD}\bar{q}_{iA}q_{jB}\bar{q}_{kC}q_{lD}\,,\\
    O_2&=Q_{CD}\delta_{AB}\bar{q}_{iA}q_{jB}\bar{q}_{kC}q_{lD}\,,\\
    O_3&=Q_{AD}\delta_{BC}\bar{q}_{iA}q_{jB}\bar{q}_{kC}q_{lD}\,,\\
    O_4&=Q_{BC}\delta_{AD}\bar{q}_{iA}q_{jB}\bar{q}_{kC}q_{lD}\,.
\end{align}
The charge operator can be expressed with Gell-Mann matrices through
\begin{equation}
    Q=\frac{\lambda ^3}{2}+\frac{\lambda ^8}{2\sqrt{3}}\,,
\end{equation}
and the third and fourth operators $O_3$ and $O_4$ can be rewritten using a Fierzlike identity:
\begin{equation}
\begin{split}
O_{3,4}=\Bigg(& \frac{\delta_{AB}}{3}Q_{CD}+Q_{AB}\frac{\delta_{CD}}{3}\pm \frac{i}{4}\left( \lambda ^1_{AB} \lambda ^2_{CD}-\lambda ^2_{AB} \lambda ^1_{CD}+\lambda ^4_{AB} \lambda ^5_{CD}-\lambda ^5_{AB} \lambda ^4_{CD} \right)\\
& +\frac{1}{12}\sum_{i=1}^5\lambda ^i_{AB}\lambda ^i_{CD}-\frac{1}{6}\sum_{i=6}^7\lambda ^i_{AB}\lambda ^i_{CD}+\frac{1}{4\sqrt{3}}\left(\lambda ^3_{AB} \lambda ^8_{CD}+\lambda ^8_{AB} \lambda ^3_{CD}\right)\\
&-\frac{1}{12}\lambda ^8_{AB}\lambda ^8_{CD}\Bigg)\bar{q}_{iA}q_{jB}\bar{q}_{kC}q_{lD}\,.
\end{split}
\end{equation}
Orthogonal operators $\tilde{O}_i$ satisfying
\begin{equation}
    \begin{split}
        &\tilde{O}_n=P^n_{ABCD}\, \bar{q}_{iA}q_{jB}\bar{q}_{kC}q_{lD}\,,\\
        &\text{with }P^n_{ABCD}P^m_{BADC}=\delta^{mn}\,,
    \end{split}
\end{equation}
can be obtained from linear combinations of the operators $O_i$ ($i=1,\dots,4$) as
\begin{align}
    \tilde{O}_1&=\frac{1}{2}\left( O_1+O_2 \right)\,,\\
    \tilde{O}_2&=\frac{1}{2}\left( O_1-O_2 \right)\,,\\
    \tilde{O}_3&=\frac{-i}{2}\left( O_3-O_4 \right)=\frac{1}{4}\left( \lambda ^1_{AB} \lambda ^2_{CD}-\lambda ^2_{AB} \lambda ^1_{CD}+\lambda ^4_{AB} \lambda ^5_{CD}-\lambda ^5_{AB} \lambda ^4_{CD}\right)\bar{q}_{iA}q_{jB}\bar{q}_{kC}q_{lD}\,, \\
    \tilde{O}_4&=\frac{3}{2\sqrt{5}}\left(  O_3+O_4-\frac{2}{3}\left( O_1+O_2 \right) \right)=\frac{3}{\sqrt{5}}\Bigg( \frac{1}{12}\sum_{i=1}^5\lambda ^i_{AB}\lambda ^i_{CD}-\frac{1}{6}\sum_{i=6}^7\lambda ^i_{AB}\lambda ^i_{CD}\nonumber\\
    &+\frac{1}{4\sqrt{3}}\left(\lambda ^3_{AB} \lambda ^8_{CD}+\lambda ^8_{AB} \lambda ^3_{CD}\right)-\frac{1}{12}\lambda ^8_{AB}\lambda ^8_{CD}\Bigg)\bar{q}_{iA}q_{jB}\bar{q}_{kC}q_{lD}\,.
\end{align}
This leads to
\begin{equation}
    \bar{q}_{iA}q_{jB}\bar{q}_{kC}q_{lD}=\sum _n P^n_{BADC}\tilde{O}_n.
\end{equation}
Finally, the fact that the vacuum expectation values of the operators must be proportional to Q implies
\begin{align}
    &\langle \bar{q}_iq_j\bar{q}_k\lambda ^3q_l\rangle =\sqrt{3}\langle \bar{q}_iq_j\bar{q}_k\lambda ^8q_l\rangle\,,\\
    &\langle \bar{q}_i\lambda ^3q_j\bar{q}_kq_l\rangle =\sqrt{3}\langle \bar{q}_i\lambda ^8q_j\bar{q}_kq_l\rangle\,,\\
    &\langle \bar{q}_i\lambda ^1 q_j\bar{q}_k\lambda ^2q_l\rangle-\langle \bar{q}_i\lambda ^2 q_j\bar{q}_k\lambda ^1q_l\rangle = \langle \bar{q}_i\lambda ^5 q_j\bar{q}_k\lambda ^4q_l\rangle-\langle \bar{q}_i\lambda ^4 q_j\bar{q}_k\lambda ^5q_l\rangle\,,\\
    &\langle \bar{q}_i\lambda ^1 q_j\bar{q}_k\lambda ^1q_l\rangle=-\langle \bar{q}_i\lambda ^8 q_j\bar{q}_k\lambda ^8q_l\rangle=\langle \bar{q}_i\lambda ^n q_j\bar{q}_k\lambda ^nq_l\rangle \, , \; \; \text{   for} \; n=2,\ldots,5\, ,\\
    &\langle \bar{q}_i\lambda ^1 q_j\bar{q}_k\lambda ^1q_l\rangle=-\frac{1}{2}\langle \bar{q}_i\lambda ^n q_j\bar{q}_k\lambda ^nq_l\rangle \, , \; \; 
    \text{   for} \; n=6,7\,,\\
    &\langle \bar{q}_i\lambda ^1 q_j\bar{q}_k\lambda ^1q_l\rangle=\frac{1}{2\sqrt{3}}\left(\langle \bar{q}_i\lambda ^3 q_j\bar{q}_k\lambda ^8q_l\rangle+\langle \bar{q}_i\lambda ^8 q_j\bar{q}_k\lambda ^3q_l\rangle \right)\,,
\end{align}
and the final flavour decomposition reads
\begin{align}\nonumber
&\bar{q}_{iA}q_{jB}\bar{q}_{kC}q_{lD}= \frac{1}{2}(\bar{q}_{i}\lambda_{1}q_{j}\bar{q}_{k}\lambda_{2}q_{l}-\bar{q}_{i}\lambda_{2}q_{j}\bar{q}_{k}\lambda_{1}q_{l})\left(\frac{\lambda_{BA}^{1}}{2}\frac{\lambda_{DC}^{2}}{2}-\frac{\lambda_{BA}^{2}}{2}\frac{\lambda_{DC}^{1}}{2}+\frac{\lambda_{BA}^{4}}{2}\frac{\lambda_{DC}^{5}}{2}-\frac{\lambda_{BA}^{5}}{2}\frac{\lambda_{DC}^{4}}{2}\right)
\\&+ \bar{q}_{i}\lambda_{1}q_{j}\bar{q}_{k}\lambda_{1}q_{l}
\left(\sum_{i=1}^5 \frac{\lambda_{BA}^{i}}{2}\frac{\lambda_{DC}^{i}}{2}-2 \sum_{i=6}^7 \frac{\lambda_{BA}^{i}}{2}\frac{\lambda_{DC}^{i}}{2}-\frac{\lambda_{BA}^{8}}{2}\frac{\lambda_{DC}^{8}}{2}+\sqrt{3}  \Bigg( \frac{\lambda_{BA}^{3}}{2}\frac{\lambda_{DC}^{8}}{2}+\frac{\lambda_{BA}^{8}}{2}\frac{\lambda_{DC}^{3}}{2}\Bigg) \right)\nonumber
\\&+ \frac{1}{2}(\bar{q}_{i}\lambda_{8}q_{j}\bar{q}_{k}q_{l}+\bar{q}_{i}q_{j}\bar{q}_{k}\lambda_{8}q_{l})\left(\frac{\lambda^{8}_{BA}+\sqrt{3}\lambda^{3}_{BA}}{2}\frac{\delta_{DC}}{3}+ \frac{\delta_{BA}}{3}\frac{\lambda^{8}_{DC}+\sqrt{3}\lambda^{3}_{DC}}{2}\right)\nonumber
\\&
+ \frac{1}{2}(\bar{q}_{i}\lambda_{8}q_{j}\bar{q}_{k}q_{l}-\bar{q}_{i}q_{j}\bar{q}_{k}\lambda_{8}q_{l})\left(\frac{\lambda^{8}_{BA}+\sqrt{3}\lambda^{3}_{BA}}{2}\frac{\delta_{DC}}{3}- \frac{\delta_{BA}}{3}\frac{\lambda^{8}_{DC}+\sqrt{3}\lambda^{3}_{DC}}{2}\right) \, .
\end{align}
Only the spinor reduction remains. As usual, one can decompose each operator above according to
\begin{equation}
\hat{O}_{i'j'k'l'}=\sum _{A,B}c_{A}c_{B} \hat{O}_{ijkl}\Gamma^{ijA}\Gamma^{klB}\;\;\Gamma_ {i'j'}^A\Gamma_{k'l'}^B \, ,
\end{equation}
where $\Gamma^{A}$ is an element in the spinor basis of~(\ref{eq:spinor}) and $c_{A}$ the corresponding normalisation defined in~(\ref{eq:spinornorm}). However, many restrictions apply. Proportionality with $F_{\mu\nu}$ leaves a small number of independent Lorentz structures possible. Moreover, since $\hat{O}_{ijkl}$ are by construction either symmetric or anti-symmetric under the exchange $(ij)\leftrightarrow(kl)$, the reduced matrix element $\hat{O}_{ijkl}\Gamma^{ijB}\Gamma^{klA}$ is trivially related to $\hat{O}_{ijkl}\Gamma^{ijA}\Gamma^{klB}$. Taking advantage of this and requiring that the reduced matrix elements should be odd under charge conjugation one finds
\begin{align}\nonumber
\frac{1}{2}&(\bar{q}_{i}\lambda_{1}q_{j}\bar{q}_{k}\lambda_{2}q_{l}-\bar{q}_{i}\lambda_{2}q_{j}\bar{q}_{k}\lambda_{1}q_{l})=\frac{1}{64}
\Big(
\gamma^{\mu}_{ji}\gamma^{\nu}_{lk}-\gamma^{\nu}_{ji}\gamma^{\mu}_{lk}
\Big)
\Big[
\bar{q}\lambda_{1}\gamma_{\mu}q\bar{q}\lambda_{2}\gamma_{\nu}q-\bar{q}\lambda_{2}\gamma_{\mu}q\bar{q}\lambda_{1}\gamma_{\nu}q
\Big]
\\ \nonumber
&+\frac{1}{64}
\Big(
(\gamma^{\mu}\gamma_{5})_{ji}(\gamma^{\nu}\gamma_{5})_{lk}-(\gamma^{\nu}\gamma_{5})_{ji}(\gamma^{\mu}\gamma_{5})_{lk}
\Big)
\Big[
\bar{q}\lambda_{1}\gamma_{\mu}\gamma_{5}q\bar{q}\lambda_{2}\gamma_{\nu}\gamma_{5}q-\bar{q}\lambda_{2}\gamma_{\mu}\gamma_{5}q\bar{q}\lambda_{1}\gamma_{\nu}\gamma_{5}q
\Big] 
\\&+\frac{1}{64}g_{\lambda \alpha}
\Big(
\sigma^{\mu\lambda}_{ji}\sigma_{lk}^{\alpha\nu}-\sigma^{\nu\lambda}_{ji}\sigma_{lk}^{\alpha\mu}
\Big)
\times \frac{1}{2} g ^{\rho \beta} 
\Big[
\bar{q}\sigma_{\mu\rho}\lambda_{1}q\,\bar{q}\sigma_{\beta\nu}\lambda_{2}q-\bar{q}\sigma_{\mu\rho}\lambda_{2}q\,\bar{q}\sigma_{\beta\nu}\lambda_{1}q
\Big]  
\end{align}

\begin{align}\nonumber
&\bar{q}_{i}\lambda_{1}q_{j}\bar{q}_{k}\lambda_{1}q_{l}=\frac{1}{32}
\Big(
\sigma^{\mu\nu}_{ji}\delta_{lk}+\delta_{ji}\sigma^{\mu\nu}_{lk}
\Big)
\Big[
\bar{q}\lambda_{1}q\bar{q}\lambda_{1}\sigma_{\mu\nu}q
\Big]
\\ \nonumber
&+\frac{1}{64}\epsilon^{\mu\nu\mu_{1}\nu_{1}}\epsilon_{\mu_{1}\nu_{1}\mu_{2}\nu_{2}}
\Big(
(\gamma_{\mu}\gamma_{5})_{ji}(\gamma_{\nu})_{lk}-(\gamma_{\mu})_{ji}(\gamma_{\nu}\gamma_{5})_{lk}
\Big)
\Big[
\bar{q}\lambda_{1}\gamma^{\mu_{2}}\gamma_{5}q\bar{q}\lambda_{1}\gamma^{\nu_{2}}q
\Big]
\\ 
&+\frac{1}{32}
\Big(
\sigma^{\mu\nu}_{ji}\gamma_{5\,lk}+\gamma_{5\,ji}\sigma^{\mu\nu}_{lk}
\Big) 
\Big[
\bar{q}\lambda_{1}\sigma_{\mu\nu}q \, \bar{q}\lambda_{1}\gamma_{5}q
\Big]
\, ,
\end{align}

\begin{align}\nonumber
\bar{q}_{i}\lambda_{8}q_{j}\bar{q}_{k}q_{l}\pm\bar{q}_{i}q_{j}\bar{q}_{k}\lambda_{8}q_{l}&=\frac{1}{32}
\Big(
\sigma^{\mu\nu}_{ji}\delta_{lk}\pm\delta_{ji}\sigma^{\mu\nu}_{lk}
\Big)
\Big[
\bar{q}\sigma_{\mu\nu}\lambda_{8}q\,\bar{q}q \pm\bar{q}\sigma_{\mu\nu}q\,\bar{q}\lambda_{8}q
\Big]
\, \\ \nonumber
&+\frac{1}{64}\epsilon^{\mu\nu\mu_{1}\nu_{1}}\epsilon_{\mu_{1}\nu_{1}\mu_{2}\nu_{2}}
\Big(
(\gamma_{\mu}\gamma_{5})_{ji}(\gamma_{\nu})_{lk}\mp(\gamma_{\mu})_{ji}(\gamma_{\nu}\gamma_{5})_{lk}
\Big)
\\ \nonumber
&
\times
\Big[
\bar{q}\lambda_{8}\gamma^{\mu_{2}}\gamma_{5}q \, \bar{q}\gamma^{\nu_{2}}q\pm \bar{q}\gamma^{\mu_{2}}\gamma_{5}q \, \bar{q}\lambda_{8}\gamma^{\nu_{2}}q
\Big]
\\ 
&+\frac{1}{32}
\Big(
\sigma^{\mu\nu}_{ji}\gamma_{5\, lk}\pm \gamma_{5\, ji}\sigma^{\mu\nu}_{lk}
\Big)
\Big[
\bar{q}\sigma^{\mu\nu}\lambda_{8}q\, \bar{q}\gamma_{5}q \pm \bar{q}\sigma^{\mu\nu}q \, \bar{q}\lambda_{8}\gamma_{5}q
\Big]
\, .
\end{align}
This reduces the original set of $1679616$ matrix elements to a basis of $12$ non-zero ones.

\section{Explicit expressions for the $\hat{\Pi}_{i}$}\label{app:explicit}
In this appendix, we seperately list the form factors $\hat{\Pi} _{i}$ for the contributions from the quark loop, one cut quark line topologies and the gluon matrix element. Note that those for the two-cut quark line topologies were given in Sec.~\ref{sec:fourquark}.
\subsection{The quark loop}
Recall~(\ref{eq:quarkloopdefs})
\begin{equation}
\hat{\Pi}^{\overline{MS}}_m=\hat{\Pi}^{0}_{m,S}+m_{q}^{2}\,\hat{\Pi}^{m_{q}^{2}}_{m,\overline{MS}}+\mathcal{O}(m_{q}^{4}) \, .
\end{equation}
The two terms above are given by
\begin{align}
\hat{\Pi}_{m,S}^{0}=\frac{N_{c}\,e_{q}^{4}}{\pi^{2}}\sum_{i,j,k,n}\left[c_{i,j,k}^{(m,n)}+f_{i,j,k}^{(m,n)}F+g^{(m,n)}_{i,j,k}\log\left(\frac{Q_{2}^2}{Q_{3}^{2}}\right)+h^{(m,n)}_{i,j,k}\log\left(\frac{Q_{1}^2}{Q_{2}^{2}}\right)\right]\lambda^{-n}\, Q_{1}^{2i}\,Q_{2}^{2j}\,Q_{3}^{2k} \, , 
\end{align}

\begin{align}\begin{split}
\hat{\Pi}^{m_{q}^{2}}_{m,\overline{MS}}&=\frac{N_{c}\,e_{q}^{4}}{\pi^{2}}\sum_{i,j,k,n}\lambda^{-n}\, Q_{1}^{2i}\,Q_{2}^{2j}\,Q_{3}^{2k}\\&\cdot\left[d_{i,j,k}^{(m,n)}+p_{i,j,k}^{(m,n)}F+q^{(m,n)}_{i,j,k}\log\left(\frac{Q_{2}^2}{Q_{3}^{2}}\right)+r^{(m,n)}_{i,j,k}\log\left(\frac{Q_{1}^2}{Q_{2}^{2}}\right)+s^{(m,n)}_{i,j,k}\log\left(\frac{Q_{3}^2}{\mu^{2}}\right)\right] \, , 
\end{split}\end{align}
where the non-zero coefficients are 
\begin{align}\begin{split}
&c^{(1,1)}_{0,0,0}= 2\,,\ 
   \end{split}
  \end{align} 
\begin{align}\begin{split}
&f^{(1,2)}_{0,1,2}= 2\,,\ f^{(1,2)}_{0,2,1}= -4\,,\ f^{(1,2)}_{0,3,0}=    2\,,\ f^{(1,2)}_{1,0,2}= 2\,,\ f^{(1,2)}_{1,1,1}= 4\,,\ f^{(1,2)}_{1,2,0}=    -2\,,\ f^{(1,2)}_{2,0,1}= -4,\\&f^{(1,2)}_{2,1,0}=    -2\,,\ f^{(1,2)}_{3,0,0}= 2,
\end{split}
\end{align} 
\begin{align}\begin{split}
&g^{(1,2)}_{0,0,2}= 2\,,\ g^{(1,2)}_{0,1,1}= 2\,,\ g^{(1,2)}_{0,2,0}=    -4\,,\ g^{(1,2)}_{1,0,1}= 2\,,\ g^{(1,2)}_{1,1,0}= 8\,,\ g^{(1,2)}_{2,0,0}=    -4,
\end{split}
\end{align} 
\begin{align}\begin{split}
&h^{(1,2)}_{0,0,2}= 1\,,\ h^{(1,2)}_{0,1,1}=    -1\,,\ h^{(1,2)}_{0,2,0}=    -1\,,\ h^{(1,2)}_{0,3,-1}=    1\,,\ h^{(1,2)}_{1,0,1}= 3\,,\ h^{(1,2)}_{1,1,0}=    4\,,\ h^{(1,2)}_{1,2,-1}=    -3,\\&h^{(1,2)}_{2,0,0}=    -3\,,\ h^{(1,2)}_{2,1,-1}=    3\,,\ h^{(1,2)}_{3,0,-1}= -1,
\end{split}
\end{align}

\begin{align}\begin{split}
&c^{(4,3)}_{-1,0,5}= 2\,,\ c^{(4,3)}_{-1,1,4}= -6\,,\ c^{(4,3)}_{-1,2,3}= 4\,,\ c^{(4,3)}_{-1,3,2}= 4\,,\ c^{(4,3)}_{-1,4,1}= -6\,,\ c^{(4,3)}_{-1,5,0}= 2\,,\ c^{(4,3)}_{0,-1,5}= 2,\\&c^{(4,3)}_{0,0,4}= 36\,,\ c^{(4,3)}_{0,1,3}=    -20\,,\ c^{(4,3)}_{0,2,2}= -64\,,\ c^{(4,3)}_{0,3,1}= 34\,,\ c^{(4,3)}_{0,4,0}= 12\,,\ c^{(4,3)}_{1,-1,4}= -6\,,\ c^{(4,3)}_{1,0,3}= -20,\\&c^{(4,3)}_{1,1,2}= 168\,,\ c^{(4,3)}_{1,2,1}= -28\,,\ c^{(4,3)}_{1,3,0}=    -66\,,\ c^{(4,3)}_{2,-1,3}= 4\,,\ c^{(4,3)}_{2,0,2}= -64\,,\ c^{(4,3)}_{2,1,1}= -28\,,\ c^{(4,3)}_{2,2,0}= 104,\\&c^{(4,3)}_{3,-1,2}= 4\,,\ c^{(4,3)}_{3,0,1}= 34\,,\ c^{(4,3)}_{3,1,0}= -66\,,\ c^{(4,3)}_{4,-1,1}=    -6\,,\ c^{(4,3)}_{4,0,0}= 12\,,\ c^{(4,3)}_{5,-1,0}= 2 ,
   \end{split}
  \end{align} 
\begin{align}
\begin{split}
&f^{(4,4)}_{0,0,7}= 12\,,\ f^{(4,4)}_{0,1,6}= -6\,,\ f^{(4,4)}_{0,2,5}=    -60\,,\ f^{(4,4)}_{0,3,4}= 66\,,\ f^{(4,4)}_{0,4,3}=    36\,,\ f^{(4,4)}_{0,5,2}= -66\,,\ f^{(4,4)}_{0,6,1}=    12,\\&f^{(4,4)}_{0,7,0}= 6\,,\ f^{(4,4)}_{1,0,6}= -6\,,\ f^{(4,4)}_{1,1,5}=    216\,,\ f^{(4,4)}_{1,2,4}= -138\,,\ f^{(4,4)}_{1,3,3}=    -360\,,\ f^{(4,4)}_{1,4,2}= 270\,,\ f^{(4,4)}_{1,5,1}=    48,\\&f^{(4,4)}_{1,6,0}= -30\,,\ f^{(4,4)}_{2,0,5}=    -60\,,\ f^{(4,4)}_{2,1,4}= -138\,,\ f^{(4,4)}_{2,2,3}=    744\,,\ f^{(4,4)}_{2,3,2}= -204\,,\ f^{(4,4)}_{2,4,1}=    -300,\\& f^{(4,4)}_{2,5,0}= 54\,,\ f^{(4,4)}_{3,0,4}=    66\,,\ f^{(4,4)}_{3,1,3}= -360\,,\ f^{(4,4)}_{3,2,2}=    -204\,,\ f^{(4,4)}_{3,3,1}= 480\,,\ f^{(4,4)}_{3,4,0}=    -30\,,\ f^{(4,4)}_{4,0,3}= 36,\\& f^{(4,4)}_{4,1,2}=    270\,,\ f^{(4,4)}_{4,2,1}= -300\,,\ f^{(4,4)}_{4,3,0}=    -30\,,\ f^{(4,4)}_{5,0,2}= -66\,,\ f^{(4,4)}_{5,1,1}=    48\,,\ f^{(4,4)}_{5,2,0}= 54\,,\ f^{(4,4)}_{6,0,1}=    12,\\& f^{(4,4)}_{6,1,0}= -30\,,\ f^{(4,4)}_{7,0,0}= 6,
      \end{split}
  \end{align}
   
\begin{align}
\begin{split}
&g^{(4,4)}_{-1,0,7}= 1\,,\ g^{(4,4)}_{-1,1,6}= -3\,,\ g^{(4,4)}_{-1,2,5}=    1\,,\ g^{(4,4)}_{-1,3,4}= 5\,,\ g^{(4,4)}_{-1,4,3}=    -5\,,\ g^{(4,4)}_{-1,5,2}= -1\,,\ g^{(4,4)}_{-1,6,1}=    3,\\&g^{(4,4)}_{-1,7,0}= -1\,,\ g^{(4,4)}_{0,-1,7}=    1\,,\ g^{(4,4)}_{0,0,6}= 50\,,\ g^{(4,4)}_{0,1,5}=    -13\,,\ g^{(4,4)}_{0,2,4}= -184\,,\ g^{(4,4)}_{0,3,3}=    119\,,\ g^{(4,4)}_{0,4,2}= 106,\\&g^{(4,4)}_{0,5,1}=    -75\,,\ g^{(4,4)}_{0,6,0}= -4\,,\ g^{(4,4)}_{1,-1,6}=    -3\,,\ g^{(4,4)}_{1,0,5}= -13\,,\ g^{(4,4)}_{1,1,4}=    486\,,\ g^{(4,4)}_{1,2,3}= -162\,,\ g^{(4,4)}_{1,3,2}=    -559,\\&g^{(4,4)}_{1,4,1}= 207\,,\ g^{(4,4)}_{1,5,0}=    44\,,\ g^{(4,4)}_{2,-1,5}= 1\,,\ g^{(4,4)}_{2,0,4}=    -184\,,\ g^{(4,4)}_{2,1,3}= -162\,,\ g^{(4,4)}_{2,2,2}=    908\,,\ g^{(4,4)}_{2,3,1}= -135,\\&g^{(4,4)}_{2,4,0}=    -124\,,\ g^{(4,4)}_{3,-1,4}= 5\,,\ g^{(4,4)}_{3,0,3}=    119\,,\ g^{(4,4)}_{3,1,2}= -559\,,\ g^{(4,4)}_{3,2,1}=    -135\,,\ g^{(4,4)}_{3,3,0}= 170\,,\ g^{(4,4)}_{4,-1,3}=    -5,\\&g^{(4,4)}_{4,0,2}= 106\,,\ g^{(4,4)}_{4,1,1}=    207\,,\ g^{(4,4)}_{4,2,0}= -124\,,\ g^{(4,4)}_{5,-1,2}=    -1\,,\ g^{(4,4)}_{5,0,1}= -75\,,\ g^{(4,4)}_{5,1,0}=    44\,,\ g^{(4,4)}_{6,-1,1}= 3,\\&g^{(4,4)}_{6,0,0}=    -4\,,\ g^{(4,4)}_{7,-1,0}= -1,     
\end{split}
\end{align}  
   
\begin{align}
\begin{split}
&h^{(4,4)}_{0,-1,7}= 1\,,\ h^{(4,4)}_{0,0,6}= 25\,,\ h^{(4,4)}_{0,1,5}=    -63\,,\ h^{(4,4)}_{0,2,4}= 1\,,\ h^{(4,4)}_{0,3,3}=    91\,,\ h^{(4,4)}_{0,4,2}= -45\,,\ h^{(4,4)}_{0,5,1}=    -29,\\&h^{(4,4)}_{0,6,0}= 19\,,\ h^{(4,4)}_{1,-1,6}=    -3\,,\ h^{(4,4)}_{1,0,5}= 50\,,\ h^{(4,4)}_{1,1,4}=    243\,,\ h^{(4,4)}_{1,2,3}= -500\,,\ h^{(4,4)}_{1,3,2}=    19\,,\ h^{(4,4)}_{1,4,1}= 258,\\&h^{(4,4)}_{1,5,0}=    -67\,,\ h^{(4,4)}_{2,-1,5}= 1\,,\ h^{(4,4)}_{2,0,4}=    -185\,,\ h^{(4,4)}_{2,1,3}= 338\,,\ h^{(4,4)}_{2,2,2}=    454\,,\ h^{(4,4)}_{2,3,1}= -563\,,\ h^{(4,4)}_{2,4,0}=    51,\\&h^{(4,4)}_{3,-1,4}= 5\,,\ h^{(4,4)}_{3,0,3}=    28\,,\ h^{(4,4)}_{3,1,2}= -578\,,\ h^{(4,4)}_{3,2,1}=    428\,,\ h^{(4,4)}_{3,3,0}= 85\,,\ h^{(4,4)}_{4,-1,3}=    -5\,,\ h^{(4,4)}_{4,0,2}= 151,\\&h^{(4,4)}_{4,1,1}=    -51\,,\ h^{(4,4)}_{4,2,0}= -175\,,\ h^{(4,4)}_{5,-1,2}=    -1\,,\ h^{(4,4)}_{5,0,1}= -46\,,\ h^{(4,4)}_{5,1,0}=    111\,,\ h^{(4,4)}_{6,-1,1}= 3\,,\ h^{(4,4)}_{6,0,0}=    -23,\\&h^{(4,4)}_{7,-1,0}= -1.
\end{split}
\end{align}   
   
\begin{align}
\begin{split}
&c^{(7,3)}_{-1,0,4}= 6\,,\ c^{(7,3)}_{-1,1,3}=    -12\,,\ c^{(7,3)}_{-1,3,1}=    12\,,\ c^{(7,3)}_{-1,4,0}=    -6\,,\ c^{(7,3)}_{0,-1,4}=    4\,,\ c^{(7,3)}_{0,0,3}=    64\,,\ c^{(7,3)}_{0,1,2}=    68,\\&c^{(7,3)}_{0,2,1}=    -112\,,\ c^{(7,3)}_{0,3,0}=    -24\,,\ c^{(7,3)}_{1,-1,3}=    -16\,,\ c^{(7,3)}_{1,0,2}=    -116\,,\ c^{(7,3)}_{1,1,1}=    100\,,\ c^{(7,3)}_{1,2,0}=    104\,,\ c^{(7,3)}_{2,-1,2}=    24,\\&c^{(7,3)}_{2,0,1}=    16\,,\ c^{(7,3)}_{2,1,0}=    -108\,,\ c^{(7,3)}_{3,-1,1}=    -16\,,\ c^{(7,3)}_{3,0,0}=    30\,,\ c^{(7,3)}_{4,-1,0}= 4,
\end{split}
\end{align}
   
\begin{align}
\begin{split}
&f^{(7,4)}_{0,0,6}= 24\,,\ f^{(7,4)}_{0,1,5}= 36\,,\ f^{(7,4)}_{0,2,4}= -156\,,\ f^{(7,4)}_{0,3,3}=    24\,,\ f^{(7,4)}_{0,4,2}= 144\,,\ f^{(7,4)}_{0,5,1}= -60\,,\ f^{(7,4)}_{0,6,0}= -12,\\&f^{(7,4)}_{1,0,5}=    -36\,,\ f^{(7,4)}_{1,1,4}= 288\,,\ f^{(7,4)}_{1,2,3}= 312\,,\ f^{(7,4)}_{1,3,2}=    -576\,,\ f^{(7,4)}_{1,4,1}= -36\,,\ f^{(7,4)}_{1,5,0}= 48\,,\ f^{(7,4)}_{2,0,4}=    -84,\\&f^{(7,4)}_{2,1,3}= -600\,,\ f^{(7,4)}_{2,2,2}= 432\,,\ f^{(7,4)}_{2,3,1}=    456\,,\ f^{(7,4)}_{2,4,0}= -60\,,\ f^{(7,4)}_{3,0,3}= 216\,,\ f^{(7,4)}_{3,1,2}=    144,\\& f^{(7,4)}_{3,2,1}= -552\,,\ f^{(7,4)}_{4,0,2}= -144\,,\ f^{(7,4)}_{4,1,1}=    180\,,\ f^{(7,4)}_{4,2,0}= 60\,,\ f^{(7,4)}_{5,0,1}= 12\,,\ f^{(7,4)}_{5,1,0}= -48\,,\ f^{(7,4)}_{6,0,0}=    12,
\end{split}
\end{align}
\begin{align}
\begin{split}
&g^{(7,4)}_{-1,0,6}= 2\,,\ g^{(7,4)}_{-1,2,4}= -18\,,\ g^{(7,4)}_{-1,3,3}= 32\,,\ g^{(7,4)}_{-1,4,2}=    -18\,,\ g^{(7,4)}_{-1,6,0}= 2\,,\ g^{(7,4)}_{0,-1,6}= 2\,,\ g^{(7,4)}_{0,0,5}= 104,\\&g^{(7,4)}_{0,1,4}=    126\,,\ g^{(7,4)}_{0,2,3}= -368\,,\ g^{(7,4)}_{0,3,2}= -58\,,\ g^{(7,4)}_{0,4,1}=    184\,,\ g^{(7,4)}_{0,5,0}= 10\,,\ g^{(7,4)}_{1,-1,5}= -8\,,\ g^{(7,4)}_{1,0,4}=    -158,\\&g^{(7,4)}_{1,1,3}= 544\,,\ g^{(7,4)}_{1,2,2}= 652\,,\ g^{(7,4)}_{1,3,1}=    -440\,,\ g^{(7,4)}_{1,4,0}= -78\,,\ g^{(7,4)}_{2,-1,4}= 10\,,\ g^{(7,4)}_{2,0,3}=    -160,\\& g^{(7,4)}_{2,1,2}= -924   \,,\ g^{(7,4)}_{2,2,1}= 208\,,\ g^{(7,4)}_{2,3,0}=    170\,,\ g^{(7,4)}_{3,0,2}= 358\,,\ g^{(7,4)}_{3,1,1}= 176\,,\ g^{(7,4)}_{3,2,0}=    -170\\& g^{(7,4)}_{4,-1,2}= -10\,,\ g^{(7,4)}_{4,0,1}= -136   \,,\ g^{(7,4)}_{4,1,0}=    78\,,\ g^{(7,4)}_{5,-1,1}= 8\,,\ g^{(7,4)}_{5,0,0}= -10\,,\ g^{(7,4)}_{6,-1,0}= -2,
\end{split}
\end{align}
  
\begin{align}
\begin{split}
&h^{(7,4)}_{0,-1,6}= 2\,,\ h^{(7,4)}_{0,0,5}= 56\,,\ h^{(7,4)}_{0,1,4}= -46\,,\ h^{(7,4)}_{0,2,3}=    -176\,,\ h^{(7,4)}_{0,3,2}= 214\,,\ h^{(7,4)}_{0,4,1}= -8\,,\ h^{(7,4)}_{0,5,0}=    -42,\\&h^{(7,4)}_{1,-1,5}= -8\,,\ h^{(7,4)}_{1,0,4}= 16\,,\ h^{(7,4)}_{1,1,3}= 536\,,\ h^{(7,4)}_{1,2,2}=    -256\,,\ h^{(7,4)}_{1,3,1}= -400\,,\ h^{(7,4)}_{1,4,0}= 112\,,\ h^{(7,4)}_{2,-1,4}=    10,\\&h^{(7,4)}_{2,0,3}= -336\,,\ h^{(7,4)}_{2,1,2}= -300\,,\ h^{(7,4)}_{2,2,1}= 800\,,\ h^{(7,4)}_{2,3,0}=    -30\,,\ h^{(7,4)}_{3,0,2}= 352\,,\ h^{(7,4)}_{3,1,1}= -360,\\& h^{(7,4)}_{3,2,0}=    -160\,,\ h^{(7,4)}_{4,-1,2}= -10\,,\ h^{(7,4)}_{4,0,1}= -40\,,\ h^{(7,4)}_{4,1,0}=    170\,,\ h^{(7,4)}_{5,-1,1}= 8\,,\ h^{(7,4)}_{5,0,0}= -48,\\& h^{(7,4)}_{6,-1,0}= -2,
\end{split}
\end{align}
   
\begin{align}
\begin{split}
&c^{(17,2)}_{0,0,1}= 16\,,\ c^{(17,2)}_{0,1,0}= -12\,,\ c^{(17,2)}_{0,2,-1}= -4\,,\ c^{(17,2)}_{1,0,0}=    -12\,,\ c^{(17,2)}_{1,1,-1}= 8\,,\ c^{(17,2)}_{2,0,-1}= -4,
\end{split}    
\end{align}   
\begin{align}
\begin{split}
&f^{(17,3)}_{0,0,4}= 4\,,\ f^{(17,3)}_{0,1,3}= -4\,,\ f^{(17,3)}_{0,2,2}= -12\,,\ f^{(17,3)}_{0,3,1}=    20\,,\ f^{(17,3)}_{0,4,0}= -8\,,\ f^{(17,3)}_{1,0,3}= -4\,,\ f^{(17,3)}_{1,1,2}= 64,\\&f^{(17,3)}_{1,2,1}=    -44\,,\ f^{(17,3)}_{1,3,0}= -16\,,\ f^{(17,3)}_{2,0,2}= -12\,,\ f^{(17,3)}_{2,1,1}=    -44\,,\ f^{(17,3)}_{2,2,0}= 48\,,\ f^{(17,3)}_{3,0,1}= 20,\\& f^{(17,3)}_{3,1,0}= -16\,,\ f^{(17,3)}_{4,0,0}=    -8,
\end{split}    
\end{align}  

\begin{align}
\begin{split}
&g^{(17,3)}_{0,0,3}= 20\,,\ g^{(17,3)}_{0,1,2}= -16\,,\ g^{(17,3)}_{0,2,1}= -28\,,\ g^{(17,3)}_{0,3,0}=    24\,,\ g^{(17,3)}_{1,0,2}= -16\,,\ g^{(17,3)}_{1,1,1}= 104,\\& g^{(17,3)}_{1,2,0}= -24 \,,\ g^{(17,3)}_{2,0,1}=    -28\,,\ g^{(17,3)}_{2,1,0}= -24\,,\ g^{(17,3)}_{3,0,0}= 24,
\end{split}    
\end{align}  

\begin{align}
\begin{split}
&h^{(17,3)}_{0,0,3}= 10\,,\ h^{(17,3)}_{0,1,2}= -28\,,\ h^{(17,3)}_{0,2,1}= 24\,,\ h^{(17,3)}_{0,3,0}=    -4\,,\ h^{(17,3)}_{0,4,-1}= -2\,,\ h^{(17,3)}_{1,0,2}= 12\,,\ h^{(17,3)}_{1,1,1}= 52,\\&h^{(17,3)}_{1,2,0}=    -68\,,\ h^{(17,3)}_{1,3,-1}= 4\,,\ h^{(17,3)}_{2,0,1}= -52\,,\ h^{(17,3)}_{2,1,0}= 44\,,\ h^{(17,3)}_{3,0,0}=    28\,,\ h^{(17,3)}_{3,1,-1}= -4\,,\ h^{(17,3)}_{4,0,-1}= 2,
\end{split}
\end{align}

\begin{align}
\begin{split}
&c^{(39,3)}_{-1,0,4}= -4\,,\ c^{(39,3)}_{-1,1,3}= 16\,,\ c^{(39,3)}_{-1,2,2}= -24\,,\ c^{(39,3)}_{-1,3,1}=    16\,,\ c^{(39,3)}_{-1,4,0}= -4\,,\ c^{(39,3)}_{0,-1,4}= -4\,,\ c^{(39,3)}_{0,0,3}= -44,\\&c^{(39,3)}_{0,1,2}=    48\,,\ c^{(39,3)}_{0,2,1}= 48\,,\ c^{(39,3)}_{0,3,0}= -44\,,\ c^{(39,3)}_{0,4,-1}= -4\,,\ c^{(39,3)}_{1,-1,3}=    16\,,\ c^{(39,3)}_{1,0,2}= 48\,,\ c^{(39,3)}_{1,1,1}= -176,\\&c^{(39,3)}_{1,2,0}= 48\,,\ c^{(39,3)}_{1,3,-1}=    16\,,\ c^{(39,3)}_{2,-1,2}= -24\,,\ c^{(39,3)}_{2,0,1}= 48\,,\ c^{(39,3)}_{2,1,0}= 48\,,\ c^{(39,3)}_{2,2,-1}=    -24\,,\ c^{(39,3)}_{3,-1,1}= 16,\\&c^{(39,3)}_{3,0,0}= -44\,,\ c^{(39,3)}_{3,1,-1}=    16\,,\ c^{(39,3)}_{4,-1,0}= -4\,,\ c^{(39,3)}_{4,0,-1}= -4,
\end{split}
\end{align}
\begin{align}
\begin{split}
&f^{(39,4)}_{0,0,6}= -20\,,\ f^{(39,4)}_{0,1,5}= 24\,,\ f^{(39,4)}_{0,2,4}= 84\,,\ f^{(39,4)}_{0,3,3}=    -176\,,\ f^{(39,4)}_{0,4,2}= 84\,,\ f^{(39,4)}_{0,5,1}= 24,\\& f^{(39,4)}_{0,6,0}= -20\,,\ f^{(39,4)}_{1,0,5}=    24\,,\ f^{(39,4)}_{1,1,4}= -288\,,\ f^{(39,4)}_{1,2,3}= 264\,,\ f^{(39,4)}_{1,3,2}=    264\,,\ f^{(39,4)}_{1,4,1}= -288,\\& f^{(39,4)}_{1,5,0}= 24\,,\ f^{(39,4)}_{2,0,4}= 84\,,\ f^{(39,4)}_{2,1,3}=    264\,,\ f^{(39,4)}_{2,2,2}= -792\,,\ f^{(39,4)}_{2,3,1}= 264\,,\ f^{(39,4)}_{2,4,0}=    84,\\& f^{(39,4)}_{3,0,3}= -176\,,\ f^{(39,4)}_{3,1,2}= 264\,,\ f^{(39,4)}_{3,2,1}=    264\,,\ f^{(39,4)}_{3,3,0}= -176\,,\ f^{(39,4)}_{4,0,2}= 84\,,\ f^{(39,4)}_{4,1,1}=    -288,\\& f^{(39,4)}_{4,2,0}= 84\,,\ f^{(39,4)}_{5,0,1}= 24\,,\ f^{(39,4)}_{5,1,0}= 24\,,\ f^{(39,4)}_{6,0,0}=    -20,
\end{split}
\end{align}
\begin{align}\begin{split}
&g^{(39,4)}_{-1,0,6}= -2\,,\ g^{(39,4)}_{-1,1,5}= 8\,,\ g^{(39,4)}_{-1,2,4}= -10\,,\ g^{(39,4)}_{-1,4,2}=    10\,,\ g^{(39,4)}_{-1,5,1}= -8\,,\ g^{(39,4)}_{-1,6,0}= 2\,,\ g^{(39,4)}_{0,-1,6}= -2,\\&g^{(39,4)}_{0,0,5}=    -76\,,\ g^{(39,4)}_{0,1,4}= 74, g^{(39,4)}_{0,2,3}= 216\,,\ g^{(39,4)}_{0,3,2}=    -302\,,\ g^{(39,4)}_{0,4,1}= 52\,,\ g^{(39,4)}_{0,5,0}= 38,\\& g^{(39,4)}_{1,-1,5}= 8\,,\ g^{(39,4)}_{1,0,4}=    74\,,\ g^{(39,4)}_{1,1,3}= -592\,,\ g^{(39,4)}_{1,2,2}= 340\,,\ g^{(39,4)}_{1,3,1}=    296\,,\ g^{(39,4)}_{1,4,0}= -126,\\& g^{(39,4)}_{2,-1,4}= -10\,,\ g^{(39,4)}_{2,0,3}=    216,\\&g^{(39,4)}_{2,1,2}= 340\,,\ g^{(39,4)}_{2,2,1}= -680\,,\ g^{(39,4)}_{2,3,0}=    86\,,\ g^{(39,4)}_{3,0,2}= -302\,,\ g^{(39,4)}_{3,1,1}= 296\,,\ g^{(39,4)}_{3,2,0}=    86,\\& g^{(39,4)}_{4,-1,2}= 10\,,\ g^{(39,4)}_{4,0,1}= 52\,,\ g^{(39,4)}_{4,1,0}=    -126\,,\ g^{(39,4)}_{5,-1,1}= -8\,,\ g^{(39,4)}_{5,0,0}= 38\,,\ g^{(39,4)}_{6,-1,0}= 2,
    \end{split}
\end{align}

\begin{align}\begin{split}
&h^{(39,4)}_{0,-1,6}= -2\,,\ h^{(39,4)}_{0,0,5}= -38\,,\ h^{(39,4)}_{0,1,4}= 126\,,\ h^{(39,4)}_{0,2,3}=    -86\,,\ h^{(39,4)}_{0,3,2}= -86\,,\ h^{(39,4)}_{0,4,1}= 126,\\& h^{(39,4)}_{0,5,0}=    -38\,,\ h^{(39,4)}_{0,6,-1}= -2\,,\ h^{(39,4)}_{1,-1,5}= 8\,,\ h^{(39,4)}_{1,0,4}= -52\,,\ h^{(39,4)}_{1,1,3}=    -296\,,\ h^{(39,4)}_{1,2,2}= 680,\\& h^{(39,4)}_{1,3,1}= -296\,,\ h^{(39,4)}_{1,4,0}=    -52\,,\ h^{(39,4)}_{1,5,-1}= 8\,,\ h^{(39,4)}_{2,-1,4}= -10\,,\ h^{(39,4)}_{2,0,3}=    302\,,\ h^{(39,4)}_{2,1,2}= -340,\\& h^{(39,4)}_{2,2,1}= -340\,,\ h^{(39,4)}_{2,3,0}=    302\,,\ h^{(39,4)}_{2,4,-1}= -10\,,\ h^{(39,4)}_{3,0,2}= -216\,,\ h^{(39,4)}_{3,1,1}=    592\,,\ h^{(39,4)}_{3,2,0}= -216,\\& h^{(39,4)}_{4,-1,2}= 10\,,\ h^{(39,4)}_{4,0,1}=    -74\,,\ h^{(39,4)}_{4,1,0}= -74\,,\ h^{(39,4)}_{4,2,-1}= 10\,,\ h^{(39,4)}_{5,-1,1}=    -8\,,\ h^{(39,4)}_{5,0,0}= 76,\\& h^{(39,4)}_{5,1,-1}= -8\,,\ h^{(39,4)}_{6,-1,0}= 2\,,\ h^{(39,4)}_{6,0,-1}=    2,
    \end{split}
\end{align}

\begin{align}
\begin{split}
&c^{(54,2)}_{-1,0,2}= -2\,,\ c^{(54,2)}_{-1,1,1}= 4\,,\ c^{(54,2)}_{-1,2,0}= -2\,,\ c^{(54,2)}_{0,-1,2}=    2\,,\ c^{(54,2)}_{0,1,0}= -14\,,\ c^{(54,2)}_{1,-1,1}= -4\,,\ c^{(54,2)}_{1,0,0}= 14,\\&c^{(54,2)}_{2,-1,0}=    2,
\end{split}
\end{align}

\begin{align}
\begin{split}
&f^{(54,3)}_{0,1,3}= -18\,,\ f^{(54,3)}_{0,2,2}= 30\,,\ f^{(54,3)}_{0,3,1}= -6\,,\ f^{(54,3)}_{0,4,0}=    -6\,,\ f^{(54,3)}_{1,0,3}= 18\,,\ f^{(54,3)}_{1,2,1}= -54\,,\ f^{(54,3)}_{1,3,0}= 12,\\&f^{(54,3)}_{2,0,2}=    -30\,,\ f^{(54,3)}_{2,1,1}= 54\,,\ f^{(54,3)}_{3,0,1}= 6\,,\ f^{(54,3)}_{3,1,0}= -12\,,\ f^{(54,3)}_{4,0,0}=    6,
\end{split}
\end{align}
\begin{align}
\begin{split}
&g^{(54,3)}_{-1,0,4}= -1\,,\ g^{(54,3)}_{-1,1,3}= 2\,,\ g^{(54,3)}_{-1,3,1}= -2\,,\ g^{(54,3)}_{-1,4,0}=    1\,,\ g^{(54,3)}_{0,-1,4}= 1\,,\ g^{(54,3)}_{0,1,2}= -48\,,\ g^{(54,3)}_{0,2,1}= 40,\\&g^{(54,3)}_{0,3,0}=    7\,,\ g^{(54,3)}_{1,-1,3}= -2\,,\ g^{(54,3)}_{1,0,2}= 48\,,\ g^{(54,3)}_{1,2,0}= -26\,,\ g^{(54,3)}_{2,0,1}=    -40\,,\ g^{(54,3)}_{2,1,0}= 26\,,\ g^{(54,3)}_{3,-1,1}= 2,\\&g^{(54,3)}_{3,0,0}= -7\,,\ g^{(54,3)}_{4,-1,0}=    -1,
\end{split}
\end{align}

\begin{align}
\begin{split}
&h^{(54,3)}_{0,-1,4}= 1\,,\ h^{(54,3)}_{0,0,3}= -10\,,\ h^{(54,3)}_{0,2,1}= 26\,,\ h^{(54,3)}_{0,3,0}=    -17\,,\ h^{(54,3)}_{1,-1,3}= -2\,,\ h^{(54,3)}_{1,0,2}= 48\,,\ h^{(54,3)}_{1,1,1}= -78,\\&h^{(54,3)}_{1,2,0}=    8\,,\ h^{(54,3)}_{2,0,1}= -14\,,\ h^{(54,3)}_{2,1,0}= 34\,,\ h^{(54,3)}_{3,-1,1}= 2\,,\ h^{(54,3)}_{3,0,0}=    -24\,,\ h^{(54,3)}_{4,-1,0}= -1,
\end{split}
\end{align}
and
\begin{align}
\begin{split}
p^{(1,1)}_{0,0,0}=-4\, ,
\end{split}
\end{align}
\begin{align}
\begin{split}
&q^{(1,3)}_{-1,-1,1}= -2\,,\ q^{(1,3)}_{-1,0,0}= 2\,,\ q^{(1,3)}_{0,-1,0}= 2,
\end{split}
\end{align}
\begin{align}
\begin{split}
&r^{(1,1)}_{-1,-1,1}= -1\,,\ r^{(1,1)}_{-1,0,0}= 3\,,\ r^{(1,1)}_{-1,1,-1}= -3\,,\ r^{(1,1)}_{-1,2,-2}=    1\,,\ r^{(1,1)}_{0,-1,0}= -1\,,\ r^{(1,1)}_{0,1,-2}= -3\,,\ r^{(1,1)}_{1,-1,-1}= 3,\\&r^{(1,1)}_{1,0,-2}=    3\,,\ r^{(1,1)}_{2,-1,-2}= -1,
\end{split}
\end{align}
\begin{align}
\begin{split}
&s^{(1,0)}_{-1,-1,-1}= -1\,,\ s^{(1,0)}_{-1,0,-2}= 1\,,\ s^{(1,0)}_{0,-1,-2}= 1,
\end{split}
\end{align}
\begin{align}
\begin{split}
&d^{(4,2)}_{-2,-1,4}= 1\,,\ d^{(4,2)}_{-2,0,3}= -3\,,\ d^{(4,2)}_{-2,1,2}= 2\,,\ d^{(4,2)}_{-2,2,1}=    2\,,\ d^{(4,2)}_{-2,3,0}= -3\,,\ d^{(4,2)}_{-2,4,-1}= 1\,,\ d^{(4,2)}_{-1,-2,4}= 1,\\&d^{(4,2)}_{-1,-1,3}=    -10\,,\ d^{(4,2)}_{-1,0,2}= 6\,,\ d^{(4,2)}_{-1,1,1}= 8\,,\ d^{(4,2)}_{-1,2,0}= 1\,,\ d^{(4,2)}_{-1,3,-1}=    -6\,,\ d^{(4,2)}_{0,-2,3}= -3\,,\ d^{(4,2)}_{0,-1,2}= 6,\\&d^{(4,2)}_{0,0,1}= -36\,,\ d^{(4,2)}_{0,1,0}=    2\,,\ d^{(4,2)}_{0,2,-1}= 15\,,\ d^{(4,2)}_{1,-2,2}= 2\,,\ d^{(4,2)}_{1,-1,1}= 8\,,\ d^{(4,2)}_{1,0,0}=    2\,,\ d^{(4,2)}_{1,1,-1}= -20,\\&d^{(4,2)}_{2,-2,1}= 2\,,\ d^{(4,2)}_{2,-1,0}= 1\,,\ d^{(4,2)}_{2,0,-1}=    15\,,\ d^{(4,2)}_{3,-2,0}= -3\,,\ d^{(4,2)}_{3,-1,-1}= -6\,,\ d^{(4,2)}_{4,-2,-1}= 1,
\end{split}
\end{align}
\begin{align}
\begin{split}
&p^{(4,3)}_{0,0,4}= -40\,,\ p^{(4,3)}_{0,1,3}= 40\,,\ p^{(4,3)}_{0,2,2}= 24\,,\ p^{(4,3)}_{0,3,1}=    -8\,,\ p^{(4,3)}_{0,4,0}= -16\,,\ p^{(4,3)}_{1,0,3}= 40\,,\ p^{(4,3)}_{1,1,2}= -208,\\&p^{(4,3)}_{1,2,1}=    8\,,\ p^{(4,3)}_{1,3,0}= 64\,,\ p^{(4,3)}_{2,0,2}= 24\,,\ p^{(4,3)}_{2,1,1}= 8\,,\ p^{(4,3)}_{2,2,0}=    -96\,,\ p^{(4,3)}_{3,0,1}= -8\,,\ p^{(4,3)}_{3,1,0}= 64,\\&p^{(4,3)}_{4,0,0}= -16,
\end{split}
\end{align}
\begin{align}
\begin{split}
&q^{(4,3)}_{-1,-1,5}= -4\,,\ q^{(4,3)}_{-1,0,4}= 16\,,\ q^{(4,3)}_{-1,1,3}= -32\,,\ q^{(4,3)}_{-1,2,2}=    40\,,\ q^{(4,3)}_{-1,3,1}= -28\,,\ q^{(4,3)}_{-1,4,0}= 8\,,\ q^{(4,3)}_{0,-1,4}= 16,\\&q^{(4,3)}_{0,0,3}=    -160\,,\ q^{(4,3)}_{0,1,2}= 40\,,\ q^{(4,3)}_{0,2,1}= 128\,,\ q^{(4,3)}_{0,3,0}= -24\,,\ q^{(4,3)}_{1,-1,3}=    -32\,,\ q^{(4,3)}_{1,0,2}= 40\,,\ q^{(4,3)}_{1,1,1}= -200,\\&q^{(4,3)}_{1,2,0}= 16\,,\ q^{(4,3)}_{2,-1,2}=    40\,,\ q^{(4,3)}_{2,0,1}= 128\,,\ q^{(4,3)}_{2,1,0}= 16\,,\ q^{(4,3)}_{3,-1,1}= -28\,,\ q^{(4,3)}_{3,0,0}=    -24\,,\ q^{(4,3)}_{4,-1,0}= 8,
\end{split}
\end{align}
\begin{align}
\begin{split}
&r^{(4,3)}_{-1,-1,5}= -2\,,\ r^{(4,3)}_{-1,0,4}= 12\,,\ r^{(4,3)}_{-1,1,3}= -30\,,\ r^{(4,3)}_{-1,2,2}=    40\,,\ r^{(4,3)}_{-1,3,1}= -30\,,\ r^{(4,3)}_{-1,4,0}= 12,\\& r^{(4,3)}_{-1,5,-1}=    -2\,,\ r^{(4,3)}_{0,-1,4}= 4\,,\ r^{(4,3)}_{0,0,3}= -80\,,\ r^{(4,3)}_{0,1,2}= 112\,,\ r^{(4,3)}_{0,2,1}=    8\,,\ r^{(4,3)}_{0,3,0}= -52\,,\ r^{(4,3)}_{0,4,-1}= 8,\\& r^{(4,3)}_{1,-1,3}= -2\,,\ r^{(4,3)}_{1,0,2}=    -72\,,\ r^{(4,3)}_{1,1,1}= -100\,,\ r^{(4,3)}_{1,2,0}= 88\,,\ r^{(4,3)}_{1,3,-1}= -10\,,\ r^{(4,3)}_{2,0,1}=    120\,,\ r^{(4,3)}_{2,1,0}= -72,\\& r^{(4,3)}_{3,-1,1}= 2\,,\ r^{(4,3)}_{3,0,0}= 28\,,\ r^{(4,3)}_{3,1,-1}=    10\,,\ r^{(4,3)}_{4,-1,0}= -4\,,\ r^{(4,3)}_{4,0,-1}= -8\,,\ r^{(4,3)}_{5,-1,-1}= 2,
\end{split}
\end{align}
\begin{align}
    \begin{split}
        &s^{(4,0)}_{-1,-1,-1}= -2,
    \end{split}
\end{align}
\begin{align}
\begin{split}
&d^{(7,2)}_{-2,-1,3}= 2\,,\ d^{(7,2)}_{-2,0,2}= -4\,,\ d^{(7,2)}_{-2,2,0}= 4\,,\ d^{(7,2)}_{-2,3,-1}=    -2\,,\ d^{(7,2)}_{-1,-2,3}= 2\,,\ d^{(7,2)}_{-1,-1,2}= -20,\\& d^{(7,2)}_{-1,0,1}=    -12\,,\ d^{(7,2)}_{-1,1,0}= 20\,,\ d^{(7,2)}_{-1,2,-1}= 10\,,\ d^{(7,2)}_{0,-2,2}=    -8\,,\ d^{(7,2)}_{0,-1,1}= 24\,,\ d^{(7,2)}_{0,0,0}= -20,\\& d^{(7,2)}_{0,1,-1}= -20\,,\ d^{(7,2)}_{1,-2,1}=    12\,,\ d^{(7,2)}_{1,-1,0}= 4\,,\ d^{(7,2)}_{1,0,-1}= 20\,,\ d^{(7,2)}_{2,-2,0}= -8\,,\ d^{(7,2)}_{2,-1,-1}=    -10,\\& d^{(7,2)}_{3,-2,-1}= 2,
\end{split}
\end{align}
\begin{align}
\begin{split}
&p^{(7,3)}_{0,0,3}= -72\,,\ p^{(7,3)}_{0,1,2}= -72\,,\ p^{(7,3)}_{0,2,1}= 120\,,\ p^{(7,3)}_{0,3,0}=    24\,,\ p^{(7,3)}_{1,0,2}= 120\,,\ p^{(7,3)}_{1,1,1}= -96\,,\ p^{(7,3)}_{1,2,0}= -72,\\&p^{(7,3)}_{2,0,1}=    -24\,,\ p^{(7,3)}_{2,1,0}= 72\,,\ p^{(7,3)}_{3,0,0}= -24,
    \end{split}
\end{align}
\begin{align}
\begin{split}
&q^{(7,3)}_{-2,0,4}= 4\,,\ q^{(7,3)}_{-2,1,3}= -16\,,\ q^{(7,3)}_{-2,2,2}= 24\,,\ q^{(7,3)}_{-2,3,1}=    -16\,,\ q^{(7,3)}_{-2,4,0}= 4\,,\ q^{(7,3)}_{-1,0,3}= -40\,,\ q^{(7,3)}_{-1,1,2}= 48,\\&q^{(7,3)}_{-1,2,1}=    24\,,\ q^{(7,3)}_{-1,3,0}= -32\,,\ q^{(7,3)}_{0,-1,3}= -8\,,\ q^{(7,3)}_{0,0,2}= -144\,,\ q^{(7,3)}_{0,1,1}=    -168\,,\ q^{(7,3)}_{0,2,0}= 64\,,\ q^{(7,3)}_{1,-1,2}= 24,\\&q^{(7,3)}_{1,0,1}= 184\,,\ q^{(7,3)}_{1,1,0}=    -40\,,\ q^{(7,3)}_{2,-1,1}= -24\,,\ q^{(7,3)}_{2,0,0}= -4\,,\ q^{(7,3)}_{3,-1,0}= 8,
    \end{split}
\end{align}
\begin{align}
\begin{split}
&r^{(7,3)}_{0,-1,3}= -8\,,\ r^{(7,3)}_{0,0,2}= -144\,,\ r^{(7,3)}_{0,1,1}= 72\,,\ r^{(7,3)}_{0,2,0}=    80\,,\ r^{(7,3)}_{1,-1,2}= 24\,,\ r^{(7,3)}_{1,0,1}= 80\,,\ r^{(7,3)}_{1,1,0}= -152,\\&r^{(7,3)}_{2,-1,1}=    -24\,,\ r^{(7,3)}_{2,0,0}= 64\,,\ r^{(7,3)}_{3,-1,0}= 8,
    \end{split}
\end{align}
\begin{align}
\begin{split}
&d^{(17,1)}_{-1,-1,0}= -4\,,\ d^{(17,1)}_{-1,0,-1}= 4\,,\ d^{(17,1)}_{0,-1,-1}= 4,
    \end{split}
\end{align}
\begin{align}
\begin{split}
&p^{(17,2)}_{0,0,1}= -24\,,\ p^{(17,2)}_{0,1,0}= 24\,,\ p^{(17,2)}_{1,0,0}= 24,
    \end{split}
\end{align}
\begin{align}
\begin{split}
&q^{(17,2)}_{-1,-1,2}= -4\,,\ q^{(17,2)}_{-1,0,1}= 8\,,\ q^{(17,2)}_{-1,1,0}= -4\,,\ q^{(17,2)}_{0,-1,1}=    8\,,\ q^{(17,2)}_{0,0,0}= -40\,,\ q^{(17,2)}_{1,-1,0}= -4,
    \end{split}
\end{align}
\begin{align}
\begin{split}
&r^{(17,2)}_{-1,-1,2}= -2\,,\ r^{(17,2)}_{-1,0,1}= 8\,,\ r^{(17,2)}_{-1,1,0}= -12\,,\ r^{(17,2)}_{-1,2,-1}=    8\,,\ r^{(17,2)}_{-1,3,-2}= -2\,,\ r^{(17,2)}_{0,0,0}= -20,\\& r^{(17,2)}_{0,1,-1}= 16\,,\ r^{(17,2)}_{0,2,-2}=    4\,,\ r^{(17,2)}_{1,-1,0}= 8\,,\ r^{(17,2)}_{1,0,-1}= -16\,,\ r^{(17,2)}_{2,-1,-1}= -8\,,\ r^{(17,2)}_{2,0,-2}=    -4\,,\ r^{(17,2)}_{3,-1,-2}= 2,
    \end{split}
\end{align}
\begin{align}
\begin{split}
&s^{(17,0)}_{-1,-1,-2}= -2,
\end{split}
\end{align}
\begin{align}
\begin{split}
&d^{(39,2)}_{-2,-1,3}= -2\,,\ d^{(39,2)}_{-2,0,2}= 8\,,\ d^{(39,2)}_{-2,1,1}= -12\,,\ d^{(39,2)}_{-2,2,0}=    8\,,\ d^{(39,2)}_{-2,3,-1}= -2\,,\ d^{(39,2)}_{-1,-2,3}= -2,\\& d^{(39,2)}_{-1,-1,2}=    18\,,\ d^{(39,2)}_{-1,0,1}= -16\,,\ d^{(39,2)}_{-1,1,0}= -16\,,\ d^{(39,2)}_{-1,2,-1}=    18\,,\ d^{(39,2)}_{-1,3,-2}= -2\,,\ d^{(39,2)}_{0,-2,2}= 8,\\& d^{(39,2)}_{0,-1,1}= -16\,,\ d^{(39,2)}_{0,0,0}=    32\,,\ d^{(39,2)}_{0,1,-1}= -16\,,\ d^{(39,2)}_{0,2,-2}= 8\,,\ d^{(39,2)}_{1,-2,1}= -12\,,\ d^{(39,2)}_{1,-1,0}=    -16,\\& d^{(39,2)}_{1,0,-1}= -16\,,\ d^{(39,2)}_{1,1,-2}= -12\,,\ d^{(39,2)}_{2,-2,0}=    8\,,\ d^{(39,2)}_{2,-1,-1}= 18\,,\ d^{(39,2)}_{2,0,-2}= 8\,,\ d^{(39,2)}_{3,-2,-1}= -2,\\& d^{(39,2)}_{3,-1,-2}=    -2,
\end{split}
\end{align}

\begin{align}
\begin{split}
&p^{(39,3)}_{0,0,3}= 48\,,\ p^{(39,3)}_{0,1,2}= -48\,,\ p^{(39,3)}_{0,2,1}= -48\,,\ p^{(39,3)}_{0,3,0}=    48\,,\ p^{(39,3)}_{1,0,2}= -48\,,\ p^{(39,3)}_{1,1,1}= 192,\\& p^{(39,3)}_{1,2,0}= -48\,,\ p^{(39,3)}_{2,0,1}=    -48\,,\ p^{(39,3)}_{2,1,0}= -48\,,\ p^{(39,3)}_{3,0,0}= 48,
\end{split}
\end{align}

\begin{align}
\begin{split}
&q^{(39,3)}_{-1,0,3}= 8\,,\ q^{(39,3)}_{-1,1,2}= -24\,,\ q^{(39,3)}_{-1,2,1}= 24\,,\ q^{(39,3)}_{-1,3,0}=    -8\,,\ q^{(39,3)}_{0,-1,3}= 8\,,\ q^{(39,3)}_{0,0,2}= 128\,,\ q^{(39,3)}_{0,1,1}= -72,\\&q^{(39,3)}_{0,2,0}=    -64\,,\ q^{(39,3)}_{1,-1,2}= -24\,,\ q^{(39,3)}_{1,0,1}= -72\,,\ q^{(39,3)}_{1,1,0}=    144\,,\ q^{(39,3)}_{2,-1,1}= 24\,,\ q^{(39,3)}_{2,0,0}= -64,\\& q^{(39,3)}_{3,-1,0}= -8,
\end{split}
\end{align}

\begin{align}
\begin{split}
&r^{(39,3)}_{0,-1,3}= 8\,,\ r^{(39,3)}_{0,0,2}= 64\,,\ r^{(39,3)}_{0,1,1}= -144\,,\ r^{(39,3)}_{0,2,0}=    64\,,\ r^{(39,3)}_{0,3,-1}= 8\,,\ r^{(39,3)}_{1,-1,2}= -24\,,\ r^{(39,3)}_{1,0,1}= 72,\\&r^{(39,3)}_{1,1,0}=    72\,,\ r^{(39,3)}_{1,2,-1}= -24\,,\ r^{(39,3)}_{2,-1,1}= 24\,,\ r^{(39,3)}_{2,0,0}=    -128\,,\ r^{(39,3)}_{2,1,-1}= 24\,,\ r^{(39,3)}_{3,-1,0}= -8,\\& r^{(39,3)}_{3,0,-1}= -8,
\end{split}
\end{align}

\begin{align}
\begin{split}
&d^{(54,1)}_{-1,0,-1}= 4\,,\ d^{(54,1)}_{0,-1,-1}= -4,
\end{split}
\end{align}
\begin{align}
\begin{split}
&p^{(54,2)}_{0,1,0}= 24\,,\ p^{(54,2)}_{1,0,0}= -24,
\end{split}
\end{align}

\begin{align}
\begin{split}
&q^{(54,2)}_{-2,-1,3}= -2\,,\ q^{(54,2)}_{-2,0,2}= 6\,,\ q^{(54,2)}_{-2,1,1}= -6\,,\ q^{(54,2)}_{-2,2,0}=    2\,,\ q^{(54,2)}_{-1,-2,3}= 2\,,\ q^{(54,2)}_{-1,0,1}= 14\,,\ q^{(54,2)}_{-1,1,0}= -16,\\&q^{(54,2)}_{0,-2,2}=    -6\,,\ q^{(54,2)}_{0,-1,1}= -14\,,\ q^{(54,2)}_{1,-2,1}= 6\,,\ q^{(54,2)}_{1,-1,0}= 16\,,\ q^{(54,2)}_{2,-2,0}=    -2,
\end{split}
\end{align}

\begin{align}
\begin{split}
&r^{(54,2)}_{-2,-1,3}= -1\,,\ r^{(54,2)}_{-2,0,2}= 4\,,\ r^{(54,2)}_{-2,1,1}= -6\,,\ r^{(54,2)}_{-2,2,0}=    4\,,\ r^{(54,2)}_{-2,3,-1}= -1\,,\ r^{(54,2)}_{-1,-2,3}= 1\,,\ r^{(54,2)}_{-1,0,1}= 2,\\&r^{(54,2)}_{-1,1,0}=    -8\,,\ r^{(54,2)}_{-1,2,-1}= 5\,,\ r^{(54,2)}_{0,-2,2}= -2\,,\ r^{(54,2)}_{0,-1,1}= -12\,,\ r^{(54,2)}_{0,0,0}=    26\,,\ r^{(54,2)}_{0,1,-1}= -4\,,\ r^{(54,2)}_{1,-1,0}= 8,\\&r^{(54,2)}_{1,0,-1}= -4\,,\ r^{(54,2)}_{2,-2,0}=    2\,,\ r^{(54,2)}_{2,-1,-1}= 5\,,\ r^{(54,2)}_{3,-2,-1}= -1,
\end{split}
\end{align}

\begin{align}
\begin{split}
&s^{(54,0)}_{-2,-1,-1}= -1\,,\ s^{(54,0)}_{-1,-2,-1}= 1\, .
\end{split}
\end{align}

\subsection{Some massless quark loop limits}
\label{app:quarklooplimits}
Here we give the explicit limits for the massless quark loop in the regimes where two momenta become much larger than the other, this is a subset of the regions where $\lambda$ becomes small. In order to find them, we change the variables to one of the large momenta, the small one and the angle between them. We then expand in the small over the large momentum. All negative powers of $\lambda$ cancel and the leading contribution in the ratio of small over large momentum is always independent of the angle and is given below. The choice of third variable is not unique, however the results to the order given are always the same.

\subsubsection{$Q_{1}\sim Q_{3}\gg Q_{2}$}
\begin{align}
\frac{\pi^2\hat{\Pi}_{1}}{e_{q}^{4}N_{c}}&=-\frac{1}{9 Q_{1}^{4}}\left[3\log\left(\frac{Q_{1}^{2}}{Q_{2}^{2}}\right)+5\right]   \, ,\\
\frac{\pi^2\hat{\Pi}_{4}}{e_{q}^{4}N_{c}}&= -\frac{1}{3 Q_{2}^{2}Q_{1}^{2}}\, ,\\
\frac{\pi^2\hat{\Pi}_{7}}{e_{q}^{4}N_{c}}&=-\frac{1}{3 Q_{2}^{2}Q_{1}^{4}} \, ,\\
\frac{\pi^2\hat{\Pi}_{17}}{e_{q}^{4}N_{c}}&=\frac{1}{18 Q_{1}^{6}}\left[6\log\left(\frac{Q_{1}^{2}}{Q_{2}^{2}}\right)-5\right] \, ,\\
\frac{\pi^2\hat{\Pi}_{39}}{e_{q}^{4}N_{c}}&=\frac{1}{3 Q_{2}^{2}Q_{1}^{4}} \, ,\\
\frac{\pi^2\hat{\Pi}_{54}}{e_{q}^{4}N_{c}}&=-\frac{1}{6 Q_{2}^{2}Q_{1}^{4}} \, .
\end{align}

\subsubsection{$Q_{1}\sim Q_{2}\gg Q_{3}$}

\begin{align}
\frac{\pi^2\hat{\Pi}_{1}}{e_{q}^{4}N_{c}}&=-\frac{1}{Q_{3}^2Q_{2}^{2}}\, ,\\
\frac{\pi^2\hat{\Pi}_{4}}{e_{q}^{4}N_{c}}&=-\frac{1}{3 Q_{2}^{4}}\, ,\\
\frac{\pi^2\hat{\Pi}_{7}}{e_{q}^{4}N_{c}}&=-\frac{1}{3 Q_{2}^{6}}\, ,\\
\frac{\pi^2\hat{\Pi}_{17}}{e_{q}^{4}N_{c}}&=\frac{1}{3Q_{3}^{2}Q_{2}^{4}}\, ,\\
\frac{\pi^2\hat{\Pi}_{39}}{e_{q}^{4}N_{c}}&=\frac{1}{3Q_{3}^{2}Q_{2}^{4}}\, ,\\
\frac{\pi^2\hat{\Pi}_{54}}{e_{q}^{4}N_{c}}&=\mathcal{O}(Q_{2}^{-7})\, .
\end{align}

\subsubsection{$Q_{2}\sim Q_{3}\gg Q_{1}$}

\begin{align}
\frac{\pi^2\hat{\Pi}_{1}}{e_{q}^{4}N_{c}}&=-\frac{1}{9 Q_{3}^{4}}\left[3\log\left(\frac{Q_{3}^{2}}{Q_{1}^{2}}\right)+5\right] \, ,\\
\frac{\pi^2\hat{\Pi}_{4}}{e_{q}^{4}N_{c}}&=-\frac{1}{3Q_{1}^{2}Q_{3}^{2}} \, ,\\
\frac{\pi^2\hat{\Pi}_{7}}{e_{q}^{4}N_{c}}&=\mathcal{O}(Q_{3}^{-5}) \, ,\\
\frac{\pi^2\hat{\Pi}_{17}}{e_{q}^{4}N_{c}}&=\frac{1}{18 Q_{3}^{6}}\left[6\log\left(\frac{Q_{3}^{2}}{Q_{1}^{2}}\right)-5\right] \, ,\\
\frac{\pi^2\hat{\Pi}_{39}}{e_{q}^{4}N_{c}}&=\frac{1}{3Q_{1}^{2}Q_{3}^{4}} \, ,\\
\frac{\pi^2\hat{\Pi}_{54}}{e_{q}^{4}N_{c}}&=\frac{1}{6Q_{1}^{2}Q_{3}^{4}} \, .
\end{align}

\subsection{Contributions from diagrams with one-cut quark lines}
The explicit expressions of the $\hat{\Pi}_{i}$ for the one-cut quark line diagrams can be written as
\begin{align}
\hat{\Pi} _{m}=e^{4}_{q}\, \sum_{i,j,k,n,p}c^{m,n,p}_{i,j,k}\, m_{q}^{n}\, X_{p} \ Q_{1}^{-2i}Q_{2}^{-2j}Q_{3}^{-2k} \, ,
\end{align}
where the non-zero coefficients are
\begin{align}
\begin{split}
&c^{1,0,8}_{-1,2,3}= 4\,,\ c^{1,0,8}_{0,1,3}= -4\,,\ c^{1,0,8}_{0,2,2}=    -\frac{8}{3}\,,\ c^{1,0,8}_{1,0,3}= -4\,,\ c^{1,0,8}_{1,2,1}=    -\frac{16}{3}\,,\ c^{1,0,8}_{2,-1,3}= 4\,,\ c^{1,0,8}_{2,0,2}=    -\frac{8}{3},\\&c^{1,0,8}_{2,1,1}= -\frac{16}{3},
\end{split}
\end{align}
\begin{align}
\begin{split}
&c^{1,0,7}_{-1,2,3}= -\frac{4}{3}\,,\ c^{1,0,7}_{0,1,3}=    \frac{4}{3}\,,\ c^{1,0,7}_{1,0,3}= \frac{4}{3}\,,\ c^{1,0,7}_{2,-1,3}=    -\frac{4}{3},
\end{split}
\end{align}
\begin{align}
\begin{split}
&c^{1,1,5}_{-1,2,3}= \frac{4}{3}\,,\ c^{1,1,5}_{0,1,3}= -4\,,\ c^{1,1,5}_{1,0,3}=    -4\,,\ c^{1,1,5}_{1,2,1}= -8\,,\ c^{1,1,5}_{2,-1,3}= \frac{4}{3}\,,\ c^{1,1,5}_{2,1,1}=    -8\,,\ c^{1,1,5}_{2,2,0}= -\frac{4}{3},
\end{split}
\end{align}

\begin{align}
\begin{split}
&c^{1,1,4}_{-1,2,3}= -\frac{4}{3}\,,\ c^{1,1,4}_{0,1,3}=    -\frac{4}{3}\,,\ c^{1,1,4}_{1,0,3}= -\frac{4}{3}\,,\ c^{1,1,4}_{1,2,1}=    -8\,,\ c^{1,1,4}_{2,-1,3}= -\frac{4}{3}\,,\ c^{1,1,4}_{2,1,1}= -8\,,\ c^{1,1,4}_{2,2,0}=    \frac{20}{3},
\end{split}
\end{align}

\begin{align}
\begin{split}
&c^{1,1,3}_{-1,2,3}= -\frac{8}{3}\,,\ c^{1,1,3}_{0,1,3}=    \frac{16}{3}\,,\ c^{1,1,3}_{1,0,3}= \frac{16}{3}\,,\ c^{1,1,3}_{1,2,1}= 8\,,\ c^{1,1,3}_{2,-1,3}=    -\frac{8}{3}\,,\ c^{1,1,3}_{2,1,1}= 8\,,\ c^{1,1,3}_{2,2,0}= \frac{4}{3},
\end{split}
\end{align}

\begin{align}
\begin{split}
c^{1,1,2}_{0,1,2}= -4\,,\ c^{1,1,2}_{1,0,2}= -4\,,\ c^{1,1,2}_{1,1,1}= 4
\end{split}
\end{align}

\begin{align}
\begin{split}
&c^{1,3,2}_{-1,2,3}= -\frac{8}{3}\,,\ c^{1,3,2}_{0,1,3}= 8\,,\ c^{1,3,2}_{0,2,2}=    8\,,\ c^{1,3,2}_{1,0,3}= 8\,,\ c^{1,3,2}_{1,2,1}= -8\,,\ c^{1,3,2}_{2,-1,3}=    -\frac{8}{3}\,,\ c^{1,3,2}_{2,0,2}= 8,\\&c^{1,3,2}_{2,1,1}= -8\,,\ c^{1,3,2}_{2,2,0}=    \frac{8}{3},
\end{split}
\end{align}

\begin{align}
\begin{split}
&c^{4,0,8}_{-1,3,2}= -\frac{4}{3}\,,\ c^{4,0,8}_{0,2,2}=    -\frac{8}{3}\,,\ c^{4,0,8}_{0,3,1}= -\frac{4}{3}\,,\ c^{4,0,8}_{1,1,2}=    -\frac{32}{3}\,,\ c^{4,0,8}_{1,2,1}= -\frac{16}{3}\,,\ c^{4,0,8}_{1,3,0}=    -\frac{8}{3}\,,\ c^{4,0,8}_{2,0,2}= -\frac{8}{3},\\&c^{4,0,8}_{2,1,1}=    -\frac{16}{3}\,,\ c^{4,0,8}_{2,2,0}= -\frac{8}{3}\,,\ c^{4,0,8}_{2,3,-1}=    -\frac{8}{3}\,,\ c^{4,0,8}_{3,-1,2}= -\frac{4}{3}\,,\ c^{4,0,8}_{3,0,1}=    -\frac{4}{3}\,,\ c^{4,0,8}_{3,1,0}= -\frac{8}{3}\,,\ c^{4,0,8}_{3,2,-1}=    -\frac{8}{3},
\end{split}    
\end{align}

\begin{align}
\begin{split}
&c^{4,0,7}_{0,2,2}= \frac{4}{3}\,,\ c^{4,0,7}_{2,0,2}= \frac{4}{3},
\end{split}    
\end{align}

\begin{align}
\begin{split}
&c^{4,1,5}_{-1,3,2}= -4\,,\ c^{4,1,5}_{0,2,2}= \frac{4}{3}\,,\ c^{4,1,5}_{0,3,1}=    -4\,,\ c^{4,1,5}_{1,1,2}= -16\,,\ c^{4,1,5}_{1,2,1}= -8\,,\ c^{4,1,5}_{1,3,0}= -4\,,\ c^{4,1,5}_{2,0,2}=    \frac{4}{3},\\&c^{4,1,5}_{2,1,1}= -8\,,\ c^{4,1,5}_{2,3,-1}= -4\,,\ c^{4,1,5}_{3,-1,2}=    -4\,,\ c^{4,1,5}_{3,0,1}= -4\,,\ c^{4,1,5}_{3,1,0}= -4\,,\ c^{4,1,5}_{3,2,-1}= -4,
\end{split}    
\end{align}

\begin{align}
\begin{split}
&c^{4,1,4}_{1,1,2}= -16\,,\ c^{4,1,4}_{1,2,1}= -8\,,\ c^{4,1,4}_{1,3,0}= -4\,,\ c^{4,1,4}_{2,1,1}=    -8\,,\ c^{4,1,4}_{2,2,0}= 8\,,\ c^{4,1,4}_{2,3,-1}= -4\,,\ c^{4,1,4}_{3,1,0}= -4,\\&c^{4,1,4}_{3,2,-1}=    -4,
\end{split}    
\end{align}

\begin{align}
\begin{split}
&c^{4,1,3}_{-1,3,2}= 4\,,\ c^{4,1,3}_{0,3,1}= 4\,,\ c^{4,1,3}_{1,1,2}= 16\,,\ c^{4,1,3}_{1,2,1}=    8\,,\ c^{4,1,3}_{1,3,0}= 4\,,\ c^{4,1,3}_{2,1,1}= 8\,,\ c^{4,1,3}_{2,3,-1}= 4,\\&c^{4,1,3}_{3,-1,2}=    4\,,\ c^{4,1,3}_{3,0,1}= 4\,,\ c^{4,1,3}_{3,1,0}= 4\,,\ c^{4,1,3}_{3,2,-1}= 4,
\end{split}    
\end{align}

\begin{align}
\begin{split}
c^{4,1,2}_{1,1,1}= 8,
\end{split}    
\end{align}

\begin{align}
\begin{split}
&c^{4,3,2}_{0,2,2}= \frac{16}{3}\,,\ c^{4,3,2}_{1,1,2}= -16\,,\ c^{4,3,2}_{1,2,1}=    -8\,,\ c^{4,3,2}_{2,0,2}= \frac{16}{3}\,,\ c^{4,3,2}_{2,1,1}= -8\,,\ c^{4,3,2}_{2,2,0}= 8,
\end{split}    
\end{align}

\begin{align}
\begin{split}
&c^{7,0,8}_{0,3,2}= -\frac{8}{3}\,,\ c^{7,0,8}_{2,3,0}=    -\frac{16}{3}\,,\ c^{7,0,8}_{3,0,2}= \frac{8}{3}\,,\ c^{7,0,8}_{3,2,0}=    -\frac{8}{3},
\end{split}    
\end{align}

\begin{align}
\begin{split}
c^{7,1,5}_{0,3,2}= -8\,,\ c^{7,1,5}_{2,3,0}= -8\,,\ c^{7,1,5}_{3,0,2}= 8\,,\ c^{7,1,5}_{3,2,0}= -8,
\end{split}    
\end{align}

\begin{align}
\begin{split}
c^{7,1,4}_{2,3,0}= -8,
\end{split}    
\end{align}

\begin{align}
\begin{split}
c^{7,1,3}_{0,3,2}= 8\,,\ c^{7,1,3}_{2,3,0}= 8\,,\ c^{7,1,3}_{3,0,2}= -8\,,\ c^{7,1,3}_{3,2,0}= 8,
\end{split}    
\end{align}

\begin{align}
\begin{split}
&c^{17,0,8}_{0,2,3}= -\frac{28}{3}\,,\ c^{17,0,8}_{0,3,2}=    \frac{4}{3}\,,\ c^{17,0,8}_{1,1,3}= -\frac{32}{3}\,,\ c^{17,0,8}_{1,2,2}=    -\frac{16}{3}\,,\ c^{17,0,8}_{2,0,3}= -\frac{28}{3}\,,\ c^{17,0,8}_{2,1,2}=    -\frac{16}{3},\\&c^{17,0,8}_{2,2,1}= \frac{16}{3}\,,\ c^{17,0,8}_{2,3,0}=    -\frac{4}{3}\,,\ c^{17,0,8}_{3,0,2}= \frac{4}{3}\,,\ c^{17,0,8}_{3,2,0}=    -\frac{4}{3},
\end{split}    
\end{align}

\begin{align}
\begin{split}
&c^{17,0,7}_{0,2,3}= \frac{8}{3}\,,\ c^{17,0,7}_{2,0,3}= \frac{8}{3},
\end{split}    
\end{align}

\begin{align}
\begin{split}
&c^{17,1,5}_{0,2,3}= -\frac{8}{3}\,,\ c^{17,1,5}_{1,1,3}= -16\,,\ c^{17,1,5}_{1,2,2}=    -8\,,\ c^{17,1,5}_{2,0,3}= -\frac{8}{3}\,,\ c^{17,1,5}_{2,1,2}= -8\,,\ c^{17,1,5}_{2,2,1}= 8,
\end{split}    
\end{align}

\begin{align}
\begin{split}
&c^{17,1,4}_{0,2,3}= -\frac{4}{3}\,,\ c^{17,1,4}_{0,3,2}= 4\,,\ c^{17,1,4}_{1,1,3}=    -16\,,\ c^{17,1,4}_{1,2,2}= -8\,,\ c^{17,1,4}_{2,0,3}= -\frac{4}{3}\,,\ c^{17,1,4}_{2,1,2}=    -8,\\&c^{17,1,4}_{2,2,1}= 8\,,\ c^{17,1,4}_{2,3,0}= -4\,,\ c^{17,1,4}_{3,0,2}= 4\,,\ c^{17,1,4}_{3,2,0}=    -4,
\end{split}    
\end{align}

\begin{align}
\begin{split}
&c^{17,1,3}_{0,2,3}= \frac{16}{3}\,,\ c^{17,1,3}_{1,1,3}= 16\,,\ c^{17,1,3}_{1,2,2}=    8\,,\ c^{17,1,3}_{2,0,3}= \frac{16}{3}\,,\ c^{17,1,3}_{2,1,2}= 8\,,\ c^{17,1,3}_{2,2,1}= -8,
\end{split}    
\end{align}

\begin{align}
\begin{split}
c^{17,1,2}_{1,1,2}= 8,
\end{split}    
\end{align}

\begin{align}
\begin{split}
&c^{17,3,2}_{0,2,3}= \frac{16}{3}\,,\ c^{17,3,2}_{1,1,3}= -16\,,\ c^{17,3,2}_{1,2,2}=    -8\,,\ c^{17,3,2}_{2,0,3}= \frac{16}{3}\,,\ c^{17,3,2}_{2,1,2}= -8\,,\ c^{17,3,2}_{2,2,1}= 8,
\end{split}    
\end{align}

\begin{align}
\begin{split}
c^{39,0,8}_{0,2,3}= 4\,,\ c^{39,0,8}_{0,3,2}= 4\,,\ c^{39,0,8}_{2,0,3}= 4\,,\ c^{39,0,8}_{2,3,0}=    4\,,\ c^{39,0,8}_{3,0,2}= 4\,,\ c^{39,0,8}_{3,2,0}= 4,
\end{split}    
\end{align}

\begin{align}
\begin{split}
c^{39,1,5}_{0,2,3}= 8\,,\ c^{39,1,5}_{0,3,2}= 8\,,\ c^{39,1,5}_{2,0,3}= 8\,,\ c^{39,1,5}_{2,3,0}=    8\,,\ c^{39,1,5}_{3,0,2}= 8\,,\ c^{39,1,5}_{3,2,0}= 8,
\end{split}    
\end{align}

\begin{align}
\begin{split}
c^{39,1,4}_{0,2,3}= 8\,,\ c^{39,1,4}_{0,3,2}= 8\,,\ c^{39,1,4}_{2,0,3}= 8\,,\ c^{39,1,4}_{2,3,0}=    8\,,\ c^{39,1,4}_{3,0,2}= 8\,,\ c^{39,1,4}_{3,2,0}= 8,
\end{split}    
\end{align}

\begin{align}
\begin{split}
c^{39,1,3}_{0,2,3}= -8\,,\ c^{39,1,3}_{0,3,2}= -8\,,\ c^{39,1,3}_{2,0,3}=    -8\,,\ c^{39,1,3}_{2,3,0}= -8\,,\ c^{39,1,3}_{3,0,2}= -8\,,\ c^{39,1,3}_{3,2,0}= -8,
\end{split}    
\end{align}

\begin{align}
\begin{split}
&c^{54,0,8}_{0,2,3}= -\frac{4}{3}\,,\ c^{54,0,8}_{0,3,2}=    \frac{4}{3}\,,\ c^{54,0,8}_{1,2,2}= \frac{16}{3}\,,\ c^{54,0,8}_{1,3,1}=    \frac{16}{3}\,,\ c^{54,0,8}_{2,0,3}= \frac{4}{3}\,,\ c^{54,0,8}_{2,1,2}=    -\frac{16}{3},\\&c^{54,0,8}_{2,3,0}= \frac{4}{3}\,,\ c^{54,0,8}_{3,0,2}=    -\frac{4}{3}\,,\ c^{54,0,8}_{3,1,1}= -\frac{16}{3}\,,\ c^{54,0,8}_{3,2,0}=    -\frac{4}{3},
\end{split}    
\end{align}

\begin{align}
\begin{split}
&c^{54,1,5}_{0,3,2}= \frac{4}{3}\,,\ c^{54,1,5}_{1,2,2}= 8\,,\ c^{54,1,5}_{1,3,1}=    8\,,\ c^{54,1,5}_{2,1,2}= -8\,,\ c^{54,1,5}_{2,3,0}= \frac{4}{3}\,,\ c^{54,1,5}_{3,0,2}=    -\frac{4}{3},\\&c^{54,1,5}_{3,1,1}= -8\,,\ c^{54,1,5}_{3,2,0}=    -\frac{4}{3},
\end{split}    
\end{align}

\begin{align}
\begin{split}
&c^{54,1,4}_{0,2,3}= -4\,,\ c^{54,1,4}_{0,3,2}=    -\frac{8}{3}\,,\ c^{54,1,4}_{1,2,2}= 8\,,\ c^{54,1,4}_{1,3,1}= 8\,,\ c^{54,1,4}_{2,0,3}=    4\,,\ c^{54,1,4}_{2,1,2}= -8,\\&c^{54,1,4}_{2,3,0}= -\frac{8}{3}\,,\ c^{54,1,4}_{3,0,2}=    \frac{8}{3}\,,\ c^{54,1,4}_{3,1,1}= -8\,,\ c^{54,1,4}_{3,2,0}=    \frac{8}{3},
\end{split}    
\end{align}

\begin{align}
\begin{split}
&c^{54,1,3}_{0,3,2}= -\frac{4}{3}\,,\ c^{54,1,3}_{1,2,2}=    -8\,,\ c^{54,1,3}_{1,3,1}= -8\,,\ c^{54,1,3}_{2,1,2}= 8\,,\ c^{54,1,3}_{2,3,0}=    -\frac{4}{3}\,,\ c^{54,1,3}_{3,0,2}= \frac{4}{3},\\&c^{54,1,3}_{3,1,1}=    8\,,\ c^{54,1,3}_{3,2,0}= \frac{4}{3},
\end{split}    
\end{align}

\begin{align}
\begin{split}
c^{54,1,2}_{1,2,1}= -4\,,\ c^{54,1,2}_{2,1,1}=    4
\end{split}    
\end{align}

\begin{align}
\begin{split}
&c^{54,3,2}_{0,3,2}= -\frac{8}{3}\,,\ c^{54,3,2}_{1,2,2}= 8\,,\ c^{54,3,2}_{1,3,1}=    8\,,\ c^{54,3,2}_{2,1,2}= -8\,,\ c^{54,3,2}_{2,3,0}= -\frac{8}{3}\,,\ c^{54,3,2}_{3,0,2}=    \frac{8}{3},\\&c^{54,3,2}_{3,1,1}= -8\,,\ c^{54,3,2}_{3,2,0}= \frac{8}{3}\, .
\end{split}    
\end{align}

\subsection{Gluon matrix element contributions}
Before taking into account the contributions coming from the mixing with other operators, the contributions have the form
\begin{equation}
\hat{\Pi}_{GG m,S}=X^{6}_S\,e_{q}^{4}\sum_{i,j,k}\left[c_{i,j,k}^{(m)}+f_{i,j,k}^{(m)}\, m_{q}^{-2}+g^{(m)}_{i,j,k}\log\left(\frac{Q_{1}^2}{Q_{2}^{2}}\right)+h^{(m)}_{i,j,k}\log\left(\frac{Q_{3}^2}{m_{q}^{2}}\right)\right] Q_{1}^{-2i}\,Q_{2}^{-2j}\,Q_{3}^{-2k} \, , 
\end{equation}
where

\begin{align}
\begin{split}
&c^{(1)}_{-1,2,3}= -\frac{5}{6}\,,\ c^{(1)}_{0,1,3}= \frac{17}{18}\,,\ c^{(1)}_{0,2,2}=    \frac{4}{9}\,,\ c^{(1)}_{1,0,3}= \frac{17}{18}\,,\ c^{(1)}_{1,1,2}= \frac{2}{3}\,,\ c^{(1)}_{1,2,1}=    \frac{7}{9},\\&c^{(1)}_{2,-1,3}= -\frac{5}{6}\,,\ c^{(1)}_{2,0,2}= \frac{4}{9}\,,\ c^{(1)}_{2,1,1}=    \frac{7}{9}\,,\ c^{(1)}_{2,2,0}= \frac{7}{18},
\end{split}
\end{align}

\begin{align}
\begin{split}
&f^{(1)}_{0,1,2}= \frac{1}{18}\,,\ f^{(1)}_{1,0,2}= \frac{1}{18}\,,\ f^{(1)}_{1,1,1}=    -\frac{1}{18},
\end{split}
\end{align}

\begin{align}
\begin{split}
&g^{(1)}_{-1,2,3}= -\frac{2}{9}\,,\ g^{(1)}_{0,1,3}= \frac{2}{9}\,,\ g^{(1)}_{1,0,3}=    -\frac{2}{9}\,,\ g^{(1)}_{2,-1,3}= \frac{2}{9},
\end{split}
\end{align}

\begin{align}
\begin{split}
&h^{(1)}_{-1,2,3}= \frac{2}{9}\,,\ h^{(1)}_{0,1,3}= -\frac{2}{9}\,,\ h^{(1)}_{1,0,3}=   - \frac{2}{9}\,,\ h^{(1)}_{2,-1,3}= +\frac{2}{9},
\end{split}
\end{align}

\begin{align}
\begin{split}
&c^{(4)}_{-1,3,2}= \frac{7}{18}\,,\ c^{(4)}_{0,2,2}= \frac{5}{6}\,,\ c^{(4)}_{0,3,1}=    \frac{7}{18}\,,\ c^{(4)}_{1,1,2}= \frac{8}{9}\,,\ c^{(4)}_{1,2,1}= \frac{8}{9}\,,\ c^{(4)}_{1,3,0}=    \frac{7}{18},\\&c^{(4)}_{2,0,2}= \frac{5}{6}\,,\ c^{(4)}_{2,1,1}= \frac{8}{9}\,,\ c^{(4)}_{2,2,0}=    \frac{5}{9}\,,\ c^{(4)}_{2,3,-1}= \frac{7}{18}\,,\ c^{(4)}_{3,-1,2}= \frac{7}{18}\,,\ c^{(4)}_{3,0,1}=    \frac{7}{18},\\&c^{(4)}_{3,1,0}= \frac{7}{18}\,,\ c^{(4)}_{3,2,-1}= \frac{7}{18},
\end{split}
\end{align}

\begin{align}
\begin{split}
&f^{(4)}_{1,1,1}= -\frac{1}{9},
\end{split}
\end{align}

\begin{align}
\begin{split}
&g^{(4)}_{0,2,2}= \frac{2}{9}\,,\ g^{(4)}_{2,0,2}= -\frac{2}{9},
\end{split}
\end{align}

\begin{align}
\begin{split}
&h^{(4)}_{0,2,2}= -\frac{2}{9}\,,\ h^{(4)}_{2,0,2}=-\frac{2}{9},
\end{split}
\end{align}

\begin{align}
\begin{split}
&c^{(7)}_{0,3,2}= \frac{7}{9}\,,\ c^{(7)}_{2,3,0}= \frac{7}{9}\,,\ c^{(7)}_{3,0,2}=    -\frac{7}{9}\,,\ c^{(7)}_{3,2,0}= \frac{7}{9},
\end{split}
\end{align}

\begin{align}
\begin{split}
&c^{(17)}_{0,2,3}= \frac{5}{3}\,,\ c^{(17)}_{1,1,3}= \frac{4}{3}\,,\ c^{(17)}_{1,2,2}=    \frac{2}{3}\,,\ c^{(17)}_{2,0,3}= \frac{5}{3}\,,\ c^{(17)}_{2,1,2}= \frac{2}{3}\,,\ c^{(17)}_{2,2,1}=    -\frac{4}{9},
\end{split}
\end{align}

\begin{align}
\begin{split}
&f^{(17)}_{1,1,2}= -\frac{1}{9},
\end{split}
\end{align}

\begin{align}
\begin{split}
&g^{(17)}_{0,2,3}= \frac{4}{9}\,,\ g^{(17)}_{2,0,3}= -\frac{4}{9},
\end{split}
\end{align}

\begin{align}
\begin{split}
&h^{(17)}_{0,2,3}= -\frac{4}{9}\,,\ h^{(17)}_{2,0,3}= -\frac{4}{9},
\end{split}
\end{align}

\begin{align}
\begin{split}
&c^{(39)}_{0,2,3}= -\frac{7}{9}\,,\ c^{(39)}_{0,3,2}= -\frac{7}{9}\,,\ c^{(39)}_{1,2,2}=    -\frac{4}{9}\,,\ c^{(39)}_{2,0,3}= -\frac{7}{9}\,,\ c^{(39)}_{2,1,2}= -\frac{4}{9}\,,\ c^{(39)}_{2,2,1}=    -\frac{4}{9},\\&c^{(39)}_{2,3,0}= -\frac{7}{9}\,,\ c^{(39)}_{3,0,2}= -\frac{7}{9}\,,\ c^{(39)}_{3,2,0}=    -\frac{7}{9},
\end{split}
\end{align}

\begin{align}
\begin{split}
&c^{(54)}_{0,3,2}= -\frac{5}{18}\,,\ c^{(54)}_{1,2,2}= -\frac{2}{3}\,,\ c^{(54)}_{1,3,1}=    -\frac{2}{3}\,,\ c^{(54)}_{2,1,2}= \frac{2}{3}\,,\ c^{(54)}_{2,3,0}= -\frac{5}{18}\,,\ c^{(54)}_{3,0,2}=    \frac{5}{18},\\&c^{(54)}_{3,1,1}= \frac{2}{3}\,,\ c^{(54)}_{3,2,0}= \frac{5}{18},
\end{split}
\end{align}

\begin{align}
\begin{split}
&f^{(54)}_{1,2,1}= \frac{1}{18}\,,\ f^{(54)}_{2,1,1}= -\frac{1}{18},
\end{split}
\end{align}

After including the mixing via~(\ref{eq:wilsreg}), the divergences exactly cancel and the renormalised form factors can be expressed as
\begin{equation}
\hat{\Pi}_{GG m}^{R}=X_{6,R}\,e_{q}^{4}\sum_{i,j,k}\left[c_{i,j,k}^{'(m)}+g^{(m)}_{i,j,k}\log\left(\frac{Q_{1}^2}{Q_{2}^{2}}\right)+h^{(m)}_{i,j,k}\log\left(\frac{Q_{3}^2}{\mu^{2}}\right)\right] Q_{1}^{-2i}\,Q_{2}^{-2j}\,Q_{3}^{-2k} \, , 
\end{equation}
where the associated coefficients $g^{(m)}_{i,j,k}$ and $h^{(m)}_{i,j,k}$ are the same as above and 
\begin{align}
\begin{split}
&c'^{(1)}_{-1,2,3}= -\frac{20}{27}\,,\ c^{'(1)}_{0,1,3}= \frac{20}{27}\,,\ c^{'(1)}_{0,2,2}=    \frac{5}{9}\,,\ c^{'(1)}_{1,0,3}= \frac{20}{27}\,,\ c^{'(1)}_{1,1,2}= \frac{2}{3}\,,\ c^{'(1)}_{1,2,1}=    \frac{1}{9},\\&c^{'(1)}_{2,-1,3}= -\frac{20}{27}\,,\ c^{'(1)}_{2,0,2}=    \frac{5}{9}\,,\ c^{'(1)}_{2,1,1}= \frac{1}{9}\,,
\end{split}
\end{align}

\begin{align}
\begin{split}
&c^{'(4)}_{-1,3,2}= -\frac{1}{18}\,,\ c^{'(4)}_{0,2,2}= \frac{61}{54}\,,\ c^{'(4)}_{0,3,1}= -\frac{1}{18}\,,\ c^{'(4)}_{1,1,2}= -\frac{4}{9}\,,\ c^{'(4)}_{1,2,1}=    \frac{2}{9}\,,\ c^{'(4)}_{1,3,0}= \frac{1}{9},\\& c^{'(4)}_{2,0,2}= \frac{61}{54}\,,\ c^{'(4)}_{2,1,1}=    \frac{2}{9}\,,\ c^{'(4)}_{2,2,0}= \frac{1}{3}\,,\ c^{'(4)}_{2,3,-1}= \frac{1}{9} \,,\ c^{'(4)}_{3,-1,2}=   - \frac{1}{18}\,,\ c^{'(4)}_{3,0,1}= -\frac{1}{18},
\\ & c^{'(4)}_{3,1,0}=    \frac{1}{9}\,,\ c^{'(4)}_{3,2,-1}= \frac{1}{9},
\end{split}
\end{align}

\begin{align}
\begin{split}
&c^{'(7)}_{0,3,2}= -\frac{1}{9}\,,\ c^{'(7)}_{2,3,0}= \frac{2}{9}\,,\ c^{'(7)}_{3,0,2}= \frac{1}{9}\,,\ c^{'(7)}_{3,2,0}= -\frac{1}{9},
\end{split}
\end{align}

\begin{align}
\begin{split}
&c^{'(17)}_{0,2,3}= \frac{89}{54}\,,\ c^{'(17)}_{0,3,2}= -\frac{1}{6}\,,\  c^{'(17)}_{2,0,3}=    \frac{89}{54}\,,\ c^{'(17)}_{2,2,1}=    \frac{2}{9},\\& c^{'(17)}_{2,3,0}= \frac{1}{6}\,,\ c^{'(17)}_{3,0,2}= -\frac{1}{6}\,,\ c^{'(17)}_{3,2,0}=    \frac{1}{6},
\end{split}
\end{align}

\begin{align}
\begin{split}
&c^{'(39)}_{0,2,3}= -\frac{1}{18}\,,\ c^{'(39)}_{0,3,2}= -\frac{1}{18}\,,\ c^{'(39)}_{1,2,2}=    -\frac{4}{9}\,,\ c^{'(39)}_{2,0,3}= -\frac{1}{18}\,,\ c^{'(39)}_{2,1,2}= -\frac{4}{9}\,,\ c^{'(39)}_{2,2,1}=    -\frac{4}{9},\\&c^{'(39)}_{2,3,0}= -\frac{1}{18}\,,\ c^{'(39)}_{3,0,2}=    -\frac{1}{18}\,,\ c^{'(39)}_{3,2,0}= -\frac{1}{18},
\end{split}
\end{align}

\begin{align}
\begin{split}
&c^{'(54)}_{0,2,3}= \frac{1}{6}\,,\ c^{'(54)}_{1,0,0}= -\frac{1}{18}\,,\  c^{'(54)}_{2,0,3}= -\frac{1}{6}\,,\  c^{'(54)}_{2,3,0}= -\frac{1}{18} \,,\ c^{'(54)}_{3,0,2}=    \frac{1}{18}\,,\ c^{'(54)}_{3,2,0}= \frac{1}{18} \, .
\end{split}
\end{align}

\section{Derivation of~(\ref{eq:onecut}) up to $n=3$}\label{app:dertrick}
Up to the order that we need, we have only contributions either without explicit gauge bosons or with one of them, which can be put together owing to~(\ref{eq:phplusgl}). Let us start by writing down the expansion coming from the contributions without gauge bosons (see Fig.~\ref{fig:onecutdiag}a). Starting from~(\ref{eq:backdyson1}) and using the decomposition~(\ref{eq:Ddec}), one trivially finds (up to permutations of the set $P=\{1,2,3\}$):
\begin{equation}
\begin{split}
\Pi^{\mu_1 \mu_2 \mu_3}_\text{NB}=&-e_q^3\sum _A \int\frac{d^4q_3}{(2\pi)^4}\left(\prod_{i=1}^3\int d^{4}x_i\, e^{-iq_i x_i}\right)\langle 0 | \bar{q}(x_1)c_A \Gamma^{A}q(x_3)|\gamma(q_4)\rangle\,  \\
&\times \mathrm{Tr}\Bigg[ \gamma^{\mu_3}\Gamma^{A}\gamma^{\mu_1} i S(x_1-x_2)\gamma^{\mu_2} 
i S(x_2-x_3)\Bigg] .
\end{split}
\end{equation}
Taking Fourier transforms for the propagators, expanding the quark fields according to~(\ref{eq:quarexp}), rewriting the outcoming space time variables $x_{i,\mu}$ as $\lim_{p_{iA}\rightarrow 0}i\frac{\partial}{\partial p_{iA}^\mu} e^{-ip_{iA}x_i}$, integrating and taking derivatives iteratively using~(\ref{eq:dertrick}), one finds:  \begin{align}\begin{split}\label{eq:nobosons}
\Pi^{\mu_1 \mu_2 \mu_3}_\text{NB}=&-e_q^3\sum _A \sum_{m,n}(-1)^{n+m}\langle 0|\bar{q}\{D^{\nu_1},\cdots,D^{\nu_n}\}\{D^{\nu_1'},\cdots,D^{\nu_m'}\}c_A \Gamma^A q |\gamma (q_4)\rangle
\\
&\times  \mathrm{Tr}\Bigg[\gamma^{\mu_3}\Gamma^A\gamma^{\mu_1} iS(-q_1)\gamma^{\nu_1}iS(-q_1)\cdots\gamma^{\nu_n}iS(-q_1)\gamma^{\mu_2}iS(q_3)\gamma^{\nu_1'}iS(q_3)\cdots \gamma^{\nu_m'}iS(q_3)\Bigg]\, ,
\end{split}\end{align}
where $\{\}$ indicates symmetrization (normalized by the number of terms) and $q_3=-q_1-q_2$.

For the topologies with one gauge boson, the only change with respect to~(\ref{eq:nobosons}) is an extra vertex in the quark chain, which can be allocated in two different positions plus the boson field itself (see Fig.~\ref{fig:onecutdiag}b):
\begin{equation}\label{eq:treelevelB}
\begin{split}
    \Pi^{\mu _1 \mu _2 \mu _3}_\text{B}=&-e_q^3\sum _A \int\frac{d^4q_3}{(2\pi)^4}\int d^4 z\left(\prod_{i=1}^3\int d^{4}x_i\, e^{-iq_i x_i}\right)\langle 0 | \bar{q}(x_1)c_A \Gamma^{A}(B_{\epsilon}(z)+ie_{q}A_{\epsilon}(z))q(x_3)|\gamma(q_4)\rangle\ \\
    &\times \left( \mathrm{Tr}\Bigg[ \gamma^{\mu_3}\Gamma^{A}\gamma^{\mu_1} iS(x_1-z)\gamma ^\epsilon iS(z-x_2 )\gamma^{\mu_2} iS(x_2-x_3)\Bigg]\right. \\
    &+ \left. \mathrm{Tr}\Bigg[ \gamma^{\mu_3}\Gamma^{A}\gamma^{\mu_1} iS(x_1-x_2)\gamma^{\mu_2} 
iS(x_2-z)\gamma ^\epsilon iS(z-x_3)\Bigg] \right) .
\end{split}    
\end{equation}
Using~(\ref{eq:phplusgl}), the matrix element in~(\ref{eq:treelevelB}) can be rewritten
\begin{align}
\begin{split}
\bar{q}(x_1)\Big( B_{\epsilon}(u)+ie_{q}A_{\epsilon}(u)\Big) q(x_3)&=\sum _{m,n} \frac{(-x_1)^{\nu_1}\cdots(-x_1)^{\nu_n}}{n!}\frac{x_3^{\nu'_1}\cdots x_3^{\nu'_m}}{m!}\\
&\times \sum_{p=1}\sum_{q=0}^p \frac{(-1)^{p-q+1}p u^{\omega_1}\cdots u^{\omega_p}}{(p+1)q!(p-q)!}
\\&\times \bar{q}D^{\nu_{1}}\cdots D^{\nu_n}D^{\omega_{1}}\cdots D^{\omega_{q}} D^{\epsilon}D^{\omega_{q+1}}\cdots D^{\omega_p} D^{\nu'_1}\cdots D^{\nu'_m}q ,
\end{split}
\end{align}
from which, following the same procedure as above, the contributions from one gauge boson can be re-expressed as
\begin{align}
\begin{split}\label{eq:contboson}
\Pi^{\mu _1 \mu _2 \mu _3}_\text{B}=& -e_q^3\lim_{p_{1A}\rightarrow 0}\lim_{p_{2A}\rightarrow 0}\lim_{p_{3A}\rightarrow 0} \sum _A \sum _{m,n} \frac{(i\partial_{p_{1A}})^{\nu_1}\cdots (i\partial_{p_{1A}})^{\nu_n}}{n!}\frac{(i\partial_{p_{3A}})^{\nu'_1}\cdots (i\partial_{p_{3A}})^{\nu'_m}}{m!}\\
&\times \sum _{p=1} \sum _{q=0}^p \frac{(i\partial_{p_{2A}})^{\omega_1}\cdots(i\partial_{p_{2A}})^{\omega_p}p(-1)^{p-q+1}}{(p+1)q!(p-q)!}
\\&\times \Bigg(\mathrm{Tr} [ \gamma^{\mu_3}\Gamma^A\gamma^{\mu_1}
iS(p_{1}^{A}-q_1)\gamma^{\epsilon}
iS(p_{1}^{A}-p_{2}^{A}-q_1)\gamma^{\mu_2}
iS(q_3+p_3^A) ] \\
&+\mathrm{Tr} [ \gamma^{\mu_3}\Gamma^A\gamma^{\mu_1}
iS(p_1^A-q_1)\gamma^{\mu_2}
iS(p_2^A+p_3^A+q_3)\gamma^{\epsilon}
iS(q_3+p_3^A)
] \Bigg) \\
&\times \langle 0 |\bar{q}D^{\nu_{1}}\cdots D^{\nu_n}D^{\omega_{1}}\cdots D^{\omega_{q}} D^{\epsilon}D^{\omega_{q+1}}\cdots D^{\omega_p} D^{\nu'_1}\cdots D^{\nu'_m}c_A \Gamma^A q | \gamma (q_4) \rangle \, .
\end{split}\end{align}
The next simplification consists in realizing that after taking the derivatives and the limits, all the traces start with $\gamma^{\mu_3}\Gamma^A\gamma^{\mu_1}$ and all the propagator on the left of $\gamma^{\mu_2}$ are 
of the form $S(-q_1)$ and all the propagators on the right are $S(q_3)$, which has a simple diagrammatic interpretation. On the other hand, we can always relabel the dummy Lorentz indices in such a way that the remaining quark current takes as indices $\bar{q}D^{\nu_1}\cdots D^{\nu_{n}}q$. Taking all this into account, any possible term in the sum can be uniquely codified as a pre-factor times a set of numbers separated by a ``wall'' term, $v$. For example, we define
\begin{equation}
\begin{split}
3(31v2)\equiv &3e_q^3\mathrm{Tr}[\gamma^{\mu_3}\Gamma^A\gamma^{\mu_1}iS(-q_1)\gamma^{\nu_3}iS(-q_1)\gamma^{\nu_1}iS(-q_1)\gamma^{\mu_2}iS(q_3)\gamma^{\nu_2}iS(q_3)]\\
&\times \langle 0 |\bar{q}D_{\nu_1}\cdots D_{\nu_{3}} c_A \Gamma^A q|\gamma (q_4)\rangle \, ,
\end{split}
\end{equation}
where we have dropped the index $A$ on the LHS since the Lorentz structure of the various traces does not depend on it. In this symbolic notation, one finds respectively, for $n=0$ ($D=3$) and $n=1$ ($D=4$), using~(\ref{eq:nobosons}),
\begin{align}
\Pi^{\mu_1\mu_2\mu_3}_{D=3}&=-(v) \, ,\\
\Pi^{\mu_1\mu_2\mu_3}_{D=4}&=(1v)+(v1) \, .
\end{align}
From the same equation, the $n=2$ ($D=5$) piece coming from the topology without gauge bosons ($NB$) is
\begin{equation}
\Pi^{\mu_1\mu_2\mu_3}_{D=5,\mathrm{NB}}=-(1v2)-\frac{1}{2}[(v12)+(21v)+(v21)+(12v)]\, ,
\end{equation}
while the bosonic piece $B$ leads, using~(\ref{eq:contboson}), to
\begin{equation}
\Pi^{\mu_1\mu_2\mu_3}_{D=5,\mathrm{B}}=\frac{1}{2}[(21v)-(12v)+(v21)-(v12)]\, .
\end{equation}
Summing, one finds
\begin{equation}
\Pi^{\mu_1\mu_2\mu_3}_{D=5}=-[(1v2)+(v12)+(12v)]\, .
\end{equation}
Finally, for $n=3$, using~(\ref{eq:nobosons}), one finds
\begin{align}\begin{split}
\Pi^{\mu_1\mu_2\mu_3}_{D=6,\mathrm{NB}}=\frac{1}{6}[&(123v)+(132v)+(213v)+(231v)+(312v)+(321v)\\&+(v123)+(v132)+(v213)+(v231)+(v312)+(v321)]\\
&+\frac{1}{2}[(12v3)+(21v3)]+\frac{1}{2}[(1v23)+(1v32)]\, .
\end{split}\end{align}
From~(\ref{eq:contboson}), the contributions of the same order that come from $p=2$ are 
\begin{align}\begin{split}
\Pi^{\mu_1\mu_2\mu_3}_{D=6,\mathrm{B1}}=\frac{1}{3}[&(123v)+(132v)+(213v)+(231v)+(312v)+(321v)\\&+(v123)+(v132)+(v213)+(v231)+(v312)+(v321)]\\
&-(213v)-(231v)-(v312)-(v132) \, ,
\end{split}\end{align}
from $p=n=1$
\begin{align}\begin{split}
\Pi^{\mu_1\mu_2\mu_3}_{D=6\mathrm{B2}}=\frac{1}{2}[(123v)+(213v)+(231v)+(1v32)-(132v)-(312v)-(321v)-(1v23)]\, ,
\end{split}\end{align}
and from $p=m=1$
\begin{align}\begin{split}
\Pi^{\mu_1\mu_2\mu_3}_{D=6,\mathrm{B3}}=-\frac{1}{2}[(v213)+(v321)+(v231)-(v123)-(v321)-(v132)-(12v3)+(21v3)]\, .
\end{split}\end{align}
Summing all of them
\begin{equation}
\Pi^{\mu_1\mu_2\mu_3}_{D=6}=\Pi^{\mu_1\mu_2\mu_3}_{D=6,\mathrm{NB+B1+B2+B3}}=[(123v)+(12v3)+(1v23)+(v123)] \, .
\end{equation}
This simplification occurs for every Dirac structure $\Gamma^A$ and therefore also for their sum. This completes the needed derivation. We conjecture that the duality holds at all dimensions\footnote{We have explicitly checked that it holds for some specific (simpler to prove) higher-order coefficients ($1234v$, $12345v$, $123456v$ and $1234567v$) when two and three gauge bosons are incorporated.}
and that its trivial generalization holds for any number of external legs, greatly simplifying calculations for this kind of topology.


\providecommand{\href}[2]{#2}\begingroup\raggedright\endgroup

\end{document}